\documentclass[11pt,a4paper]{article}
\pdfoutput=1

\usepackage{jcappub}
\usepackage[margin=0.95in]{geometry}

\usepackage[utf8]{inputenc}
\usepackage{graphicx}
\usepackage{verbatim}
\usepackage{epsfig}
\usepackage{subfigure}
\usepackage{color}
\usepackage{multirow}
\usepackage{comment}
\usepackage{slashed}
\usepackage{amsmath}
\usepackage{ulem}
\usepackage{framed}

\usepackage{amssymb,amsmath,amsfonts,mathrsfs}
\usepackage[dvipsnames]{xcolor}
\usepackage{graphicx}
\usepackage{longtable}
\usepackage{verbatim}
\usepackage{color}
\usepackage[mathscr]{euscript}
\usepackage{framed}
\usepackage{mdframed} 

\usepackage{cancel}
\usepackage{color}
\usepackage{hyperref}
\hypersetup{colorlinks, citecolor=bluscuro, linkcolor=black, urlcolor=bluscuro}
\definecolor{rossos}{cmyk}{0,1,1,0.55}
\definecolor{bluscuro}{rgb}{0.15, 0.2, .85}
\definecolor{bluchiaro}{cmyk}{1,.3,0.,0.1}

\numberwithin{equation}{section}
\renewcommand\theequation{\arabic{section}.\arabic{equation}}

\usepackage{tikz}

\usepackage{tcolorbox}

\newcommand{\lp }[0]{\left (}
\newcommand{\rp }[0]{\right )}
\newcommand{\llp }[0]{\left [}
\newcommand{\rrp }[0]{\right ]}

\def\PBH{\text{\tiny PBH}}

\newcommand{\be}{\begin{equation}\begin{aligned}}
\newcommand{\ee}{\end{aligned}\end{equation}}

\newcommand{\bbe}{\begin{align}}
\newcommand{\eee}{\end{align}}

\newcommand{\bea}{\begin{eqnarray}}
\newcommand{\eea}{\end{eqnarray}}

\def\beq{\begin{equation}}
\def\eeq{\end{equation}}

\def\d{{\rm d}}

\def\beqa{\begin{eqnarray}}

	\def\eeqa{\end{eqnarray}}
	
\def\lsim{\mathrel{\rlap{\lower4pt\hbox{\hskip0.5pt$\sim$}}
		\raise1pt\hbox{$<$}}}     
\def\gsim{\mathrel{\rlap{\lower4pt\hbox{\hskip0.5pt$\sim$}}
		\raise1pt\hbox{$>$}}}     

\def\d{{\rm d}}

\def\d{{\rm d}}

\def\PBH{\text{\tiny PBH}}

\newcommand{\vtl}  {\vert \boldsymbol{    {\ell}} \vert}
\newcommand{\vta}[1]{\vert \boldsymbol{\chi}_{#1}    \vert}

\def\eeqa{\end{eqnarray}}

\def\bq{\begin{quote}}
\def\eq{\end{quote}}

 at 10truept
\newcommand{\arXiv}[2]{\href{http://arxiv.org/pdf/#1}{{\tt [#2/#1]}}}
\newcommand{\arXivold}[1]{\href{http://arxiv.org/pdf/#1}{{\tt [#1]}}}

\def\eeqa{\end{eqnarray}}
\def\lsim{\mathrel{\rlap{\lower4pt\hbox{\hskip0.5pt$\sim$}}
  \raise1pt\hbox{$<$}}}     
\def\gsim{\mathrel{\rlap{\lower4pt\hbox{\hskip0.5pt$\sim$}}
  \raise1pt\hbox{$>$}}}     

\title{The Evolution of  Primordial Black Holes and their Final Observable Spins}
\author[a]{V. De Luca,}
\author[a]{G. Franciolini,}
\author[b,c]{P. Pani,}
\author[a,c]{A. Riotto}

\affiliation[a]{
	Department of Theoretical Physics and Center for Astroparticle Physics (CAP) \\
			24 quai E. Ansermet, CH-1211 Geneva 4, Switzerland}
\affiliation[b]{Dipartimento di Fisica, “Sapienza” Università di Roma, Piazzale Aldo Moro 5, 00185, Roma, Italy}
\affiliation[c]{INFN, Sezione di Roma, Piazzale Aldo Moro 2, 00185, Roma, Italy}

\abstract{Primordial black holes in the mass range of ground-based gravitational-wave detectors can comprise a 
significant fraction of the dark matter. Mass and spin measurements from coalescences 
can be used to distinguish between an astrophysical or a primordial origin of the binary black holes. In 
standard  scenarios the spin of primordial black holes is very small at formation. However, the mass and spin can evolve 
through the cosmic history due to accretion. We show that the mass and spin of primordial black holes are correlated 
in a redshift-dependent fashion, in particular primordial black holes with masses below ${\cal O}(30)M_\odot$ are likely 
non-spinning at any redshift, whereas heavier black holes can be nearly extremal up to redshift $z\sim10$. The 
dependence of the mass and spin distributions on the redshift can be probed with future detectors such as the Einstein 
Telescope. The mass and spin evolution affect the gravitational waveform parameters, in 
particular the distribution of the final mass and spin of the merger remnant, and that of the effective spin of the 
binary. We argue that, compared to the astrophysical-formation scenario, a primordial origin of black hole binaries 
might better explain the spin distribution of merger events detected by LIGO-Virgo, in which the effective spin 
parameter of the binary is compatible to zero except possibly for few high-mass events. Upcoming results from LIGO-Virgo third
observation run might reinforce or weaken  these predictions.}
\emailAdd{valerio.deluca@unige.ch}
\emailAdd{gabriele.franciolini@unige.ch}
\emailAdd{paolo.pani@uniroma1.it}
\emailAdd{antonio.riotto@unige.ch}
\begin{document}

\maketitle
\flushbottom

\section{Introduction}
The LIGO-Virgo detection of gravitational waves~(GWs) generated by the coalescence of rather massive binary black 
holes~(BHs)~\cite{Abbott:2016blz, TheLIGOScientific:2016pea, LIGOScientific:2018jsj, LIGOScientific:2018mvr} has 
renewed the interest in understanding the 
physical origin (either astrophysical or primordial~\cite{Bird:2016dcv}) of these binaries.
This has also motivated the idea that a fraction (or all) of the dark 
matter~(DM) in the universe may be composed by Primordial Black Holes~(PBHs) whose formation takes place at primordial 
epochs~\cite{Carr:2009jm,Carr:2016drx, Blinnikov:2016bxu, juan} (see also Refs.~\cite{sasaki, revPBH1,Carr:2020gox} for some 
reviews). For instance, one of the most common scenarios for PBH formation is through the collapse of sufficiently 
sizeable overdensities in the early universe originated by an enhancement in the comoving curvature perturbation power 
spectrum at small scales during the inflationary era~\cite{s1,s2,s3}. In  the absence of  primordial non-Gaussianity, PBHs are  also not initially clustered \cite{cl1,cl2,cl3,dizgah}.

Besides their mass, a particularly relevant property of BHs is their spin. 
Larger spins imply smaller orbital separations at merger, longer inspiral phases, and therefore 
larger integrated fluxes of GWs. Furthermore, the mass and spin distributions carry the footprint of the BH formation 
channels and allow performing ``BH archaeology" by tracing back the physical origin of BHs from their 
observed properties.
For the GW events so far detected~\cite{LIGOScientific:2018mvr} it is possible to 
measure the individual masses, the final mass and spin of the merger remnant, as well as the effective spin of the 
BH binary  (we use $G=c=1$ henceforth)
\be
\chi_{\text{\tiny eff}}=\frac{\vec J_1 /M_1  + \vec J_2 /M_2 }{M_1+M_2}\cdot \hat L, \label{chieffdef}
\ee
where $M_1$ and $M_2$ are the individual BH masses, $\vec J_1$ and $\vec J_2$  are the corresponding 
angular-momentum vectors, and $\hat L$ is the direction of the orbital angular momentum.

From the observational point of view, there is a general tendency for the effective spin parameter 
of merger events detected so far to be compatible with zero~\cite{LIGOScientific:2018jsj,LIGOScientific:2018mvr}, with 
possible few exceptions~\cite{LIGOScientific:2018mvr,Zackay:2019tzo,Venumadhav:2019lyq,Huang:2020ysn} of high-mass 
events (total mass larger than $50\,M_\odot$).
The available data indicates that the dispersion of $\chi_\text{\tiny eff}$ around zero grows with the 
mass~\cite{Safarzadeh:2020mlb}.
Several attempts have been made in order to explain such a tendency. As far as BHs of astrophysical origin are 
concerned, there is still not a firm prediction for their spin distributions. There are currently two major models 
which try to address how astrophysical BHs spins are distributed. The first considers the formation of binaries in a 
shared envelope evolution within galactic fields, for which the final spin and the orbital angular momentum are nearly
aligned~\cite{bl}, although some astrophysical models predict spin misalignment (see, e.g., 
Refs.~\cite{Gerosa:2013laa,Gerosa:2018wbw}).
The second mechanism concerns binaries originated in globular or stellar clusters by dynamical 
capture in the proximity of active galactic nuclei and leads to an isotropic spin distribution centered around 
zero. However, in both scenarios the spin magnitude is not necessarily small (see  Ref. \cite{Safarzadeh:2020mlb} for a 
discussion).
The information on the effective spin parameter of the binary can be used to disentangle between these two 
astrophysical 
regimes~\cite{Rodriguez:2016vmx,Gerosa:2017kvu,Fishbach:2017dwv,Farr:2017uvj}.
In particular it has been claimed that, as long as the spins of the binary BHs are not small, the aligned angular 
distribution is disfavoured~\cite{Farr:2017gtv}, but see Ref.~\cite{Belczynski:2017gds} for some caveats.
Other attempts have also been made to distinguish between the isotropic and aligned distributions
even regardless of the distribution of the intrinsic spins \cite{Vitale:2017cfs}.

As far as a primordial origin is concerned, PBHs produced during a radiation-dominated epoch through the collapse of 
large overdensities are born with dimensionless spins at the percentage level \cite{Mirbabayi:2019uph, DeLuca:2019buf}, 
even though other formation mechanisms 
may lead to larger initial spins, see for instance Refs.~\cite{Cotner:2017tir, Harada:2017fjm}. The fact that 
PBHs are likely produced with small initial spins might suggest a natural explanation for the small observed 
values of the effective spin parameters in binary mergers. 
However, this would be true only if PBHs do not change their spin considerably during their cosmological evolution. 

The goal of this paper is to investigate the evolution of the spins and masses of PBHs which have merged to provide a GW signal and therefore
have formed a binary either in the early or late universe.
We will therefore provide the 
probability distributions for the binary parameters, including $\chi_{\text{\tiny eff}}$, and the final mass and spin 
of the BH merger remnant. 

Several phenomena might affect the spin evolution of PBHs \cite{Berti:2008af}\footnote{Notice that the spin evolution is not accounted for 
in Ref. \cite{profumo}.}. First, PBHs might accrete efficiently during the cosmic history~\cite{Ricotti:2007jk}.  The accretion rate depends strongly on the velocity of the accreting system relative to the 
surrounding gas: it therefore depends on whether the PBH accretes when isolated (i.e., with  
relative velocities smaller than or comparable to the speed of sound in the gas) or in a binary (i.e., with much larger 
relative velocities), and hence on whether the PBH binary forms~\cite{sasaki} (see Appendix \ref{app:formation}) before or after the peak of 
the accretion history, which occurs at $z\lesssim 100$~\cite{Ricotti:2007jk,Ricotti:2007au}.
Furthermore, if PBHs do not comprise the whole DM (as 
recent bounds suggest~\cite{sasaki}) they might accrete an ordinary DM halo which increases their gravitational 
potential, enhancing ordinary gas accretion~\cite{Ricotti:2007au}. This latter effect has been neglected in previous 
studies, but it might be important since the accretion rate can be super-Eddington at redshift $z\sim(10\div 100)$, depending on 
the PBH mass~\cite{Ricotti:2007jk,Ricotti:2007au}. Second, PBHs might undergo multiple mergers during their cosmic history, so that 
the spin distribution of the detected events might be determined by the spin of second-generation mergers rather than 
being natal~\cite{Gerosa:2017kvu}. Finally, the spin of PBHs might decrease due to plasma-driven superradiant 
instabilities~\cite{Pani:2013hpa,Brito:2015oca,Conlon:2017hhi}. This effect depends strongly on the geometry of the 
plasma around the BH and is negligible for realistic systems~\cite{Dima:2020rzg}. For this reason we shall neglect 
plasma-driven superradiant instabilities, whereas in Appendix~\ref{appmer} we show that second-generation mergers 
constitute a negligible fraction of the total PBH binaries in the relevant mass and redshift ranges.
Thus, the main effect driving mass and spin evolution of PBHs is gas accretion, which is reviewed and discussed in 
Sec.~\ref{secacc}.
Based on the results of Refs.~\cite{Ricotti:2007jk,Ricotti:2007au} we discuss how mass accretion can occur at 
super-Eddington rates at redshifts $(10\div 100)$, depending on the PBH mass. This implies that the mass distribution of PBHs 
at low redshift might be significantly different from that at high redshift, and it implies that constraints on the 
PBH abundance based on local measurements should take accretion into account. We investigate this point in a 
forthcoming work~\cite{followup}, whereas here we focus on the evolution of the spin.

Our main result is the computation of a redshift-dependent, sharp, mass-spin distribution for PBHs. In particular, we 
show that PBHs with masses below ${\cal O}(30)M_\odot$ are likely non-spinning at any redshift, whereas heavier BHs can 
be extremal up to redshift $z\sim10$. In Sec.~\ref{secres} we discuss how this affects the distribution of the final 
mass and spin of the BH remnant and the effective spin parameter of the binary. 
Finally, in Sec.~\ref{secconcl} we draw our conclusions and discuss future work.

 
\section{Accretion onto PBHs}
\label{secacc}

Once a PBH is formed in the early universe, one needs to keep track of the mass and angular momentum accretion 
during the cosmological history to describe how the spin evolves up to the present epoch. 
Indeed, depending on the angular momentum of the infalling material, not only the mass, but also the spin changes.

In simple terms, a spherical accretion onto the PBH, where the accreting gas possesses small or negligible angular 
momentum, does not change the initial spin $\vec J$ of the PBH significantly, while the (dimensionless) Kerr 
parameter 
\be
\chi = \frac{|\vec J|}{M^2}
\ee
decreases due to the increase of the PBH mass $M$.
In the opposite case in which the formation of a disk leads to an efficient non-spherical accretion flow, the 
angular momentum of the infalling material induces an enhancement of the PBH spin. 
It is therefore crucial to quantify the mass accretion rate together with the geometry of the accretion flow. In order 
to do so we follow the description of the accretion process provided in Ref.~\cite{Ricotti:2007au} (and references 
therein) and in Ref.~\cite{Ali-Haimoud:2017rtz}. 
The knowledgeable reader can skip this section. Notice that the radiation emitted during accretion and its impact on 
the CMB is used to constrain the PBHs abundance at higher redshifts and higher masses than the ones we consider in this 
paper~\cite{Ali-Haimoud:2016mbv,Horowitz:2016lib, Poulin:2017bwe, Serpico:2020ehh}.

Since the accretion rate depends strongly on the velocity of the accreting system relative to the surrounding gas, we 
need to distinguish two cases: (i)~accretion onto an isolated PBH, where the relative velocity is smaller than or 
comparable to the speed of sound in the gas; (ii)~accretion onto a PBH binary, in which the orbital velocities are 
typically much larger than the speed of sound. Since the accretion rate peaks at $z\lesssim 100$~\cite{Ricotti:2007jk,Ricotti:2007au}, 
the former case is relevant for those PBH binaries formed at smaller redshifts, whereas the latter case is relevant in 
the more likely scenario in which PBH binaries formed at $z\gg100$. In Appendix~\ref{app:formation} we review these two 
formation scenarios for PBH binaries~\cite{sasaki}. 
In the following we shall discuss accretion onto isolated and binary PBHs, separately.

\subsection{Accretion onto isolated PBHs}
\label{secaccisolated}
 
In the case in which binaries form in the present-day halos after accretion is over, one has to consider the evolution of isolated PBHs.
PBHs immersed in the intergalactic medium can experience accretion processes which can change their mass 
considerably. 
An isolated PBH with mass $M$, moving with a relative velocity $v_\text{\tiny rel}$ with respect to the surrounding matter, 
described as an hydrogen gas with number density $n_{\rm gas}$ and sound speed $c_s$, will accrete at the Bondi-Hoyle 
rate given by~\cite{ShapiroTeukolsky,Ricotti:2007jk,Ricotti:2007au}
\be
\label{R1}
\dot{M}_\text{\tiny B} = 4 \pi \lambda m_H n_{\rm gas} v_\text{\tiny eff} r_\text{\tiny B}^2
\ee
in terms of the effective velocity $v_\text{\tiny eff} = \sqrt{v_\text{\tiny rel}^2 
+ c_s^2}$, the Bondi-Hoyle radius
\be
r_\text{\tiny B} \equiv \frac{M}{v_\text{\tiny eff}^2} \simeq 1.3 \times 10^{-4}\lp \frac{M}{M_\odot} \rp 
\lp \frac{v_\text{\tiny eff}}{5.7 {\rm\, km \, s^{-1}}} \rp^{-2}\,  {\rm pc} , \label{rB}
\ee
and the cosmic gas density
\be
n_{\rm gas} \simeq 200 \, {\rm cm}^{-3} \lp \frac{1+z}{1000} \rp^3.
\ee
The sound speed of the gas in equilibrium at the temperature of the intergalactic medium is given by 
\be
c_s \simeq 5.7 \,  \left ( \frac{1+z}{1000}\right)^{1/2}
\llp \left (\frac{1+ z_{\rm dec}}{1+z} \right)^\beta +1 \rrp^{-1/2 \beta}\, {\rm km \, s^{-1}},
\ee
with $\beta = 1.72$, and $z_{\rm dec} \simeq 130$ being the redshift at which the baryonic matter decouples from the 
radiation fluid.
The accretion eigenvalue $\lambda$ keeps into account the effects of the Hubble expansion, the coupling of the CMB 
radiation to the gas through Compton scattering, and the gas viscosity. Its analytical expression can be found in 
Ref.~\cite{Ricotti:2007jk} and, for the reader's convenience, is reported in Appendix~\ref{accretionapp}. 

As we shall see in the following, the accretion effects become noticeable for masses larger than ${\cal O}(10) 
M_\odot$. 
In such a mass range, there are already stringent bounds on the fraction of DM composed by PBHs, see for example 
\cite{Carr:2020gox}. Thus one is forced to consider the accretion onto PBHs in the presence of an additional DM 
component which forms a dark halo of mass $M_h$, truncated at a radius $r_h$ given by (assuming a power law density 
profile $\rho \propto r^{-\alpha}$, with approximately $\alpha \simeq 2.25$)~\cite{Mack:2006gz, Adamek:2019gns}
\be
\label{halo mass}
M_h(z) = 3 M \lp \frac{1+z}{1000} \rp^{-1}, \quad r_h = 0.019 \, {\rm pc} \lp \frac{M}{M_\odot}\rp^{1/3} \lp 
\frac{1+z}{1000} \rp^{-1}.
\ee
The mass $M_h$ grows with time as long as the PBHs are isolated and eventually stops when all the available DM
has been accreted, i.e. approximately when $3 f_\PBH (1+z/1000)^{-1} = 1$.
This DM clothing basically acts as a catalyst enhancing the gas accretion rate. On the other hand, the amount of mass 
accreted due to the infall of the surrounding DM component is 
negligible~\cite{Ricotti:2007jk,zhang}.

One can define a dimensionless accretion rate normalised to the Eddington one
\be
\dot m = \frac{\dot{M}_\text{\tiny B}}{\dot{M}_\text{\tiny Edd}}
\quad\text{with}\quad 
 \dot{M}_\text{\tiny Edd}= 1.44 \times 10^{17} \lp \frac{M}{M_\odot}\rp \rm{g \, s^{-1}}, \label{mdot}
\ee
which is plotted in Fig.~\ref{dotm} and obtained following the procedure reported in Appendix~\ref{accretionapp}, 
including the relevant  estimate of the relative velocity $v_\text{\tiny rel}$. Later on we will discuss in more detail 
the implications of having phases of super-Eddington accretion ($\dot m\gsim 1$) for the formation of a thin 
accretion disk.
\begin{figure}[t!]
\centering
\includegraphics[width=0.55\linewidth]{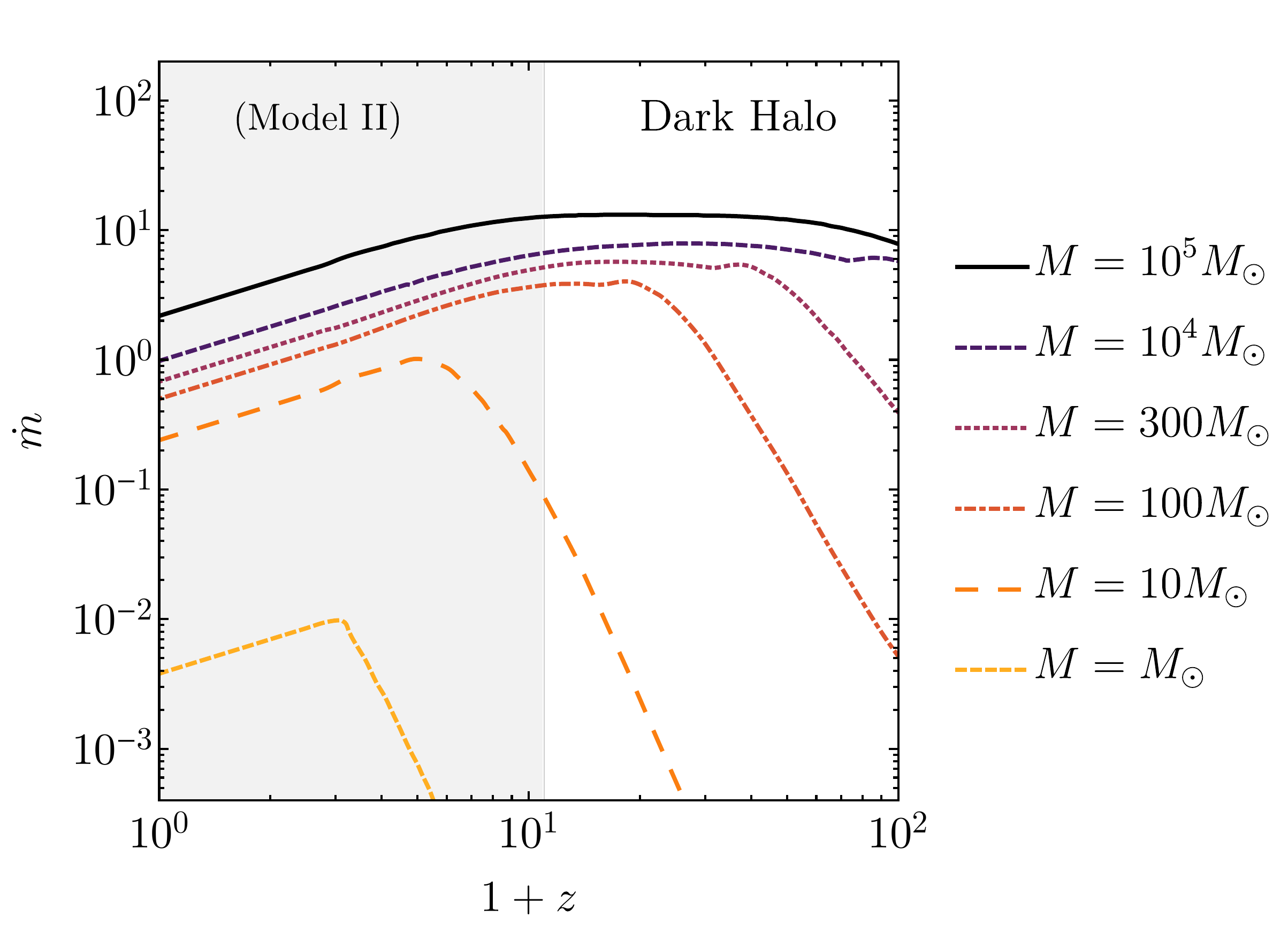}
\caption{\it The dimensionless accretion rate as a function of the redshift for various PBHs masses in the 
scenario in which PBHs are isolated and each of them is surrounded by a dark halo, see 
also~\cite{Ricotti:2007au}. 
The shaded gray region shows the critical redshift ($z=10$) below which the two accretion models discussed in this 
work start differing from each other. This plot shows $\dot m$ for Model~II, whereas in Model~I $\dot m$ sharply 
decreases for $z<10$.
}
\label{dotm}
\end{figure}

\begin{figure}[t!]
    \centering
    \includegraphics[width=0.49 \linewidth]{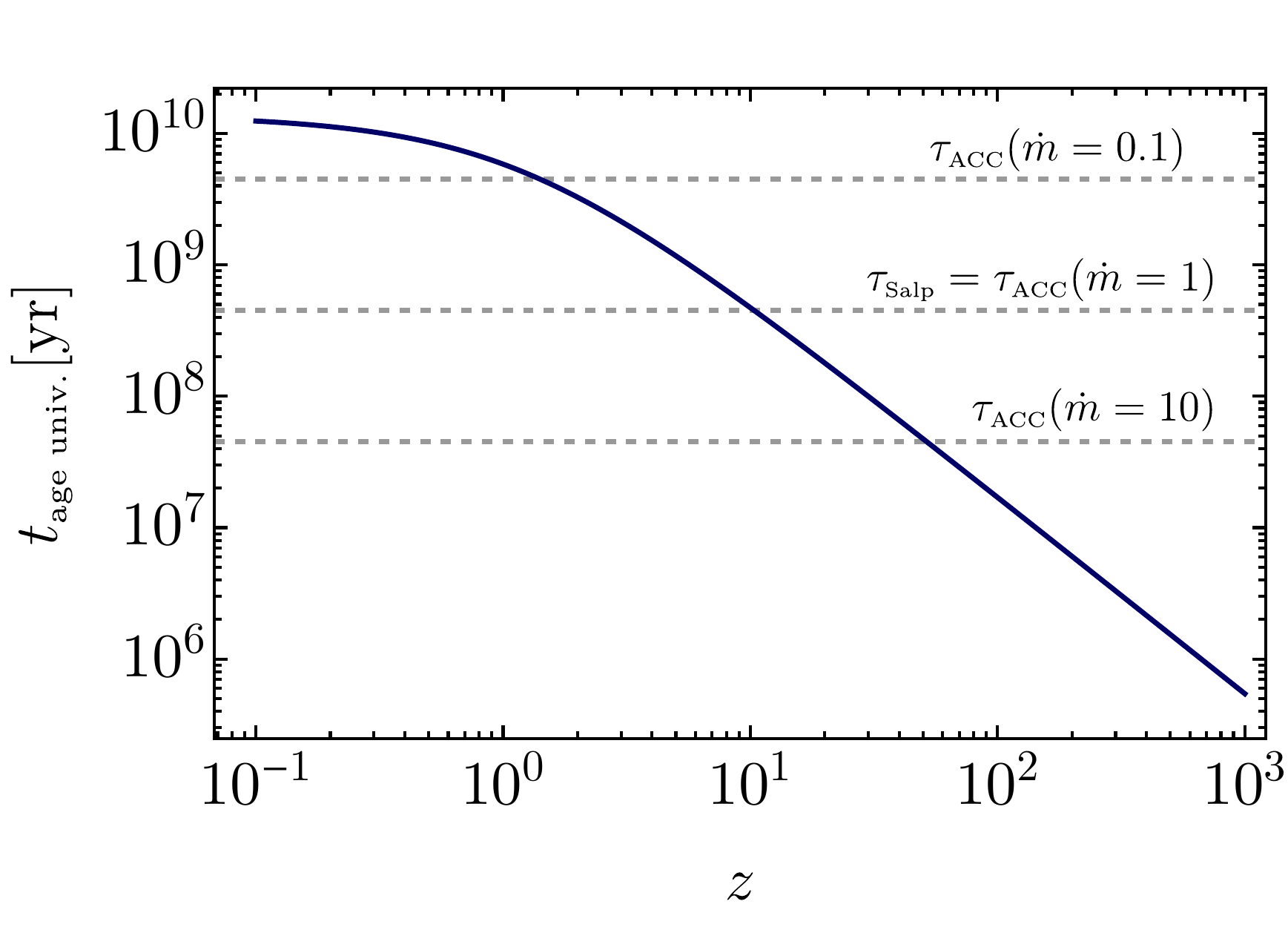}
    \caption{\it Typical accretion time scale $\tau_\text{\rm \tiny ACC} \equiv \tau_\text{\rm \tiny 
Salp}/\dot m $ compared to the age of the universe at a given redshift $z$. The relation between the 
redshift and the age of the universe $t = t(z)$ has been derived within the $\Lambda$-CDM universe.}
    \label{tausalp}
\end{figure}

From Eq.~\eqref{mdot}, the mass accretion can be written as, see for example Ref. \cite{Barausse:2014tra},
\be
\dot M \sim 0.002\, \dot m\,\left( \frac{M(t)}{10^6 M_\odot} \right)M_\odot \,{\rm yr}^{-1}\,.
\ee
The typical time scale for the accretion process is of the order of the Salpeter time, defined as $\tau_\text{\tiny 
Salp}  =\sigma_\text{\tiny T} /4 \pi 
m_\text{\tiny p} =4.5 \times 10^8 \,{\rm yr}$ where $\sigma_\text{\tiny T}$ is the Thompson cross section and 
$m_\text{\tiny p}$ is the proton mass, and it is given by $\tau_\text{\tiny ACC} \equiv \tau_\text{\tiny 
	Salp}/\dot m $. This is compared to the age of the universe at a given 
redshift in Fig.~\ref{tausalp}. 

We focus our attention on redshifts smaller than $z 
\lesssim100$, since at earlier times the typical age of the universe is orders of magnitude smaller than 
$\tau_\text{\tiny ACC}$ even for the largest accretion rates, see 
Fig.~\ref{tausalp}.
At smaller intermediate redshifts, changes in the ionization fraction do not impact 
significantly on the accretion rate and we can neglect the local thermal feedback induced by the X-ray emission 
on the gas temperature and ionization~\cite{Ricotti:2007au}. 

The relative velocity between the PBHs and the baryonic matter starts increasing significantly with the beginning of 
structure formation. A large part of the population of PBHs starts falling in the gravitational potential well of 
large-scale structures after redshift around $z\simeq 10$ and experiences an increase of the relative velocity up to 
one order of magnitude, with a consequent potentially large suppression of the accretion rate~\cite{Ricotti:2007au, Ali-Haimoud:2017rtz,raidalsm}. 
Given the fact that a detailed description of the dynamics of the PBHs population is still lacking (see however Ref. 
\cite{Inman:2019wvr} for a recent attempt in this direction), 
and due to the uncertainties in modelling the global thermal feedback and the change in the relative velocity due to 
the structure formation\footnote{We thank M. Ricotti for several discussions about this point.}, we have decided to 
consider two opposite and extreme scenarios. In one case (dubbed as Model~I) the accretion drastically decreases after 
redshift $z\sim 10$, to model the fact that structure formation as well as reionization may strongly suppress the accretion rate.
 In the other case (dubbed as Model~II) we assume a different scenario, and maybe somewhat extreme, where the impact of structure formation and reionization is limited: a moderate 
accretion is maintained up to very low redshifts. In this case we compute $\dot m$ down to redshift $z\lsim 3$ by neglecting 
structure formation and following the procedure of Ref.~\cite{Ricotti:2007au} (see Appendix~\ref{accretionapp}) until 
the evolution is monotonic, and then extrapolate the behaviour of $\dot m$ down to lower redshifts, see Fig.~\ref{dotm}. 
In the following, we will evaluate the PBH spin evolution in both models, although we remark that Model~I is more realistic.

\subsubsection{Formation of an accretion disk around an isolated PBH}
The infalling accreting gas onto a PBH can carry angular momentum which crucially determines the geometry of the 
accreting flow, the possible formation of an accreting disk, and eventually the
PBH spin evolution.

One can start from the expression for the baryon velocity variance provided in Ref.~\cite{Ricotti:2007au} as
\be
\label{fit50}
\sigma_{\text{\tiny b}} \simeq \sigma_{\text{\tiny b},0} \xi^{-1.7}(z) \lp \frac{1+z}{1000} \rp^{-1} \lp \frac{M_h}{M_\odot} \rp^{0.85},
\ee
where $\sigma_{\text{\tiny b},0} = 3.8 \times 10^{-7} {\rm km \, s^{-1}}$ and $\xi(z)= {\rm Max}[1, \langle 
v_\text{\tiny eff} 
\rangle/c_s]$ describes the effect of a (relatively small) PBH proper motion in reducing the Bondi radius. Then,
if the typical gas velocity is smaller than the Keplerian velocity close to the PBH, the accretion geometry is 
quasi-spherical. In other words, a disk can form only in the opposite regime, i.e. if
\be
\sigma_{\text{\tiny b}} \gsim 2 D \xi^2 (z) c_s^2.
\ee
The constant $D \sim \mathcal{O}(1) \div {\cal O}(10)$ takes into account relativistic corrections.
Using the above condition and Eq.~\eqref{halo mass}, we can estimate the minimum PBH mass for which the accreting gas 
acquires a disk geometry,
\be
\label{critM}
M \gsim 6\times 10^2 M_\odot \,D^{1.17} \xi^{4.33}(z) \frac{\lp 1+z/1000\rp^{3.35}}{\llp 1 + 0.031 
\lp1+z/1000\rp^{-1.72} \rrp^{0.68}}.
\ee
The angular momentum of the accreting DM is typically much smaller than the one of the gas and thus does not 
lead to the formation of a DM disk, while it has an impact on the density profile of the dark halo which 
envelops the PBH.

The condition in Eq.~\eqref{critM} is still not sufficient to describe the formation of a thin disk. Indeed, such a 
formation happens in the case accretion is sufficiently efficient. Following Ref.~\cite{Ricotti:2007au}, we assume that 
a thin disk forms when Eq.~\eqref{critM} is satisfied and
\be
\label{a}
\dot m \gsim 1.
\ee
The formation of a disk, while enhancing the mass accretion, leads also to an efficient 
spin-up of the PBH, as we will discuss in the following.

One can numerically check that the condition~\eqref{a} is always more stringent than 
condition~\eqref{critM}. This implies that $\dot m\gsim1$ can be considered as the sufficient condition for the formation of a thin disk around an isolated PBH.

\subsection{Accretion onto binary PBHs}
\label{secaccbinary}
In the case in which binaries form in the early universe before accretion starts, one has to consider the evolution of PBHs within a binary.
In such a case, one has to take into account both {\it global} accretion processes 
(i.e., of the binary as a whole) and {\it local} accretion processes (i.e., onto the individual components of the 
binary). 

The center of mass of the binary moves with typical velocity $v_\text{\tiny rel}$ as defined in the previous section. 
Therefore, the Bondi-Hoyle radius of the binary is the same as that defined in Eq.~\eqref{rB} simply replacing $M$ with 
$M_\text{\tiny tot}$, where $M_\text{\tiny tot}$ is the total mass of the binary.

When the orbital separation $r_\text{\tiny orb}$ is larger than the Bondi-Hoyle radius of the binary, accretion onto 
the binary is negligible. However, when the binary is contained in its own Bondi-Hoyle radius, it will accrete at a 
rate given by Eq.~\eqref{R1} (with the replacement $M\to M_\text{\tiny tot}$ in the definition of $r_\text{\tiny B}$), 
namely
\be
\label{R1bin}
\dot{M}_\text{\tiny B}^\text{\tiny bin} = 4 \pi \lambda m_H n_{\rm gas} v_\text{\tiny eff} (r_\text{\tiny B}^\text{\tiny 
bin})^2\,, \qquad r_\text{\tiny B}^\text{\tiny 
bin}\equiv \frac{M_\text{\tiny tot}}{v_\text{\tiny eff}^2}.
\ee
Within the Bondi-Hoyle radius the matter falling in the gravitational potential of the object per unit time is 
constant~\cite{Ali-Haimoud:2017rtz}, i.e. locally each PBH accretes at the rate given in Eq.~\eqref{R1bin}.

Therefore, in this case the normalized accretion rate is the same as that defined in Eq.~\eqref{mdot}, modulo a factor 
$(M_\text{\tiny tot}/M)^2$ which accounts for the fact that the accretion flow is driven by the binary.
For equal mass binaries, this correction would increase the accretion rate by a factor $4$. For unequal mass 
binaries, it would increase much more the accretion rate onto the smaller binary component relative to the larger 
one.
In order to perform a common analysis of the cases of accretion onto isolated and binary PBHs, in the following we 
shall neglect this correction factor and assume that each PBH (either isolated or in a binary) accretes at the rate 
given in Eq.~\eqref{mdot}. This is a conservative assumption, since in the binary case the accretion rate can be larger and 
in any case within the uncertainties of the physics of the accretion.

\subsubsection{Formation of an accretion disk around a  PBH in a binary}

Locally, each PBH in a binary has a typical velocity given by orbital one, $v_\text{\tiny orb}=\sqrt{M_\text{\tiny tot}/r_\text{\tiny 
orb}}$. For orbital separations smaller than the Bondi-Hoyle radius of the binary, the orbital velocity is always much 
larger than $v_\text{\tiny rel}$ and $c_s$. This has an important impact in the angular momentum transferred during 
the accretion.

Indeed, let us consider an element of gas at the Bondy-Hoyle radius of one of the individual PBHs of the binary. This 
is given by Eq.~\eqref{rB} with the substitution $v_\text{\tiny eff}\to v_\text{\tiny orb}$, i.e.
\begin{equation}
 r_\text{\tiny B}^\text{\tiny local}\equiv \frac{M}{v_\text{\tiny orb}^2}\,, \label{rBlocal}
\end{equation}
since now the relative velocity is $\sim v_\text{\tiny orb}\gg c_s$. For the same reason, in the reference frame of the 
accreting BH, the typical velocity of the gas element is of the order of the Keplerian velocity at $r_\text{\tiny 
B}^\text{\tiny local}$, i.e. $v_\text{\tiny B}=\sqrt{M_\text{\tiny tot}/r_\text{\tiny B}^\text{\tiny local}}$. Thus, the angular 
momentum per unit mass of the gas element is $v_\text{\tiny B} r_\text{\tiny B}^\text{\tiny local} = \sqrt{M_\text{\tiny tot} 
r_\text{\tiny B}^\text{\tiny local}}$, and is conserved along the accretion flow.
As previously discussed, a necessary condition for the formation of the accretion disk is that the specific 
angular momentum of the gas element be larger than the specific angular momentum at the innermost stable circular 
orbit~(ISCO) of the PBH~\cite{ShapiroTeukolsky}. The latter is $v_\text{\tiny ISCO} r_\text{\tiny ISCO}= 
\sqrt{M_\text{\tiny tot} r_\text{\tiny ISCO}}$. Since $r_\text{\tiny B}^\text{\tiny local}\gg r_\text{\tiny ISCO}$ the necessary 
condition for the formation of a disk is always satisfied in this case.

Thus, compared to the case of an isolated PBH discussed above, in this case condition~\eqref{critM} is absent, and the 
only condition for the formation of a thin accretion disk is Eq.~\eqref{a}.

To summarize, for both isolated and binary PBHs we can assume that a thin accretion disk forms whenever $\dot m\gtrsim 
1$ along the cosmic history.

\section{The spin evolution}
In order to estimate the PBH spin at the present epoch, we 
need to consider the spin inherited by the formation dynamics and how it evolves throughout the cosmological history.
In this section we first briefly review the initial conditions in a standard formation scenario during the 
radiation-dominated era in which the PBH is formed through the collapse
of the perturbations generated during an inflationary epoch upon horizon re-entry~\cite{s1,s2,s3}. We will also review 
the spin evolution in the presence of a thin accretion disk. Indeed, once a thin disk of accreting gas is formed around 
the PBH, the accreted baryonic material significantly affects the spin of the object over a time scale $\tau_\text{\tiny 
ACC}$.

\subsection{Initial conditions}
In this subsection we review the physics underlying the formation of the spin of the PBHs from collapse of density 
perturbations in the radiation-dominated epoch\footnote{During the radiation-dominated phase, the relation between the 
PBH mass and the radiation temperature is  $M\simeq M_\odot(T_\text{\tiny QCD}/T_\text{\tiny GeV})^2$ where $T_\text{\tiny 
QCD}\simeq 10^2$ 
MeV and $T_\text{\tiny GeV}$ is the temperature measured in GeV.}~\cite{s1,s2,s3},  following the results of 
Refs.~\cite{Mirbabayi:2019uph,DeLuca:2019buf}.
The PBHs mass fraction is bounded by the requirement that the cosmological abundance is less than the DM 
abundance. This requires the collapse of density perturbations generating a PBH to be a rare event. Using the peak 
theory formalism \cite{bbks} one finds that high (and rare) peaks in the density contrast, which eventually 
collapse to form PBHs, tend to possess a spherical shape. However, at first order in perturbation theory, the presence 
of small asphericities allows for the action of torques induced by the surrounding matter perturbations, which leads to 
the generation of a small angular momentum before collapse. The action of the torque moments is 
indeed limited in time due to the small time scales characterising the overdensity collapse.

The estimated PBH spin at formation is~\cite{DeLuca:2019buf} 
\be
\chi_{\text{\tiny form}} = \frac{\Omega_\text{\tiny m}}{\pi} \sigma_\delta \sqrt{1-\gamma^2} \sim 10^{-2} 
\sqrt{1-\gamma^2},
\ee
where $\Omega_\text{\tiny m}$ represents the DM abundance, $\sigma_\delta$ indicates the variance of the 
density perturbations at the horizon crossing time, and $\gamma$ parametrises the shape of the power spectrum of the 
density perturbations in terms of its variances (being $\gamma=1$ for a monochromatic power spectrum). The initial spin of the PBHs is therefore expected to be below the percent level.  PBH formation in non-standard scenarios, like during an early matter-dominated epoch~\cite{Harada:2017fjm} 
following inflation or from Q-balls~\cite{Cotner:2017tir}, may lead to larger values of the initial spin.

\subsection{The spin dynamics}
When the conditions for formation of a thin accretion disk are satisfied -- namely, the inequalities (\ref{critM}) and 
(\ref{a})-- mass accretion is accompanied by an increase of the PBH spin. 
The accreting disk is responsible for the angular momentum acquired by the initially slowly rotating PBH and thus the 
PBH spin can be safely assumed to be rapidly aligned perpendicularly to the disk plane. In such a configuration, 
one can use a geodesic model to describe the disk~\cite{Bardeen:1972fi}. The gas of rest 
mass $\d M_0$ falling in the PBH from the last stable orbit gives rise to an increase in total gravitational mass $\d M 
= E(M,J) \d M_0$, together with an increase in the magnitude of the angular momentum $\d J = L(M,J) \d M_0 $, where the 
energy and angular momentum for unit mass are given by~\cite{Bardeen:1972fi}
\be
E(M,J) = \sqrt{1- 2 \frac{M}{3 r_\text{\tiny ISCO}}}
\qquad 
\text{and}
\qquad
L(M,J) = \frac{2 M }{3 \sqrt{3} } \lp 1+ 2 \sqrt{ 3 \frac{r_\text{\tiny ISCO} }{M }-2}\rp\,,
\ee
where the ISCO radius 
$r_\text{\tiny ISCO}(M,J)$ is written in terms of the mass $M$ and dimensionless Kerr parameter $\chi$ as 
\be
r_\text{\tiny ISCO}(M,J) = M \llp 3 + Z_2 - \sqrt{\lp 3-Z_1\rp \lp 3+Z_1+2 Z_2\rp } \rrp,
\ee
with
\be
Z_1= 1+ \lp 1- \chi^2 \rp ^{1/3} \llp \lp 1+\chi\rp^{1/3}+\lp 1-\chi\rp^{1/3} \rrp 
\qquad 
\text{and}
\qquad
Z_2= \sqrt{3 \chi^2 + Z_1^2}.
\ee
Finally, for circular disk motion the rate of change of $|\vec J|$ is related to the mass accretion rate by the 
relation (see also Refs. \cite{thorne,Brito:2014wla,volo})
\be\label{spinchange}
\dot {J} = \frac{L(M,J)}{E(M,J)} \dot {M},
\ee
which can be re-arranged to describe the time evolution of the Kerr parameter
\be
\dot \chi = \lp {\cal F} (\chi) - 2 \chi \rp \frac{\dot {M}}{M},
\ee
where we have defined the combination 
\be
{\cal F} (\chi) \equiv \frac{L(M,J)}{M E(M,J)},
\ee
which is only a function of the dimensionless Kerr parameter.
\begin{figure}[t!]
    \centering
    \includegraphics[width=0.45\linewidth]{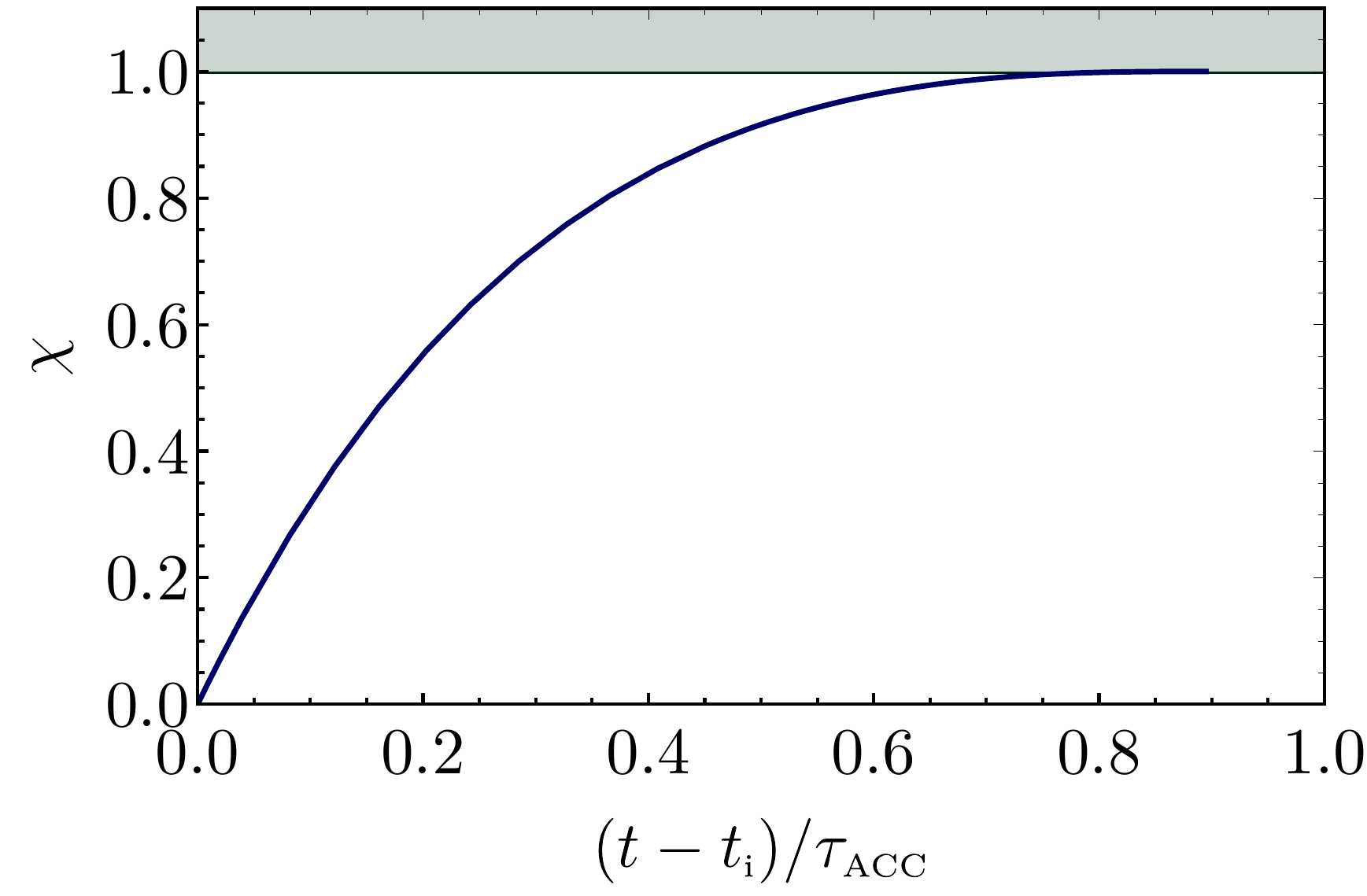}
    \caption{\it
    Evolution of the dimensionless Kerr parameter $\chi$ as a function of time (normalised by a typical accretion time 
scale, $\tau_\text{\rm \tiny ACC}$) within a thin-disk accretion model. The plot is cut at $\chi = 0.998$, the 
maximum allowed value when radiation effects are taken into consideration~\cite{thorne}. The value of the spin attained 
at the present epoch critically depends on the time scale of the process, $\tau_\text{\rm \tiny ACC} \equiv 
\tau_\text{\tiny Salp}/\dot m$.
    }
    \label{rp}
   \end{figure}
   As one can see from Fig.~\ref{rp}, the evolution of $\chi$ is quite rapid in terms of the typical accretion 
time scales and reaches the maximum value (allowed if one considers radiation effects~\cite{thorne}) of $\chi \simeq 
0.998$ in less than an $e$-folding time $\tau_\text{\tiny ACC}$\footnote{Magnetohydrodynamic simulations of accretion 
disks around Kerr BHs suggest that the maximum spin might be slightly smaller, $\chi \simeq 
0.9$~\cite{Gammie:2003qi}. However, this limit may not apply to geometrically thin disks and --~in any case~-- the spin 
evolution time scale does not change significantly in more realistic models~\cite{Gammie:2003qi}.}. 
   
Thus, whenever a thin accretion disk is formed, the dimensionless Kerr parameter grows 
efficiently from a small initial value until it reaches (almost) extremality. As we shall discuss, this gives rise to a 
rapid transition between the two regimes (of small and large values of $\chi$, respectively), depending on 
whether the conditions for thin-disk accretion are satisfied during the cosmological evolution of the PBH.

\subsection{Imprints of second-generation mergers}
Before turning our attention to the quantitative results which will be outlined in the following section, we point out the irrelevance of second-generation mergers for the PBH spin evolution. 
Secondary mergers are those in which at least one of the two components of the binary results from the merger of a 
previous binary system~\cite{Gerosa:2017kvu}.
In this case the spin of the secondary binary is determined 
by the masses and spins of the older binary (one can think about the simple case in which the merger of two PBHs with 
zero spin produces a PBH with $\chi\approx 0.68$, which eventually forms another binary that merges  in the LIGO-Virgo band).
In such a case, the spin of the PBHs participating to the observed merger is mostly determined by the previous merger, 
rather than by the dynamics studied in this paper.
However, as explicitly investigated in Appendix~\ref{appmer}, the probability of occurrence of such a secondary merger 
process is almost negligible for the range of masses and redshifts of interest (the probability of third- or 
higher-generation mergers is even smaller), and therefore we can safely ignore the 
impact of secondary mergers on the spin of PBHs. 
 

\section{Results}
\label{secres}
With the theoretical framework described in the previous sections in hands, we can now discuss the evolution of the 
masses and spins of PBHs and their impact for GW astronomy. We will first address the spin evolution from its initial 
value up to the present epoch and we will subsequently compute the probability distributions of the effective spin 
parameter of the 
binary, as well as the distribution of the mass and spin of the merger remnant.

At this point we stress that when presenting the results, the mass of the PBHs will always refer to the mass at detection --~which is the only 
measurable quantity~-- not to the initial mass at formation.

\subsection{The spin of PBHs as a function of mass and redshift}

In Figs.~\ref{Spin-I} and~\ref{Spin-II} we present the evolution of the mass accretion rate (left panels) and spin 
(right panels) in the $(M,z)$ plane, starting from the initial conditions set at high 
redshifts (we assume initial conditions at $z\sim 100$) for Model~I and Model~II, respectively. 
On the contour plots we also superimpose the curves representing the maximum 
distance current and future GW experiments like aLIGO and Einstein Telescope~(ET)~\cite{Hild:2010id} may reach at a 
given redshift. We take such maximum distance to be the corresponding visible horizon. 

In the mass range of interest no significant evolution of the mass (and, correspondingly, of the spin) takes place 
before redshift $z\sim 30$. This is due to the long time scales characterising the accretion compared to the age of the 
universe up to that epoch. 
After $z\sim 30$, PBHs masses start evolving rapidly for $M\gsim {\cal O}(30) M_\odot$.
In the left panels of Figs.~\ref{Spin-I} and~\ref{Spin-II} we show the trajectories (black dashed 
lines) that a PBH with a certain initial mass would follow during the cosmic history. Correspondingly, we 
observe that the spins of PBHs with masses $M\gsim {\cal O}(30) M_\odot$, even if they start 
from an initial value at the percent level, make a rapid transition to extremality if, during its evolution, the system 
enters a region where $\dot m \gsim 1$.
It is worth noting that the transition region is sensitive to the magnitude of $\dot m$ and to the actual value of the 
redshift at which structure formation and reionization take place. In particular, increasing $\dot m$ or delaying the 
reionization epoch would push the transition region to lower masses.

In the following, we describe in details the evolution of mass and spin after redshift $z\sim 10$ separately for 
Model~I and Model~II. The implications regarding such a prediction for GW detections with aLIGO and ET are discussed in 
Sec.~\ref{sec:GWs}.

\subsubsection{Model~I}
\begin{figure}[h!]
    \centering
    \includegraphics[width=0.499 \linewidth]{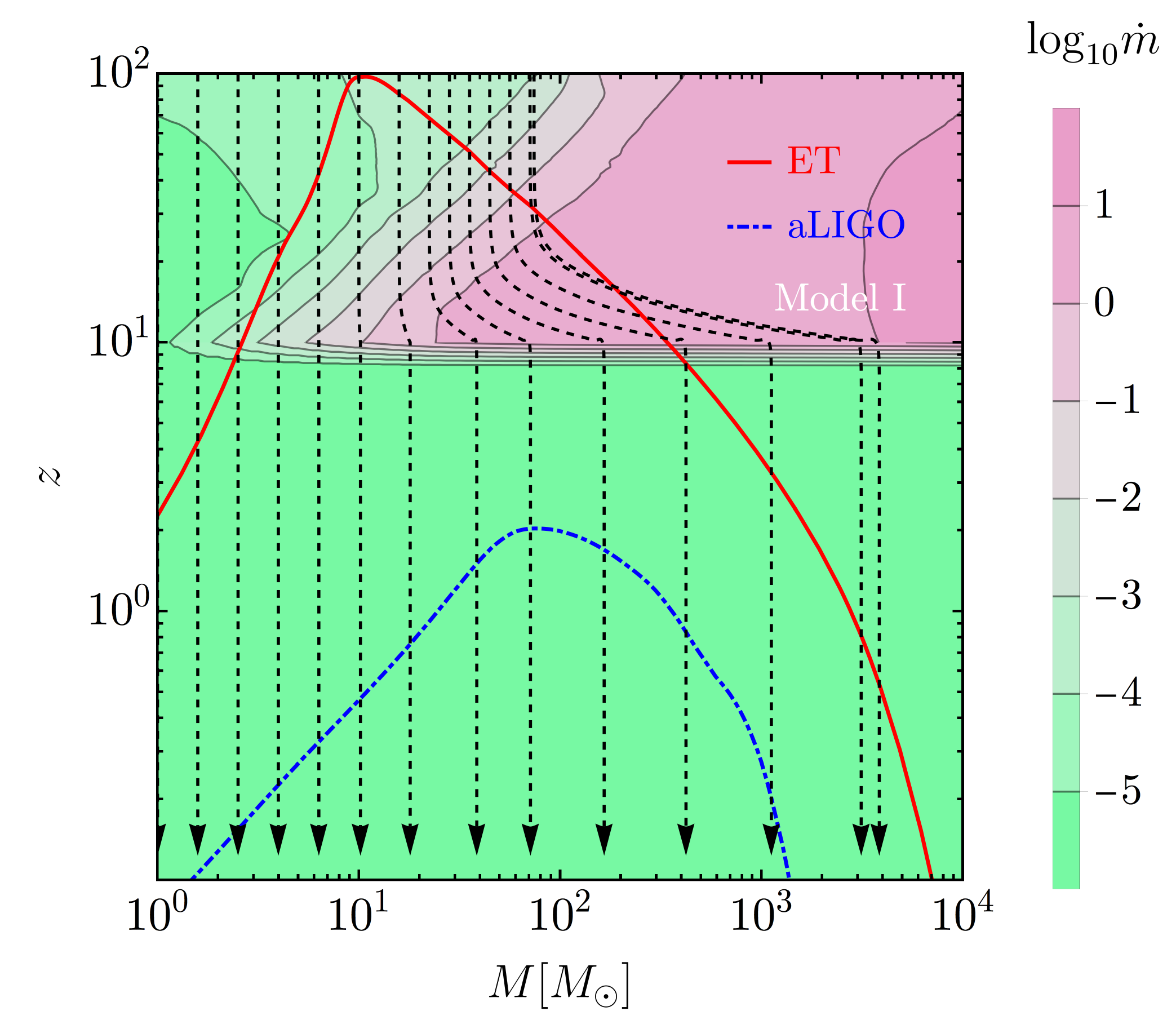}
    \hspace{.1 cm}
    \includegraphics[width=0.479 \linewidth]{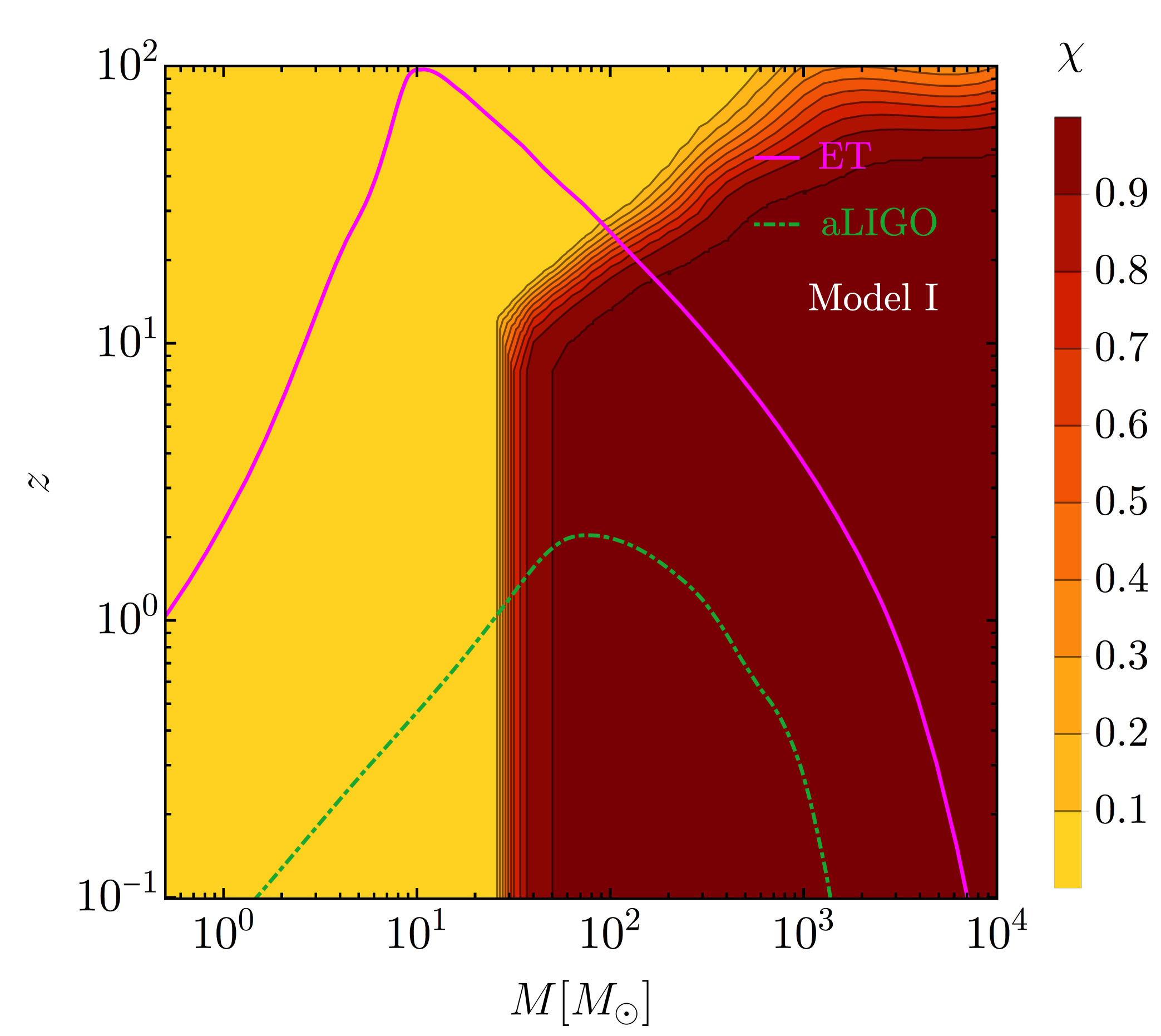}
    \caption{\it 
    { Left:} The accretion rate parameter $\dot m$ as a function of the mass of PBHs and redshift for our Model~I. In 
black we show the trajectories of individual PBHs in the $(M,z)$ plane. 
Model I assumes a sharp transition to a regime of negligible accretion after redshift $z=10$ due to the effect of 
structure formation. 
    { Right:} Evolution of the spin $\chi$ in the $(M,z)$ plane. In both panels we have superimposed the corresponding 
horizons for aLIGO and ET.}
    \label{Spin-I}
   \end{figure}

This model assumes a sharp decrease of the mass accretion rate after $z\sim 10$. As one can appreciate from 
Fig.~\ref{Spin-I} (left panel), each individual PBH starts following vertical trajectories in the $(M,z)$ plane after 
that redshift. This shows that the mass evolution is negligible in that region. Correspondingly, the spin stops 
evolving after that epoch, see right panel of Fig.~\ref{Spin-I}. 
We note a correlation between low (high) values of the masses and low 
(high) spins, with a sharp transition around $M\sim {\cal O} (30) M_\odot$.
More specifically, PBHs with masses below ${\cal O}(30)M_\odot$ are non-spinning, whereas heavier PBHs can 
be nearly extremal up to redshift $z\sim10$ for $M\sim 200\,M_\odot$, and even to higher redshifts for heavier PBHs, 
although that region will be outside the horizon of ET. One can notice that in this region high values of the spin are 
not reached for values of redshift higher than $z \gsim 10^2$ due to the large accretion time scales with respect to the 
age of the universe at that epoch.

\subsubsection{Model~II}
This model assumes a sustained accretion after redshift $z\sim 10$. 
At variance with Model~I, the evolution of the mass and spin proceeds after that redshift, with a 
significant increase also of the smaller masses. 

The transition region between small and high values of the spin after redshift $z\sim 10$ is pushed to higher masses respect to model~I as 
now  PBHs with those masses have never experienced a period of thin disk accretion. 
Also, for masses smaller than $\sim 10\, M_\odot$, the spherical accretion, while leaving $|\vec J|$ unaffected, 
decreases the Kerr parameter $\chi$, thus erasing any memory of the initial spin. We finally note that --~in the 
region in which $\dot m$ is bigger than unity~-- an extremal value of the spin is always rapidly attained.
One can also appreciate that in this model PBHs within the aLIGO horizon are expected to be slowly spinning with small 
values of $\chi.$
\begin{figure}[h!]
    \centering
    \includegraphics[width=0.499 \linewidth]{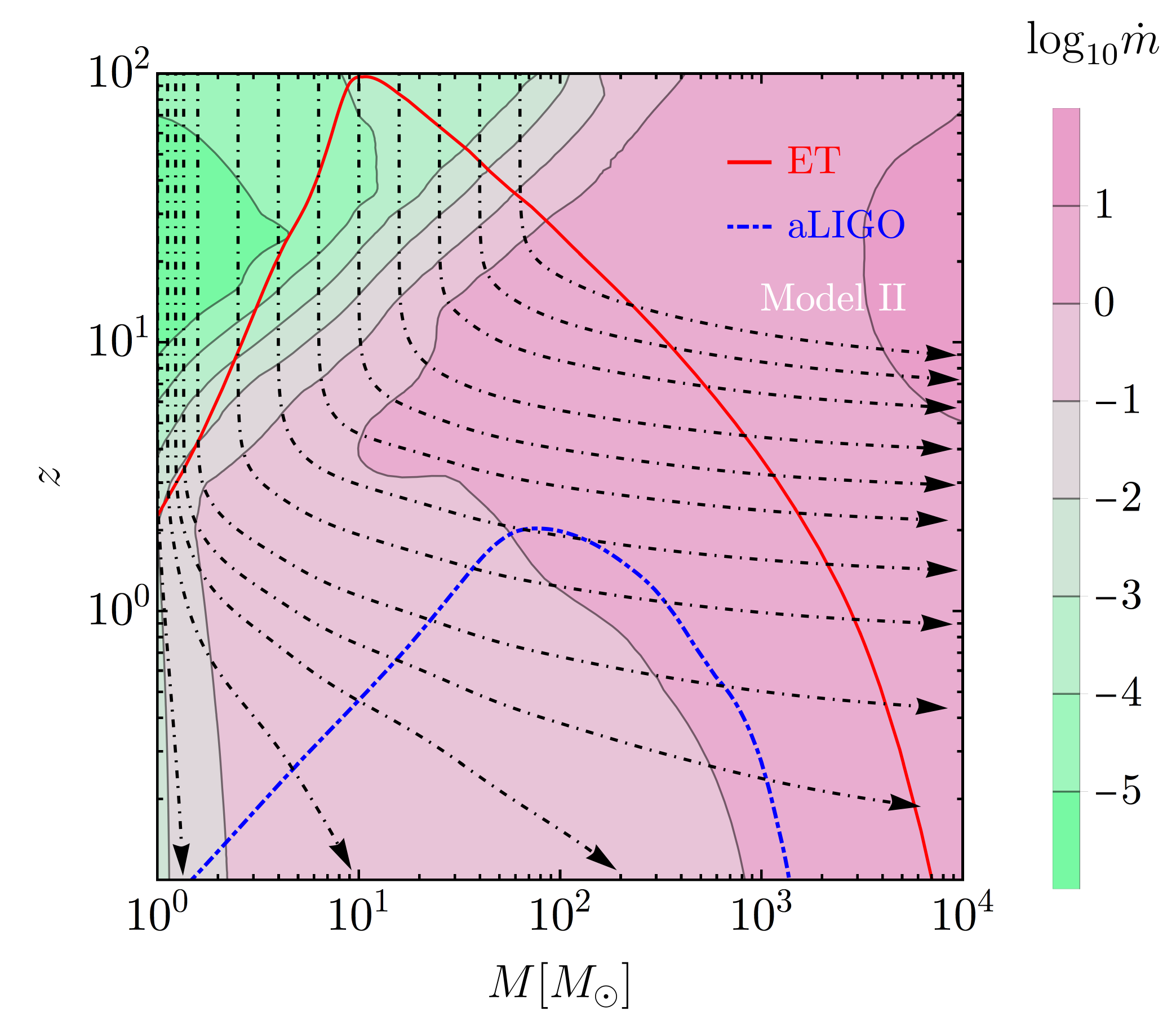}
        \hspace{.1 cm}
    \includegraphics[width=0.479 \linewidth]{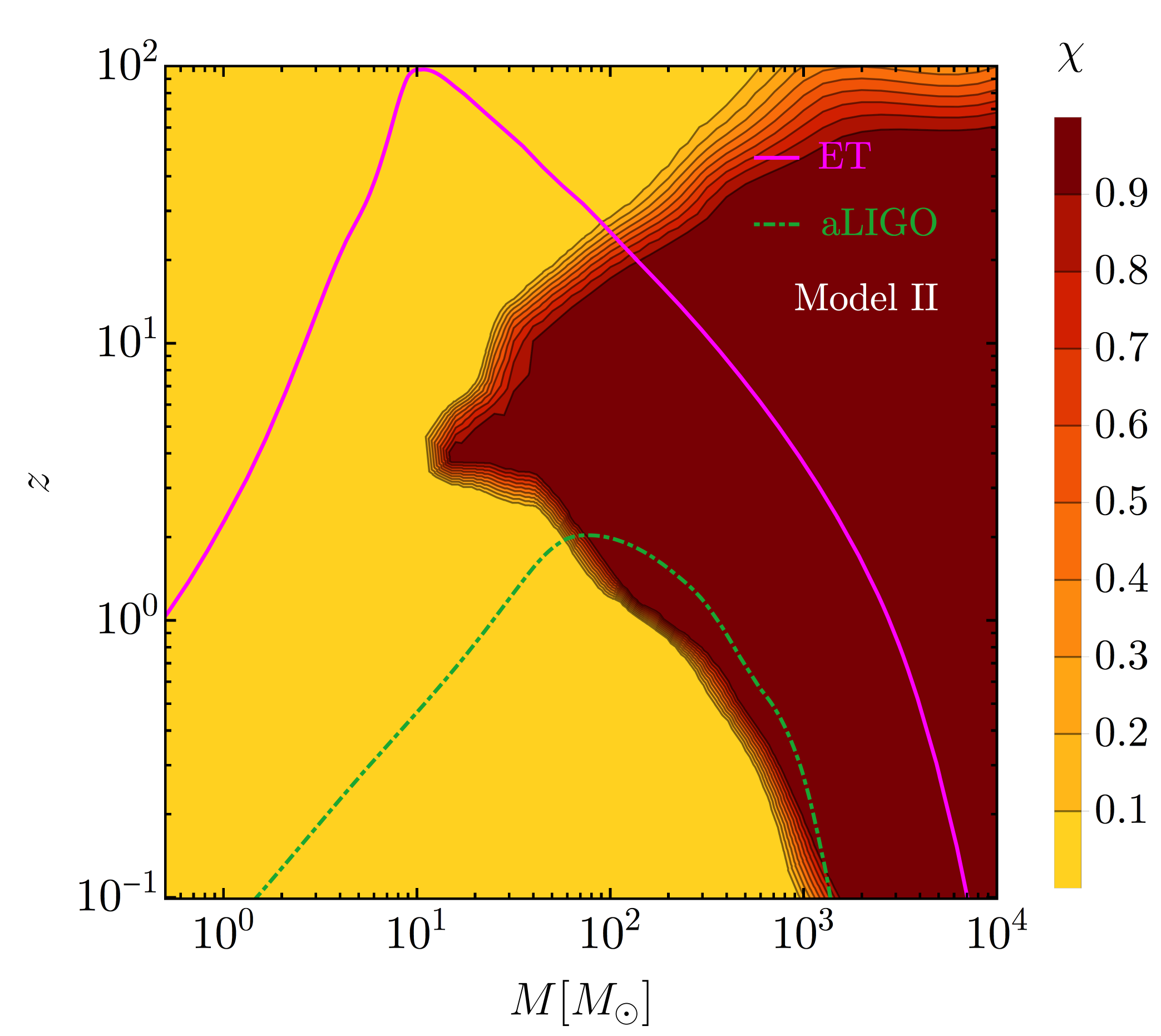}
    \caption{\it Same as in Fig.~\ref{Spin-I} but for Model~II, i.e. assuming a sustained accretion also when $z<10$.}
    \label{Spin-II}
   \end{figure}

\subsection{Implications for GW events}\label{sec:GWs}
Ultimately, we are interested in giving the prediction for the key observables which can be measured in GW coalescence 
events with current (LIGO/Virgo) and future (e.g., ET) detectors\footnote{We are neglecting the possible effect of 
accretion during the coalescence as we expect the increase in velocity to happen only during the last stages of inspiral 
and thus at much smaller characteristic time scales.}.

In a merger event of two PBHs with masses $M_{1}$ and $M_{2}<M_1$ with a binary mass ratio defined as 
$q=M_{2}/M_{1}$, and 
dimensionless spin vectors $\boldsymbol{\chi}_1$ and $\boldsymbol{\chi}_2$, one can estimate the final spin of the 
PBH resulting from the merger as~\cite{Barausse:2009uz,Kesden:2010yp,Barausse:2012fy,Hofmann:2016yih} 
\begin{align}
	\label{eq:general}
\chi_f=
	\frac{1}{(1+q)^2}\Big[ \vta{1}^2 + \vta{2}^2 q^4+
	2 {\vert \boldsymbol{\chi}_2\vert}{\vert 
		\boldsymbol{\chi}_1\vert} q^2 \cos \alpha\,+
	2\left(
	{\vert \boldsymbol{\chi}_1\vert}\cos \beta +
	{\vert \boldsymbol{\chi}_2\vert} q^2 \cos \gamma
	\right) {\vert \boldsymbol{{\ell}} \vert}{q}+\vert \boldsymbol{{\ell}}\vert^2 q^2
	\Big]^{1/2},
\end{align}
with 
\begin{align}
	\label{eq:L2}
	\vtl &= 2 \sqrt{3}+ t_2 \nu + t_3 \nu^2 +
	\frac{s_4}{(1+q^2)^2} \left(\vta{1}^2 + \vta{2}^2 q^4 
	+ 2 \vta{1} \vta{2} q^2 \cos\alpha\right) \nonumber \\
	& + 
	\left(\frac{s_5 \nu + t_0 + 2}{1+q^2}\right)
	\left(\vta{1}\cos{\beta} + 
	\vta{2} q^2 \cos{\gamma}\right),
\end{align}
in terms of the numerical parameters $s_4 = -0.1229\pm0.0075$, $s_5 = 0.4537\pm0.1463$, $t_0=-2.8904\pm0.0359$,
$t_3 = 2.5763\pm0.4833$ and $t_2=-3.5171 \pm 0.1208$.
Here $\nu=q/(1+q)^2$ is the symmetric mass ratio and 
\begin{align}
	\cos \alpha = {\hat{\boldsymbol{\chi}}}_2 \cdot {\hat{\boldsymbol{\chi}}}_1,
	\qquad
	\cos \beta = {\hat{\boldsymbol{\chi}}}_1 \cdot {\hat{\boldsymbol{L}}},
	\qquad
	\cos \gamma = {\hat{\boldsymbol{\chi}}}_2 \cdot {\hat{\boldsymbol{L}}}
\end{align}
are the angles (at large separation) between the two spins and between each individual spin and the direction of the 
orbital angular momentum ${\hat{\boldsymbol{L}}}$, respectively. 
The mass of the final PBH is~\cite{Tichy:2008du} 
\begin{align}
\label{finalmass}
	M_f=(M_1 + M_2) \times [1 + 4\nu(m_0-1)+ 16 m_1 \nu^2 (\vta{1} \cos\beta+ \vta{2} \cos\gamma)],
\end{align}
where $m_0=0.9515\pm 0.001$ and $m_1=-0.013\pm 0.007$ are numerical coefficients.

The effect of the spin of the binary components mostly affects the gravitational waveform to leading post-Newtonian 
order 
through the effective spin parameter, defined as the mass weighted projection of the effective spin of the binary to 
the orbital angular momentum (see Eq.~\eqref{chieffdef}), 
\be
\label{chieff}
\chi_\text{\tiny eff} = \frac{M_{1} \vta{1} \cos \beta + M_{2} \vta{2} \cos \gamma}{M_{1} + M_{2}}
= \frac{ \vta{1} \cos \beta + q\, \vta{2} \cos \gamma}{1+q},
\ee
where, being $\vta{i} < 1$, the possible range of values is $|\chi_\text{\tiny eff}|< 1$.\footnote{The occurrence of 
merging events with highly-spinning components may also increase the stochastic GW background signal resulting from 	
the coalescences, with a consequent change in the deduced bounds from its non-observation \cite{Raidal:2017mfl}, see 
Ref.~\cite{Hemberger:2013hsa} for details about the radiated energy from a merging event in terms of the BHs spin.} 

The effective spins measured so far with GWs are 
affected by large uncertainties and are compatible to zero for almost all sources~\cite{LIGOScientific:2018jsj}. Only 
few high-mass events have been detected so far for which $|\chi_\text{\tiny 
eff}|>0$~\cite{LIGOScientific:2018mvr,Zackay:2019tzo,Venumadhav:2019lyq}, although for two 
low-significance events --~namely GW151216 and GW170403~-- the measured value of $\chi_\text{\tiny eff}$ is 
significantly affected by the prior on the spin angles~\cite{Huang:2020ysn}.
Furthermore, future detections will provide measurements of the individual spins with $30\%$ 
accuracy~\cite{TheLIGOScientific:2016pea}, also alleviating the degeneracy between the individual spins and other 
binary parameters such as the mass ratio.

To obtain the probability distribution functions (PDFs) of the spin~\eqref{eq:general} and mass~\eqref{finalmass} of 
the merger remnant, along with the effective spin of the binary~\eqref{chieff}, one has to perform a 
statistical ensemble over the masses of the binary components and the relevant angles of the spin vectors. We have 
assumed a uniform distribution for the spin vectors orientations on a unit two-sphere~\cite{LIGOScientific:2018mvr,Gerosa:2017kvu} 
and several shapes of the mass 
functions $\psi (M)$ (see Appendix~\ref{appmer}) at the redshift of observation $z$. Such shapes are assumed to result 
from the evolution of an 
initial mass function due to mass accretion, see Appendix~\ref{appmf} for details about this time evolution.

\begin{figure}[t!]
 \centering
 \includegraphics[width=0.3 \linewidth]{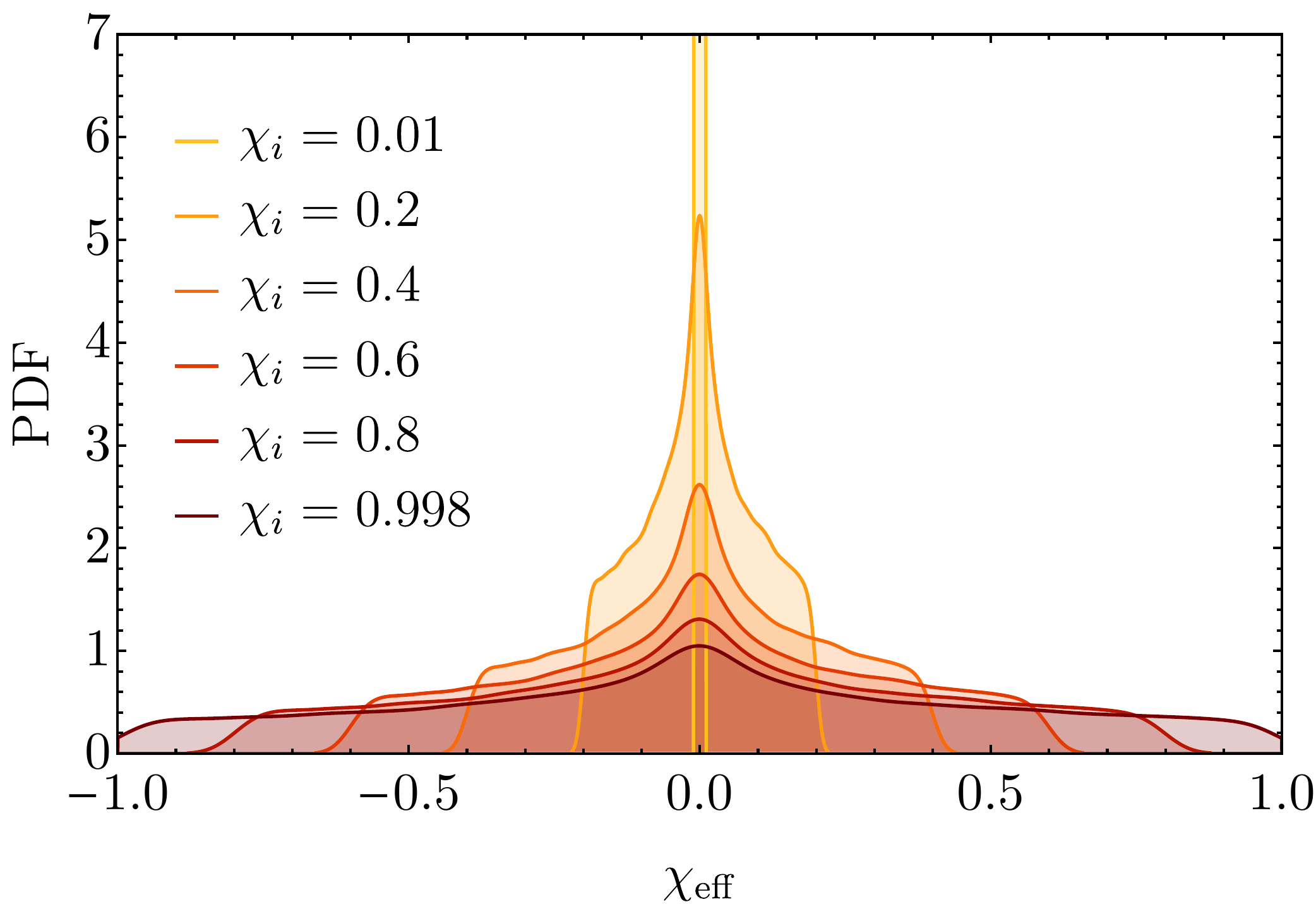}
    	\hspace{.2cm}
	\includegraphics[width=0.3 \linewidth]{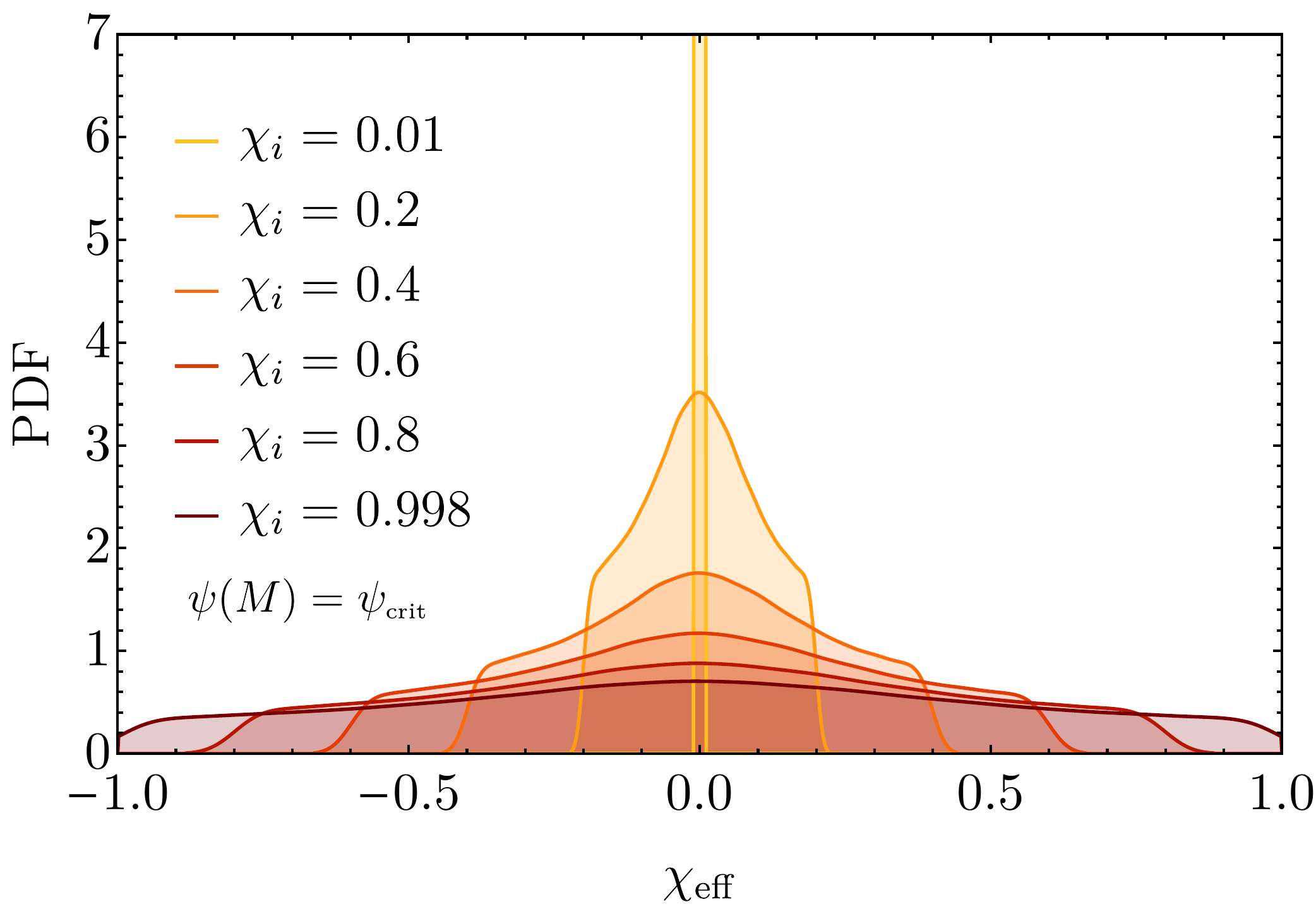}
    	\hspace{.2cm}
	\includegraphics[width=0.3 \linewidth]{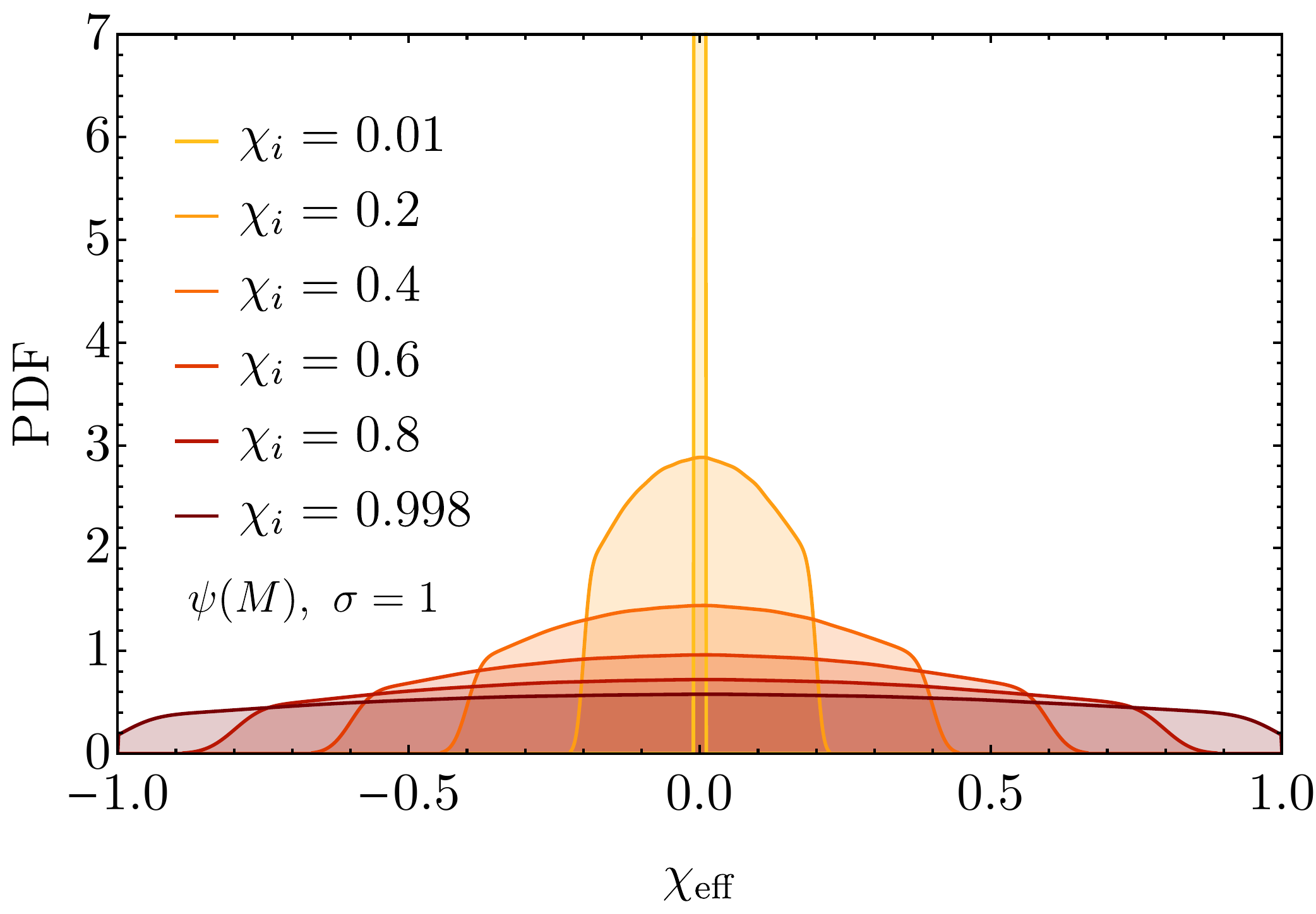}

	\vspace{.2cm}

	\includegraphics[width=0.3 \linewidth]{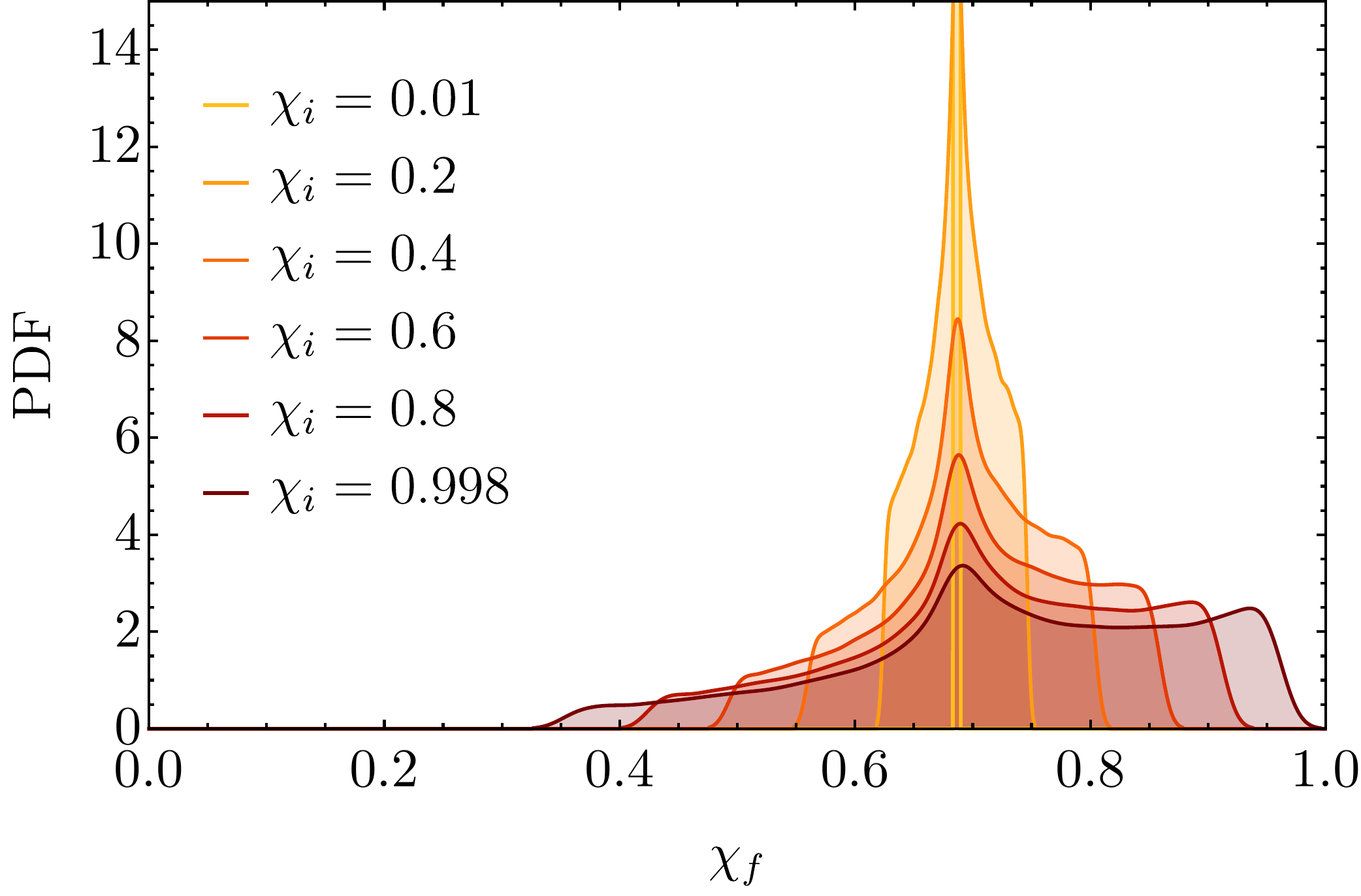}
    	\hspace{.2cm}
	\includegraphics[width=0.3 \linewidth]{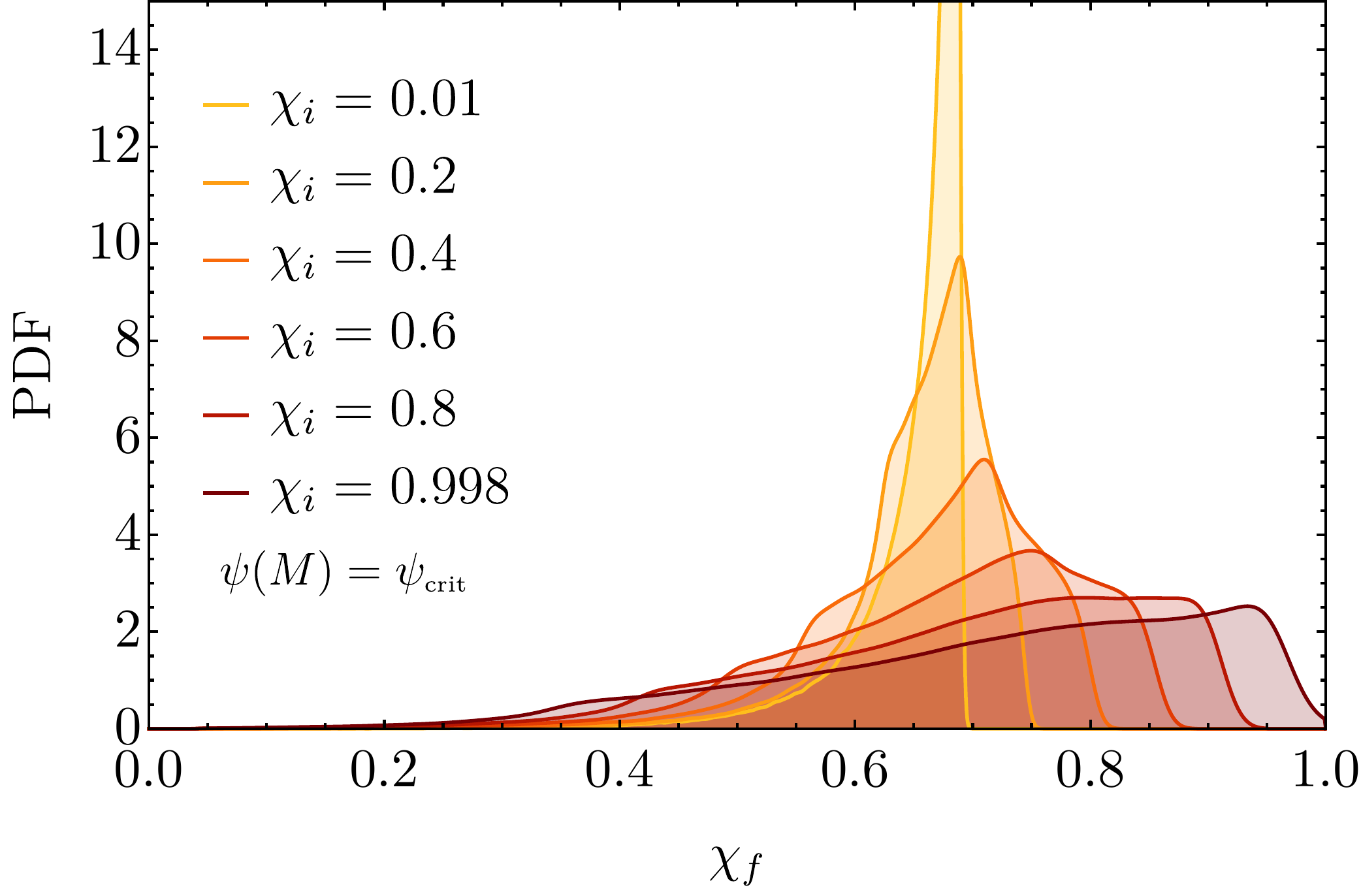}
    	\hspace{.2cm}
	\includegraphics[width=0.3 \linewidth]{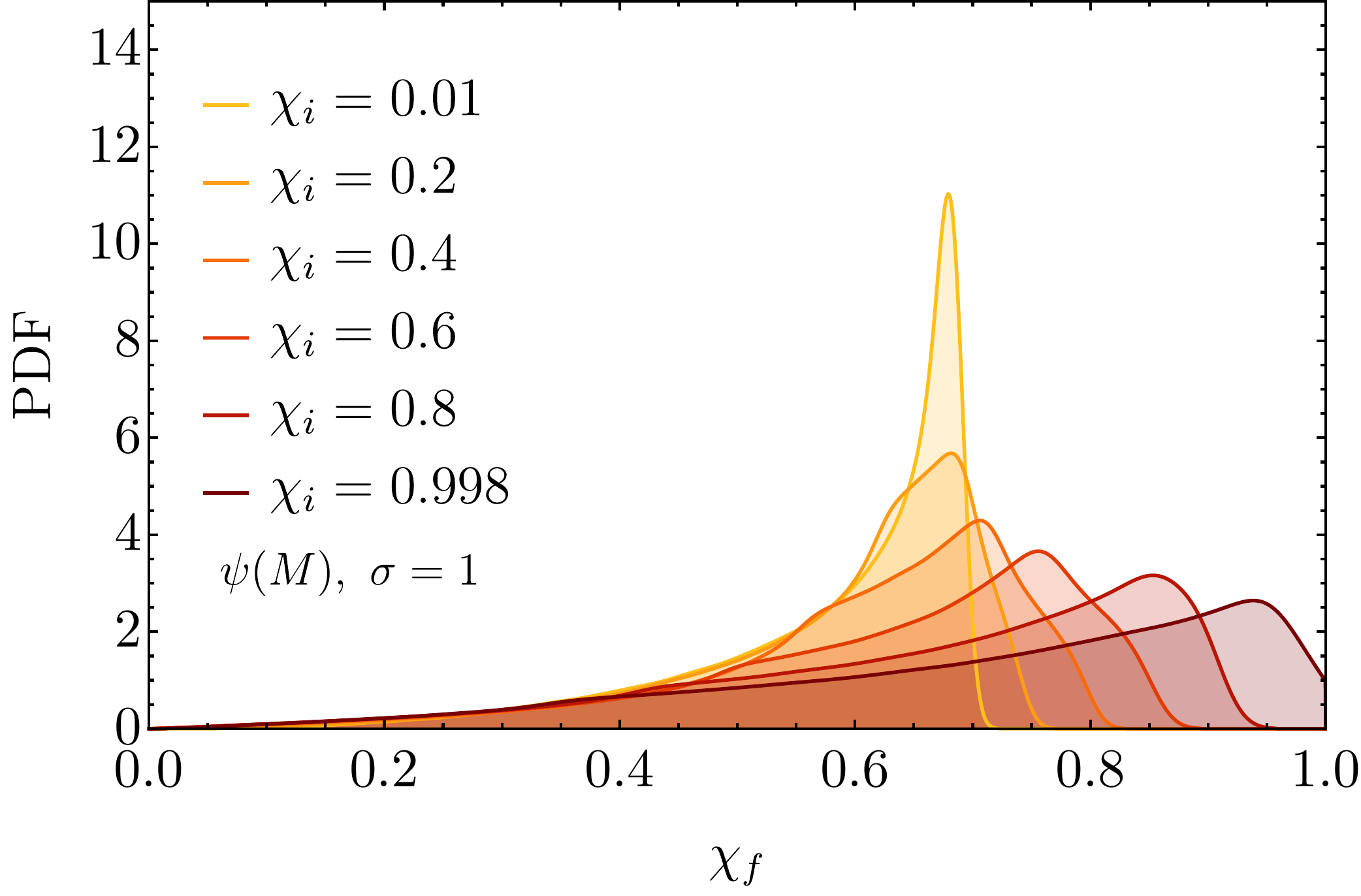}

	\vspace{.2cm}
	
	\includegraphics[width=0.3 \linewidth]{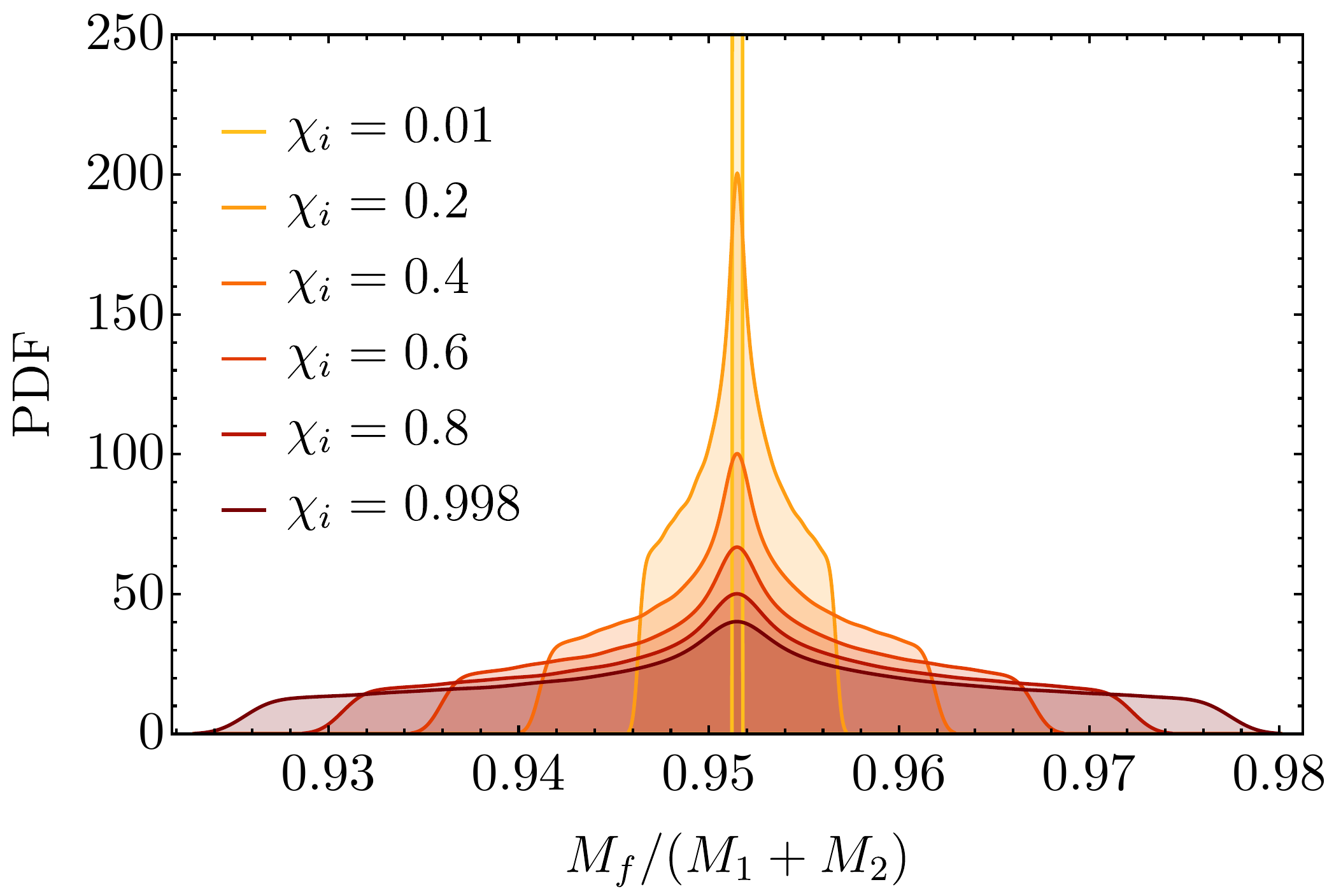}
    	\hspace{.2cm}
	\includegraphics[width=0.3 \linewidth]{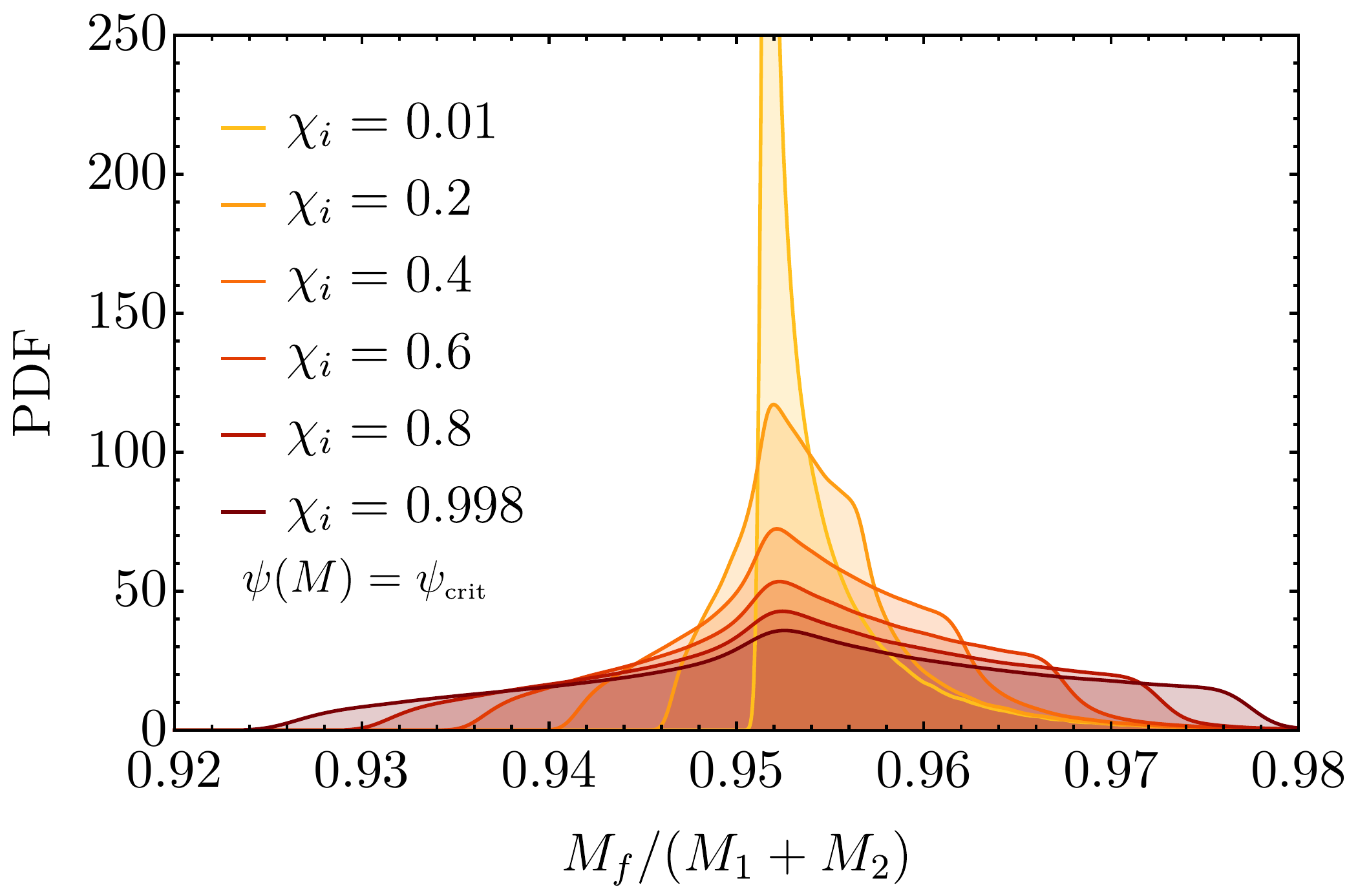}
    	\hspace{.2cm}
	\includegraphics[width=0.288 \linewidth]{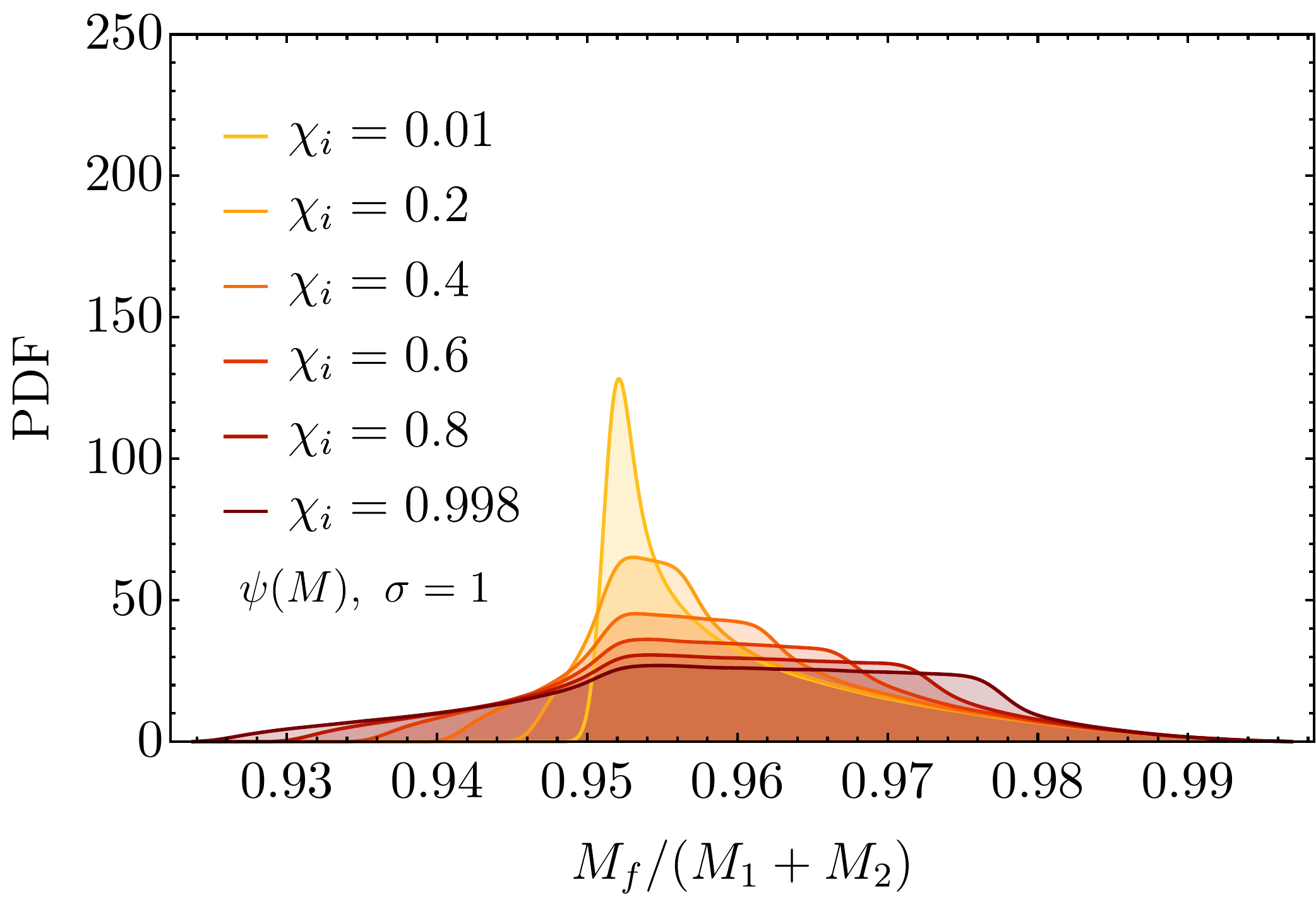}
    \caption{\it 
	Top:
	 PDF of the effective spin parameter of the binary assuming the coalescence of two PBHs with the same mass 
	 and same binary-component spin, $\chi_1=\chi_2=\chi_i$, at merging time.
	Center: 
	 PDF of the spin of the final BH remnant assuming the coalescence of two PBHs with an isotropic 
distribution of the spins directions on a sphere.
	Bottom: 
	PDF for the final mass as a function of the spins of the PBHs, $\chi_1=\chi_2=\chi_i$.
	From Left to Right: monochromatic mass function ($q=1$), critical mass 
function $\psi_\text{\rm \tiny crit}$, and lognormal with width $\sigma=1$.  
Notice that we have used the same colour code of the right panels of Figs.~\ref{Spin-I} and \ref{Spin-II}.
    }
    \label{pdf}
   \end{figure}

Results are shown in Fig.~\ref{pdf}, where for simplicity we have assumed that both components of the binary have the 
same spin before merger, $\chi_1=\chi_2=\chi_i$.
We stress that the colour code used to plot the probability distributions corresponding to a particular spin $\chi_i$, 
as shown in the legend,
has been chosen to match the one used in Figs.~\ref{Spin-I} and \ref{Spin-II} (right panels). In other words, one can identify the expected PDFs
for the relevant parameters in each point of the parameter space $(M,z)$ of the contour plots by looking at the 
corresponding colour in Fig.~\ref{pdf}.
In particular, the distributions shown in Fig.~\ref{pdf} only depend on the value of the binary component spins 
at coalescence. They are therefore similar to those computed in other astrophysical 
scenarios~\cite{volo,Gerosa:2017kvu,Fishbach:2017dwv}. However, there are crucial differences in our case. One is the effect that the value of 
the individual spins is correlated with the mass of the binary components and with the redshift at coalescence. Another one is the presence of a given
PBH mass function.

The first column shows the PDFs for a monochromatic shape of the mass function, for which the mass ratio of the binary 
is $q = 1$, while the second and third columns show the result for more realistic and broader shapes of the mass 
distribution, namely a critical mass and lognormal with width $\sigma=1$, respectively, for which the mass ratio 
distribution is peaked at smaller values (see Appendices~\ref{appmer} and~\ref{appmf} for details about the mass 
functions).

Since the individual spins are isotropically oriented, the PDF for the effective spin parameter of the binary 
(first row) is peaked around the central value $\chi_{\text{\tiny eff}}\simeq 0$ for small initial spins, maintaining 
the peak also for broader mass functions. However, for higher initial spins the distribution of $\chi_\text{\tiny 
eff}$ is much broader for all choices of the mass function.

The PDF of the final spin (second row) is peaked at 
the value $0.68$ for low initial spins of the binary (since in this case the distribution is almost independent on the 
values of the spins angles), and the distribution becomes broader for bigger values of the PBH spins before the 
merger. For broader mass functions the PDF tends to become broader (see panels of Fig.~\ref{pdf} from left to right).

Finally, the probability distribution for the final mass peaks at the value $M_f \simeq 0.96 (M_1+M_2)$ for low initial spins; the peak values decrease for broader mass 
functions, and has a flatter shape for higher spins. One can also notice how the distribution tends to be asymmetric 
with respect to  the centre for broader mass functions.

\begin{figure}[t!]
 \centering
	\includegraphics[width=0.39 \linewidth]{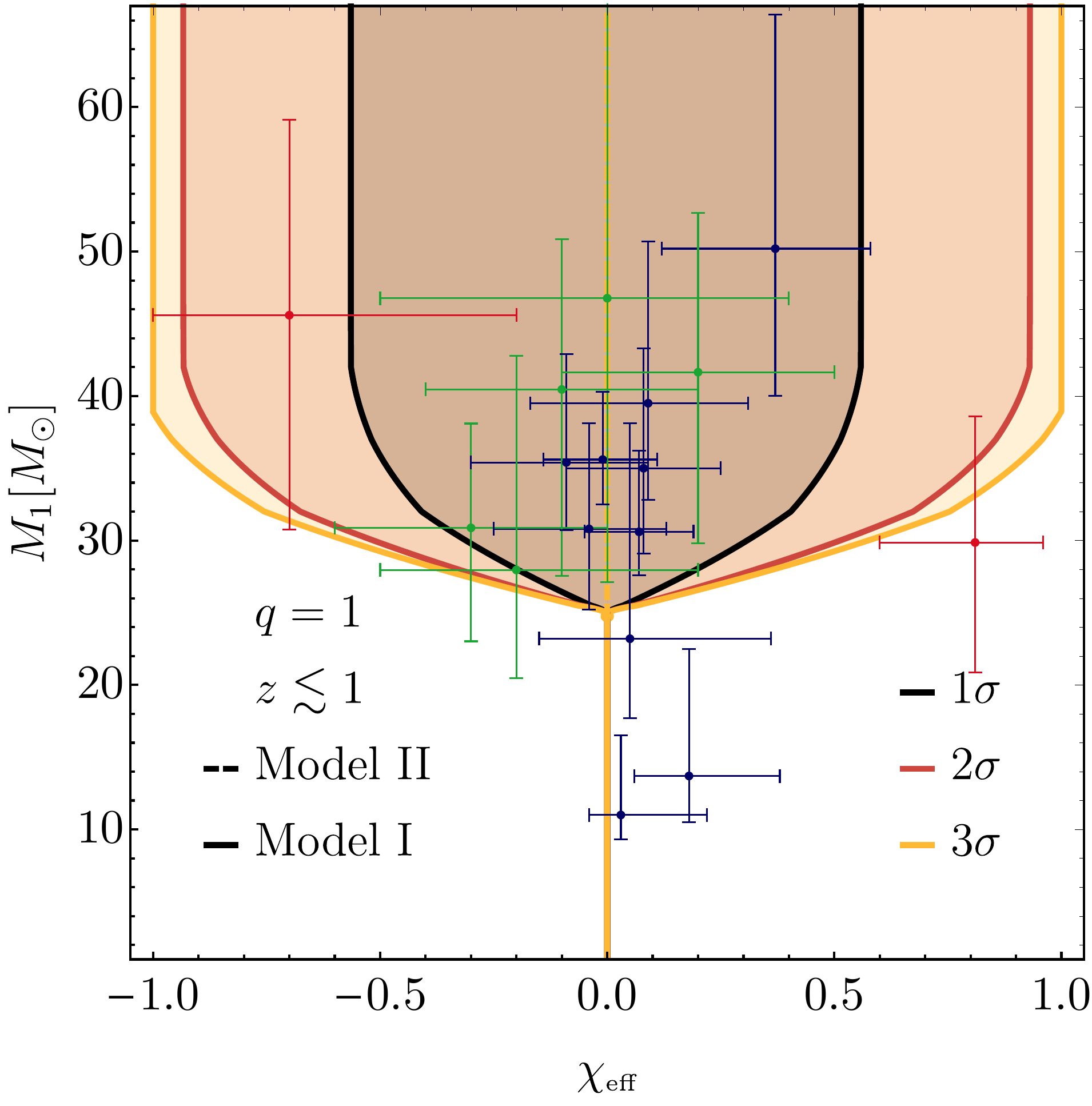}
    	\hspace{.25cm}
	\includegraphics[width=0.39 \linewidth]{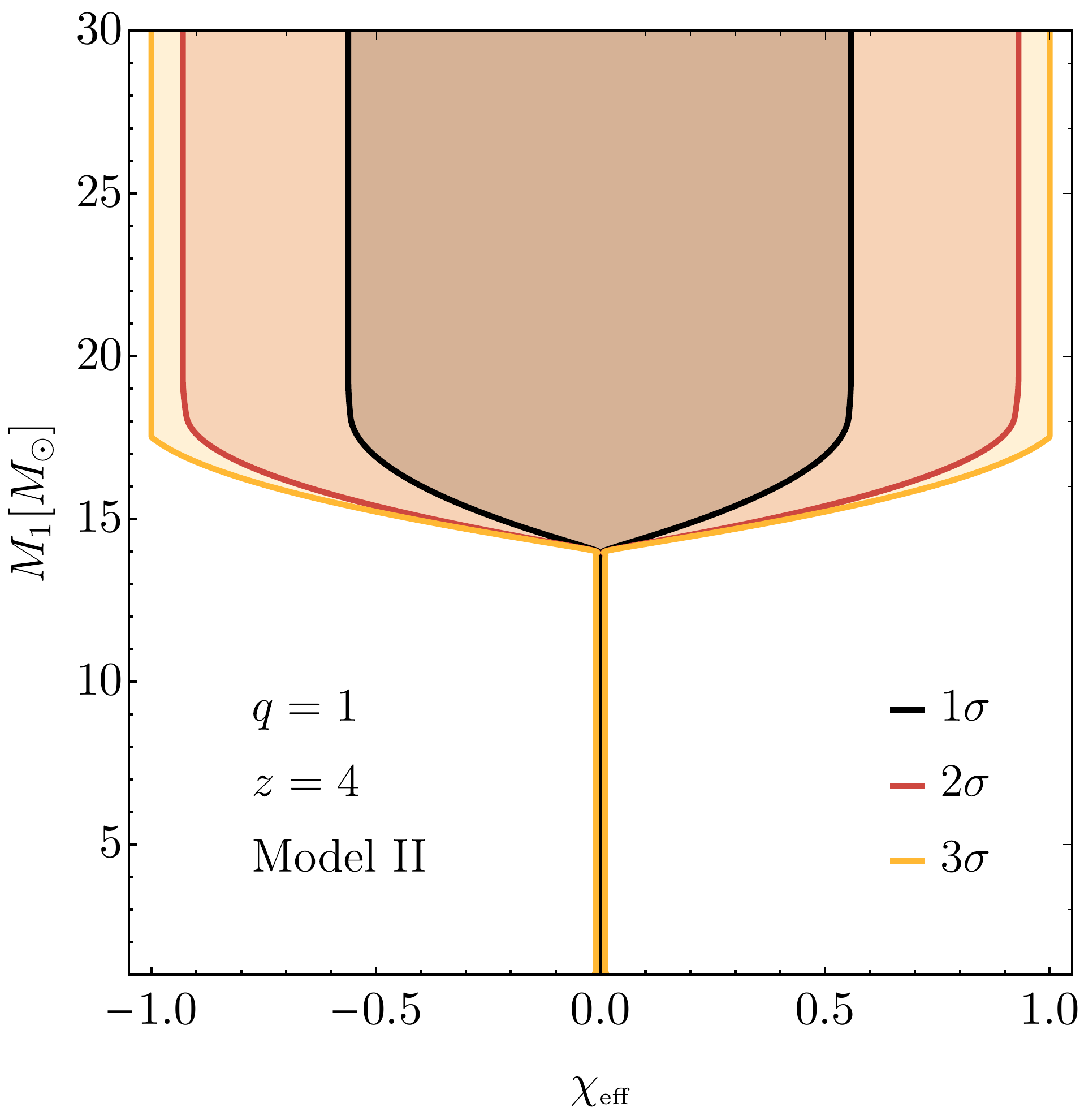}
	
	\vspace{.2cm}

	\includegraphics[width=0.39 \linewidth]{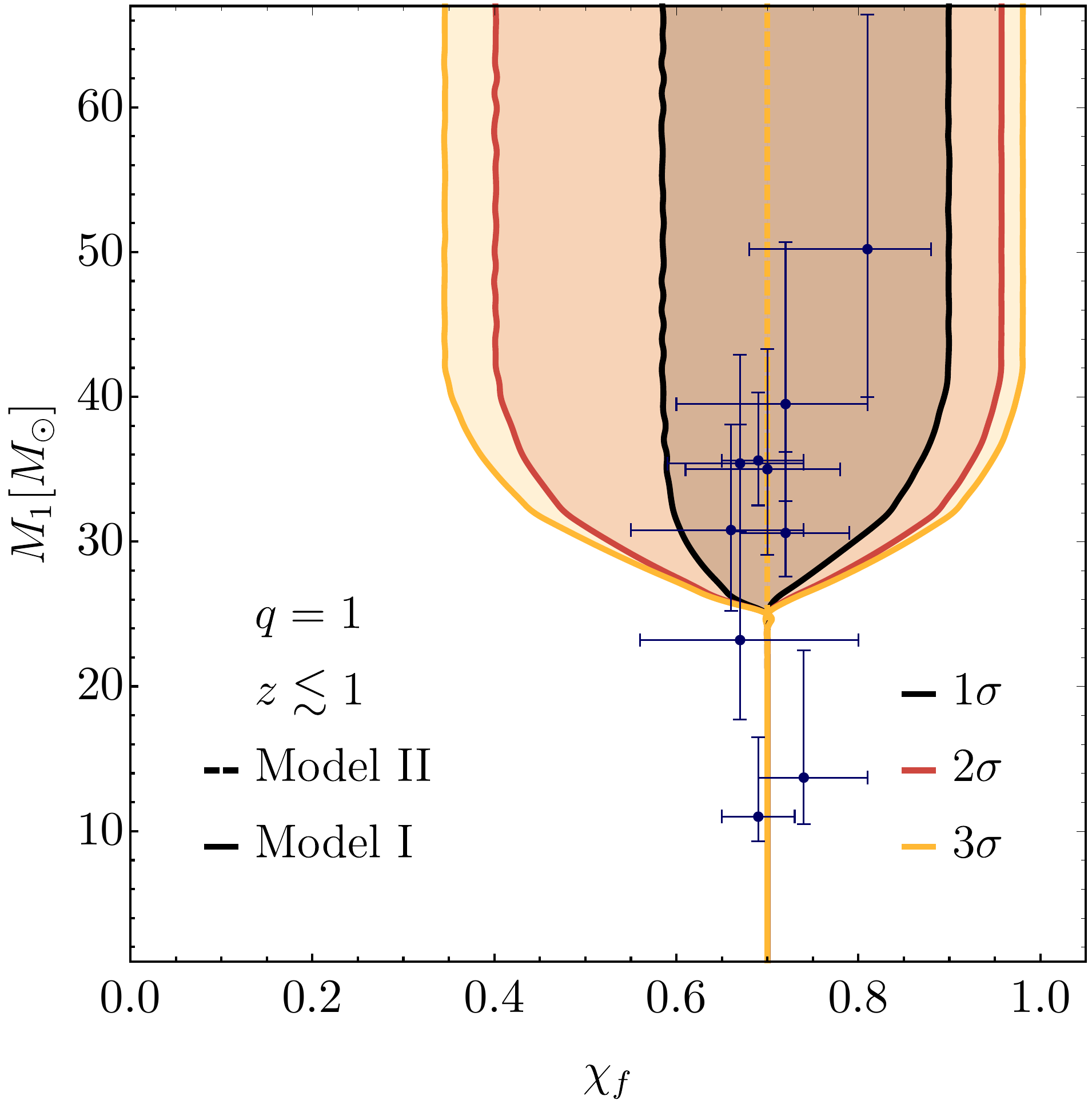}
    	\hspace{.25cm}
	\includegraphics[width=0.39 \linewidth]{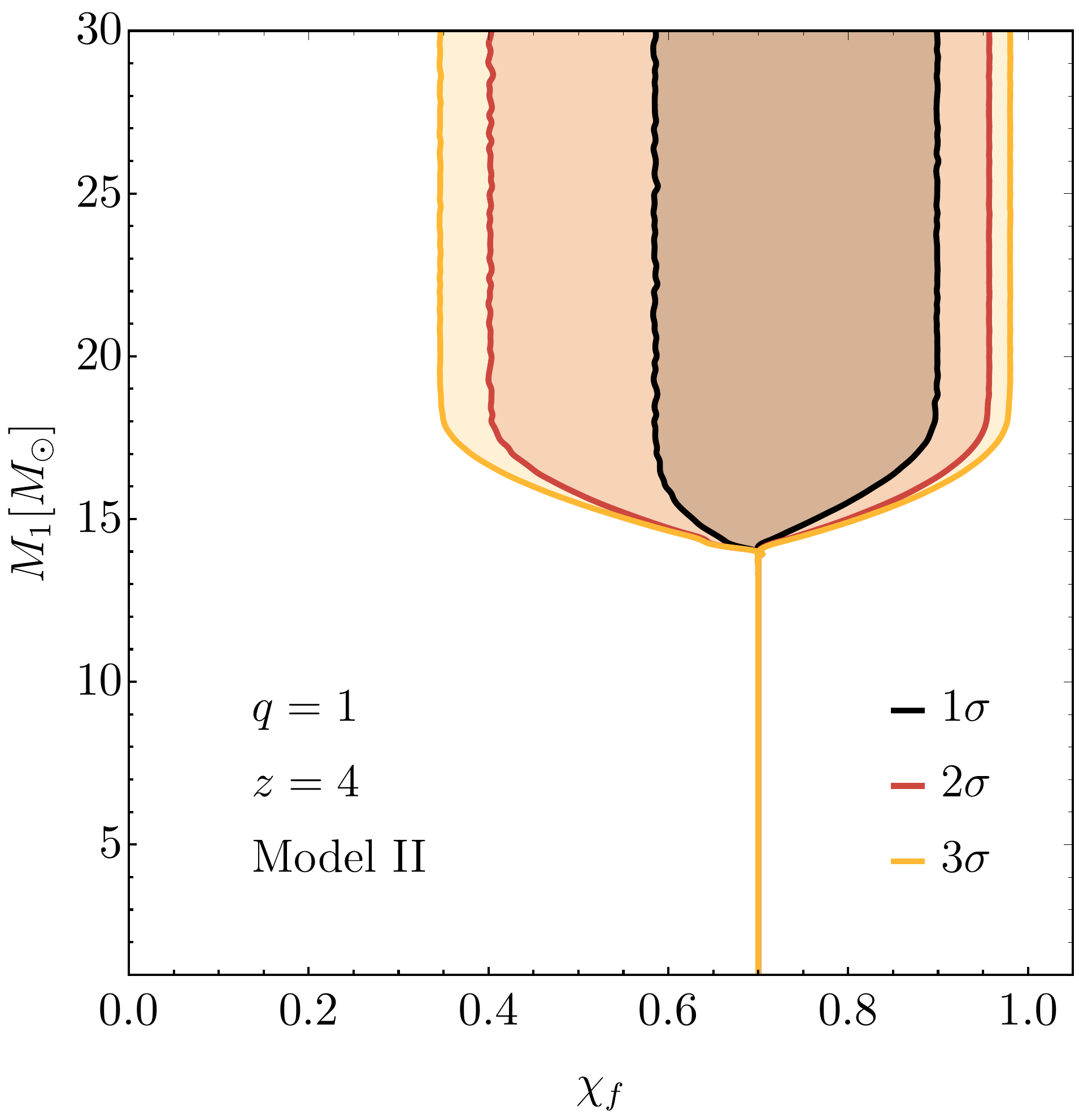}

	\vspace{.2cm}
	
	\includegraphics[width=0.39 \linewidth]{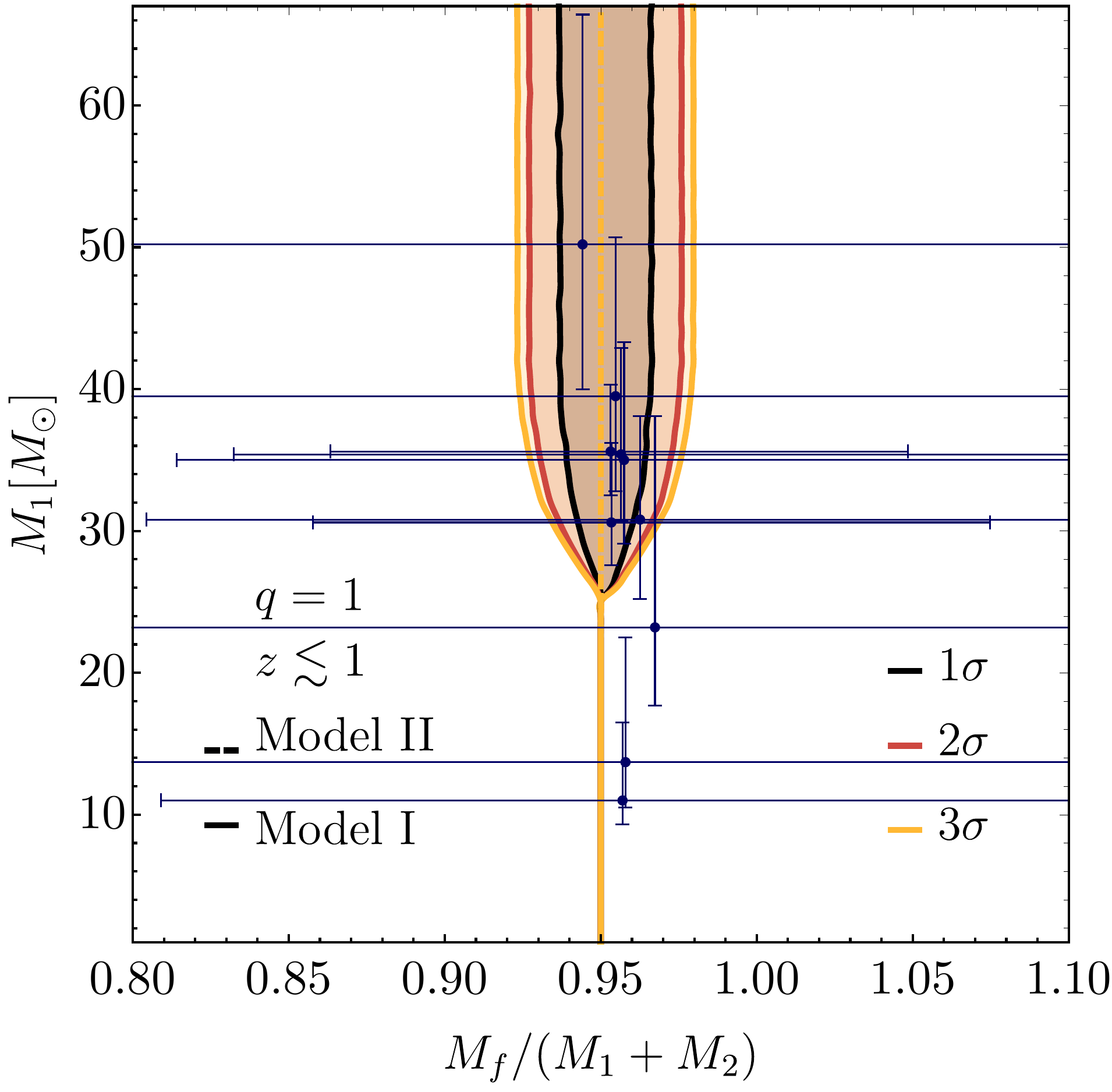}
    	\hspace{.25cm}
	\includegraphics[width=0.39 \linewidth]{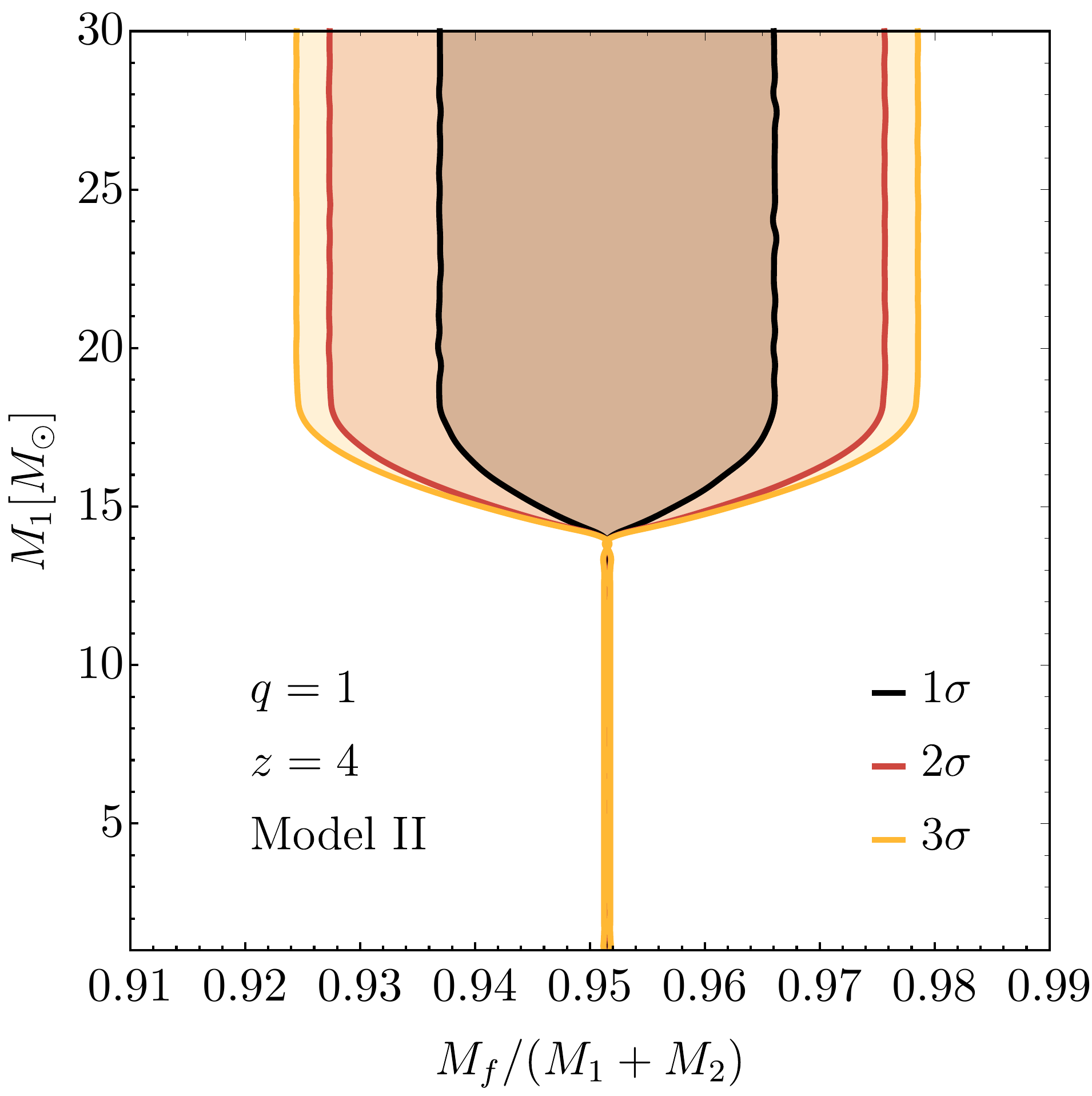}
    \caption{\it
   Confidence intervals (from top to bottom) for the parameters $\chi_\text{\tiny \rm  eff}$, $\chi_f$ and 
$M_f$ for given source-frame masses $M_1=M_2$, and for Model~I (Model~II) in the left 
(right) panels. 
On the scale of the left panels the different confidence levels for Model~II appear as straight vertical lines.
We notice that in  the left panels there is no evolution for $z \lsim 1$ and therefore we have plotted 
the current observed data: blue data points refer to the events listed in Ref.~\cite{LIGOScientific:2018mvr}, whereas 
green and red data points refer to the events discovered in Refs.~\cite{Zackay:2019tzo,Venumadhav:2019lyq}; the 
red data points refer to GW151216 and GW170403, for which the measured value of $\chi_\text{\tiny \rm eff}$ is 
significantly affected by the prior on the spin angles~\cite{Huang:2020ysn}. In the right 
panels there are not yet observed data at redshift $z=4$. 
	    }
    \label{pdf-Mchi}
   \end{figure}

\begin{figure}[t!]
 \centering
	\includegraphics[width=0.39 \linewidth]{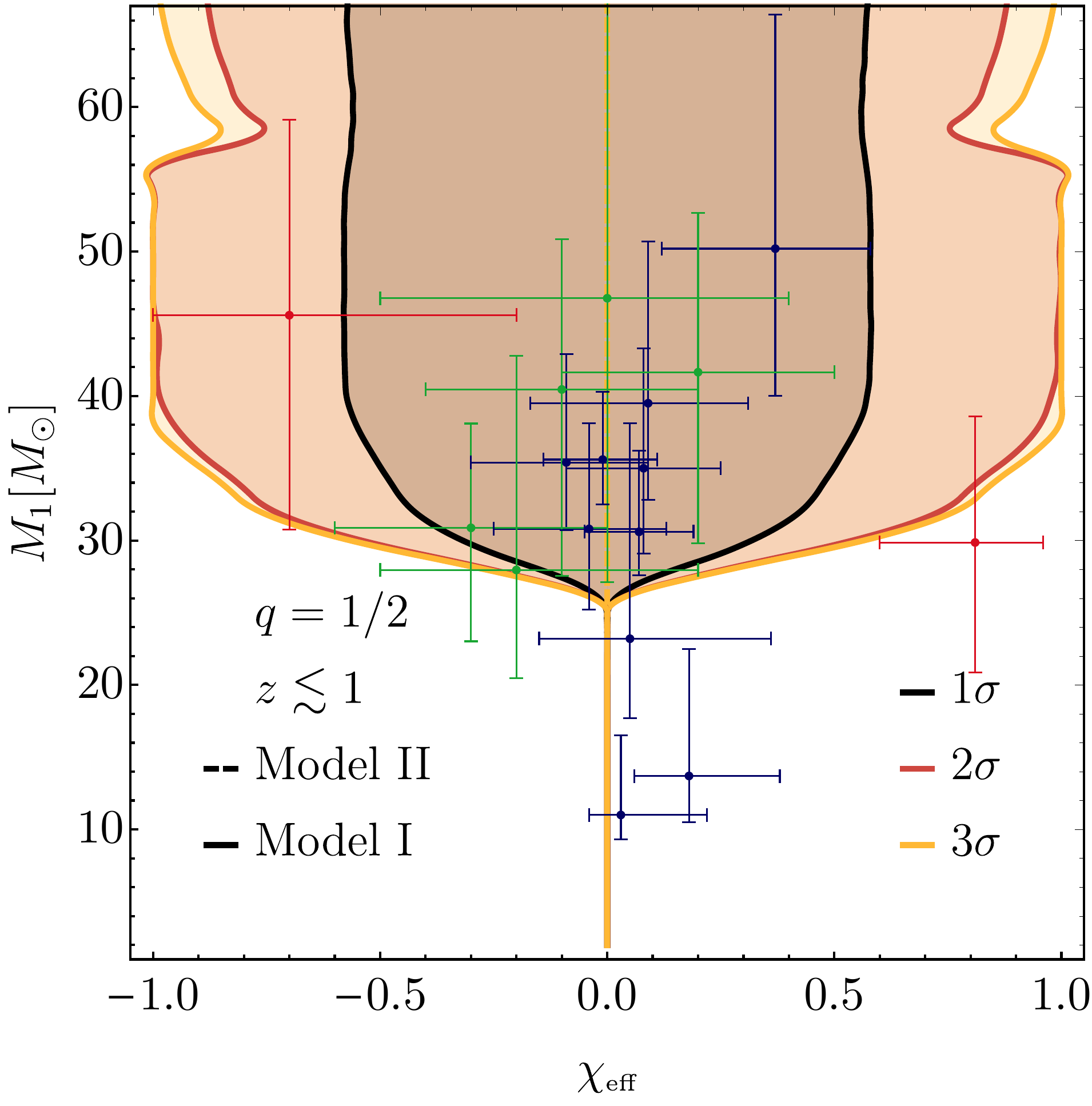}
    	\hspace{.25cm}
	\includegraphics[width=0.39 \linewidth]{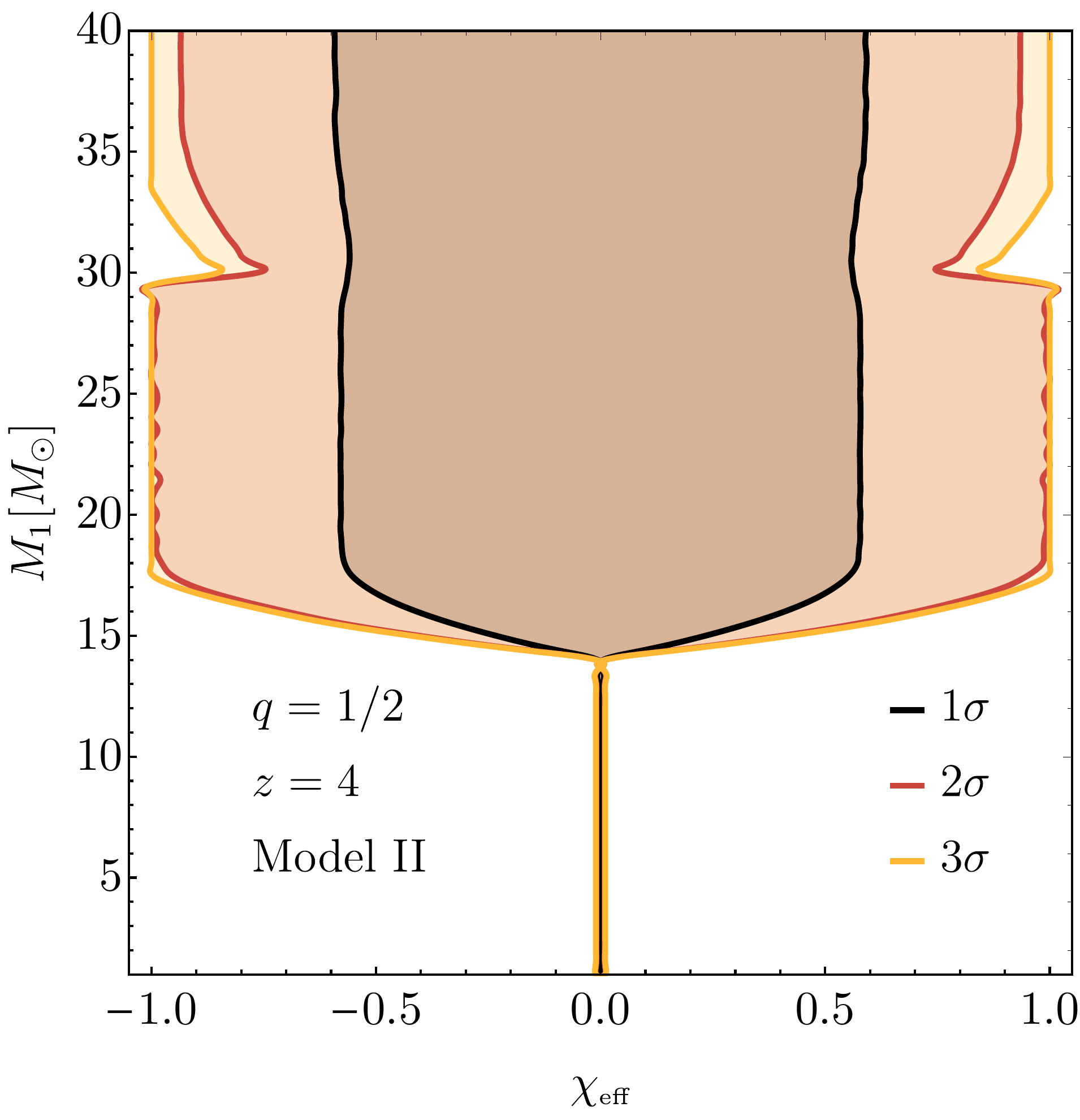}

	\vspace{.2cm}
	
	\includegraphics[width=0.39 \linewidth]{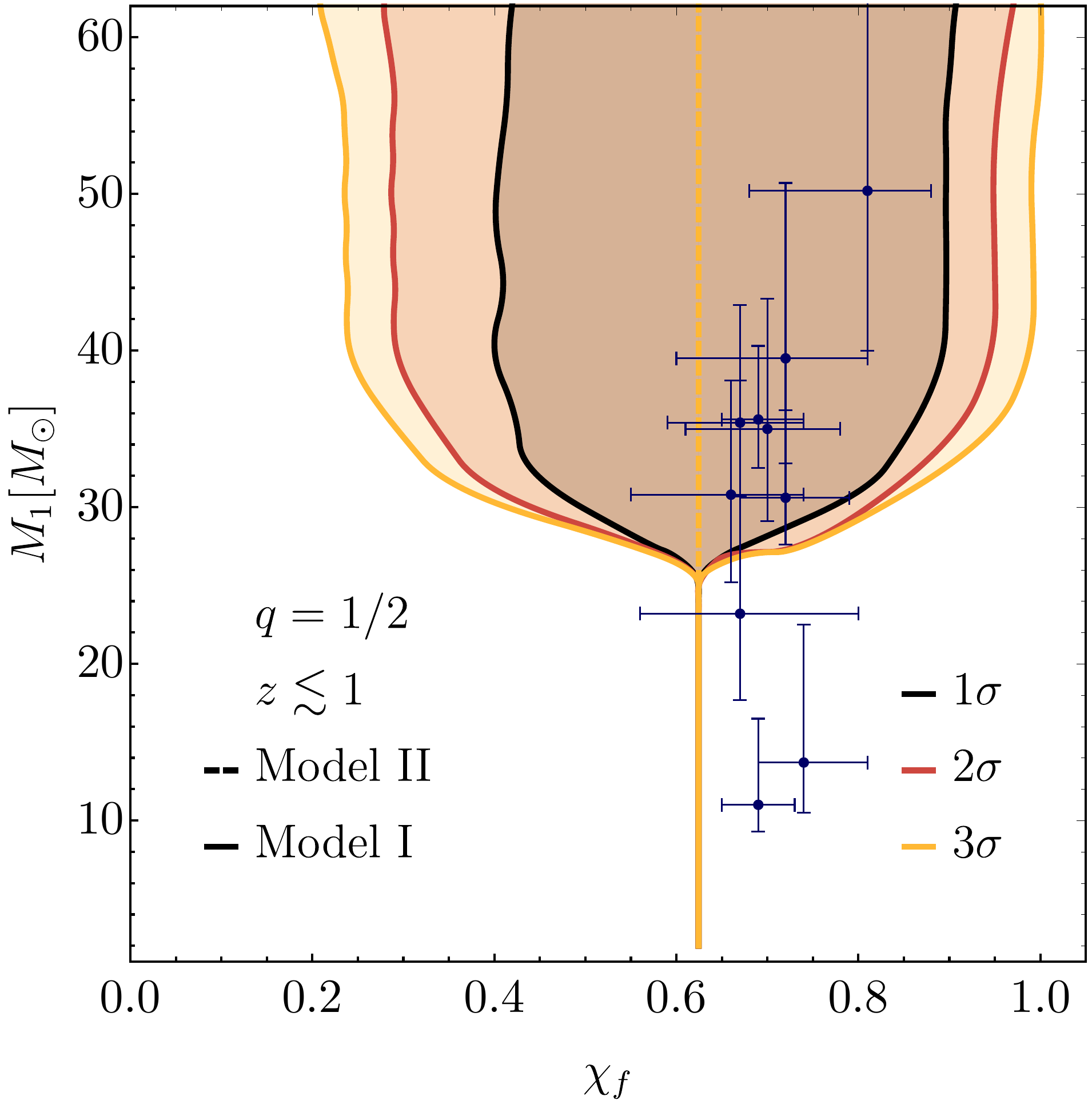}
    	\hspace{.25cm}
	\includegraphics[width=0.39 \linewidth]{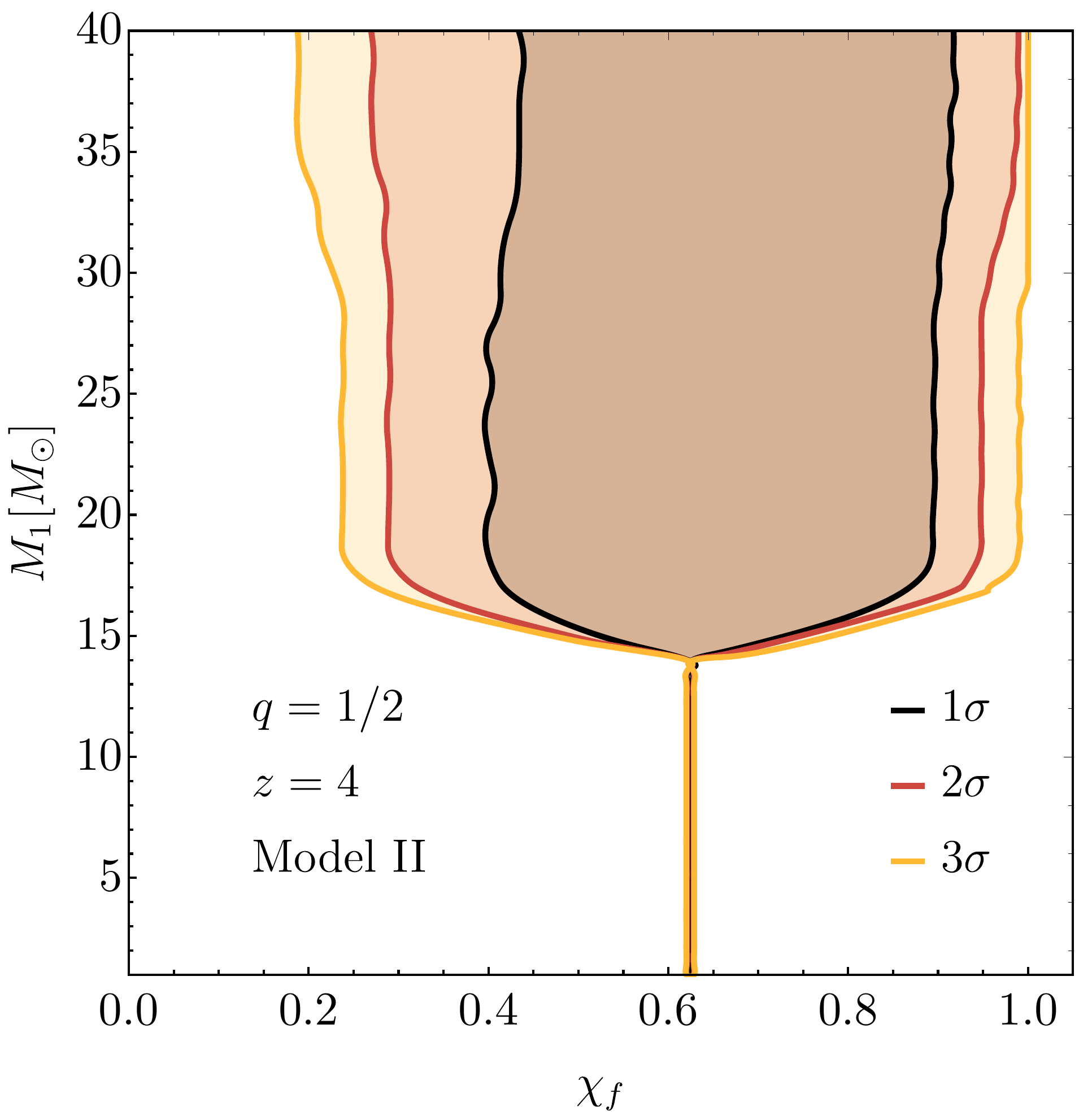}

	\vspace{.2cm}
	
	\includegraphics[width=0.39 \linewidth]{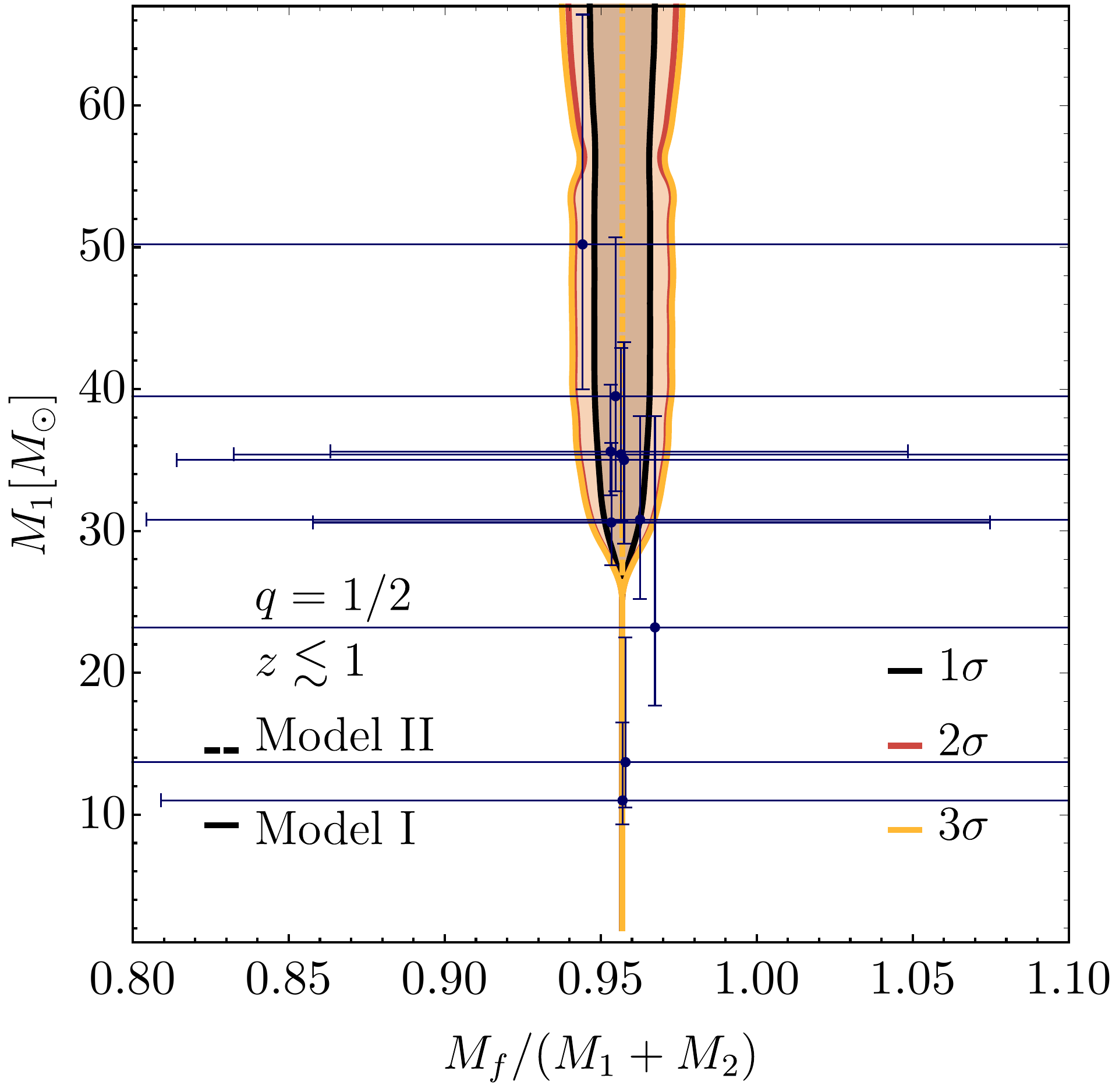}
    	\hspace{.25cm}
	\includegraphics[width=0.39 \linewidth]{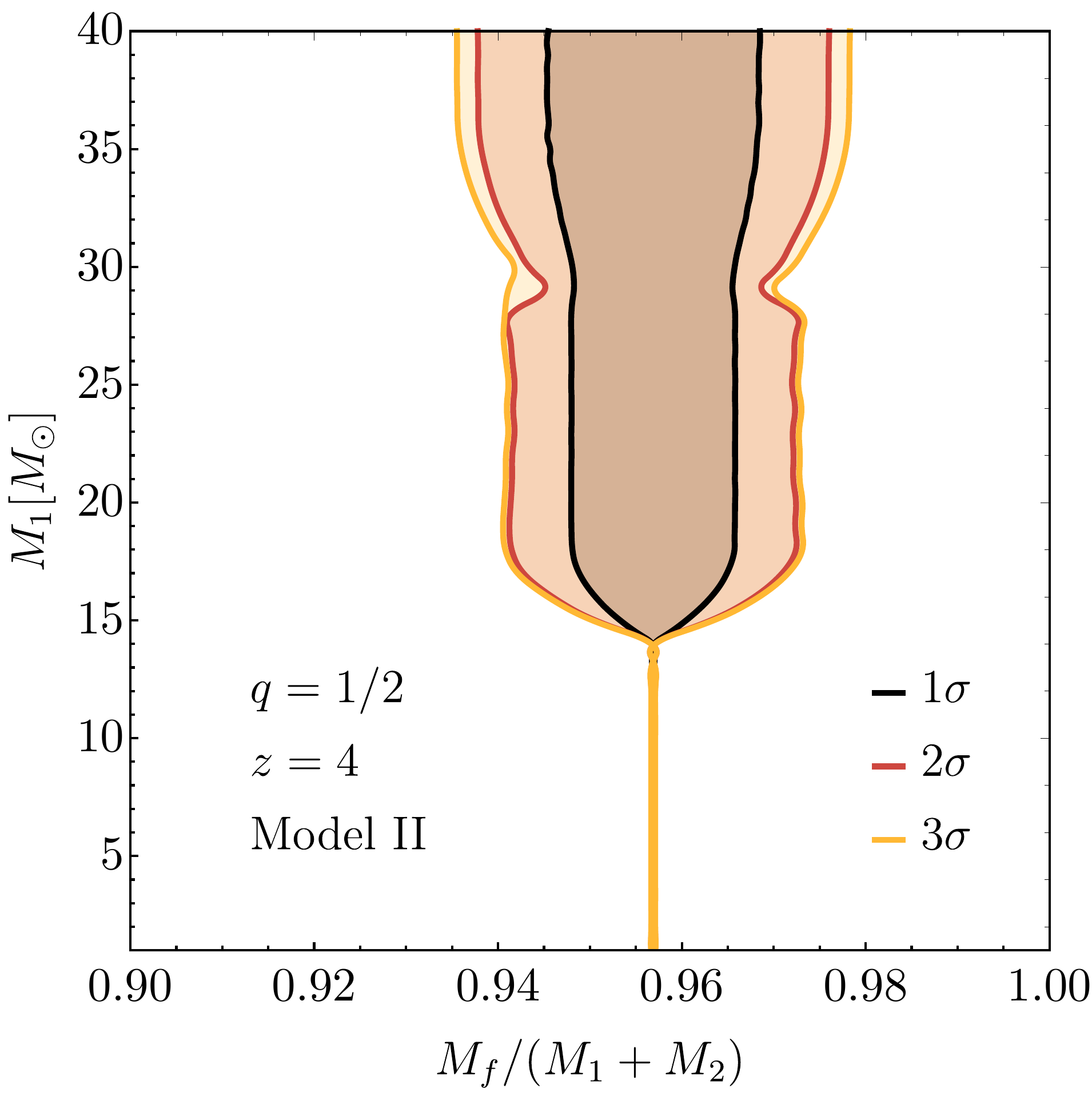}
    \caption{\it
    Same as in Fig.~\ref{pdf-Mchi} but for $M_1= 2 M_2$ (i.e., mass ratio $q=1/2$). 
	    }
    \label{pdf-Mchiq2}
   \end{figure}
   
   \begin{figure}[t!]
 \centering
	\includegraphics[width=0.39 \linewidth]{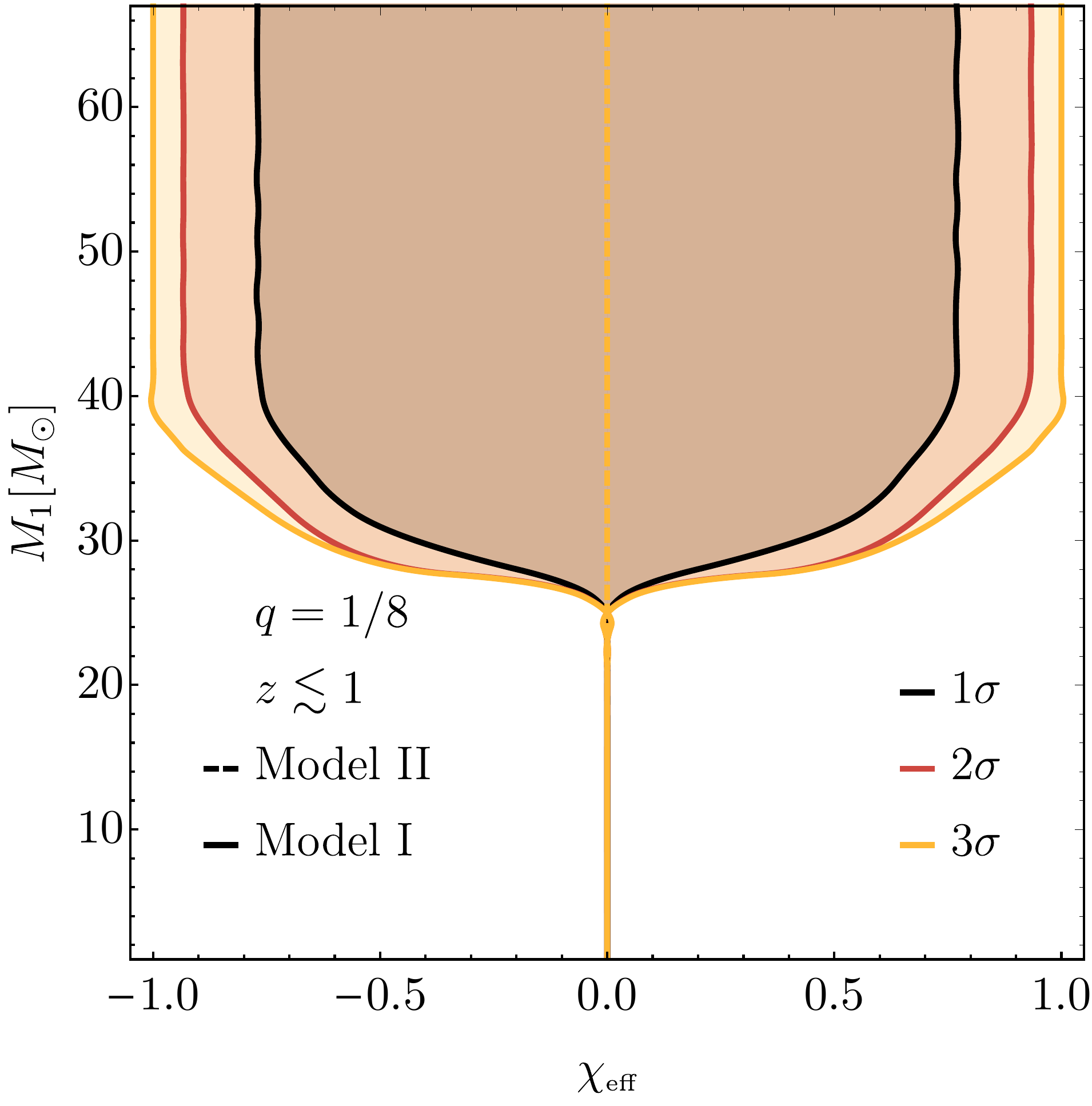}
    	\hspace{.25cm}
	\includegraphics[width=0.39 \linewidth]{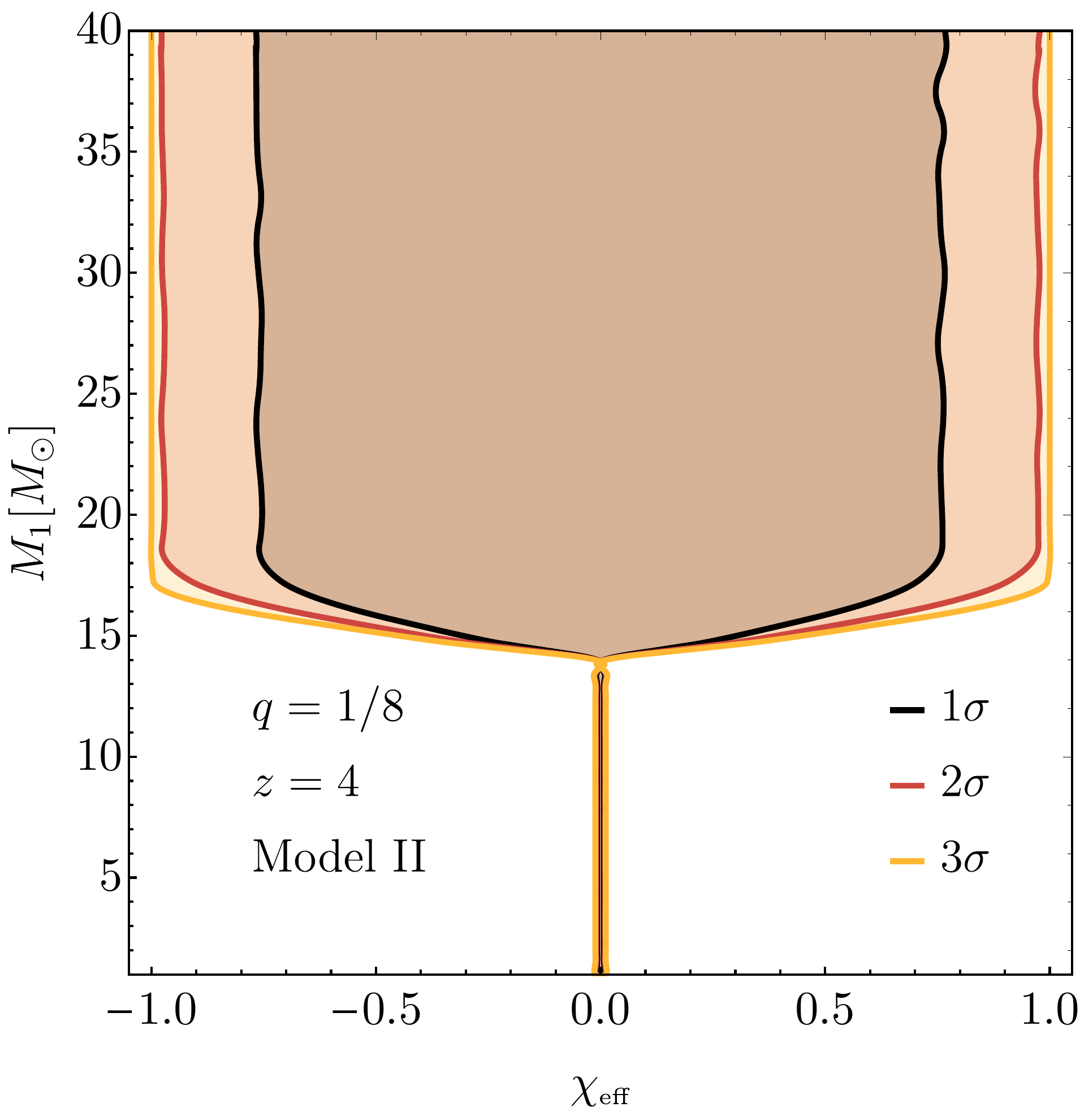}

	\vspace{.2cm}
	
	\includegraphics[width=0.39 \linewidth]{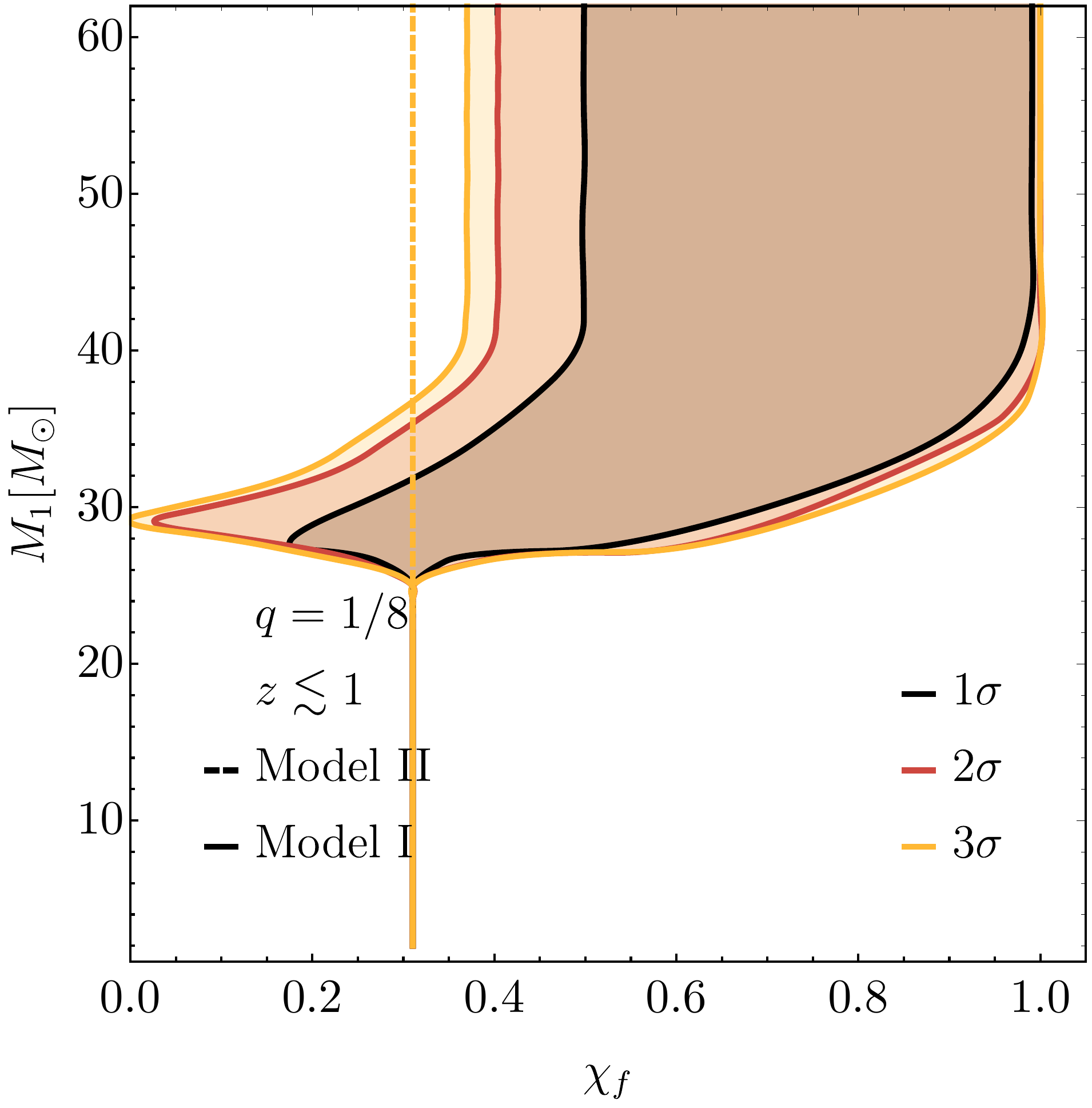}
    	\hspace{.25cm}
	\includegraphics[width=0.39 \linewidth]{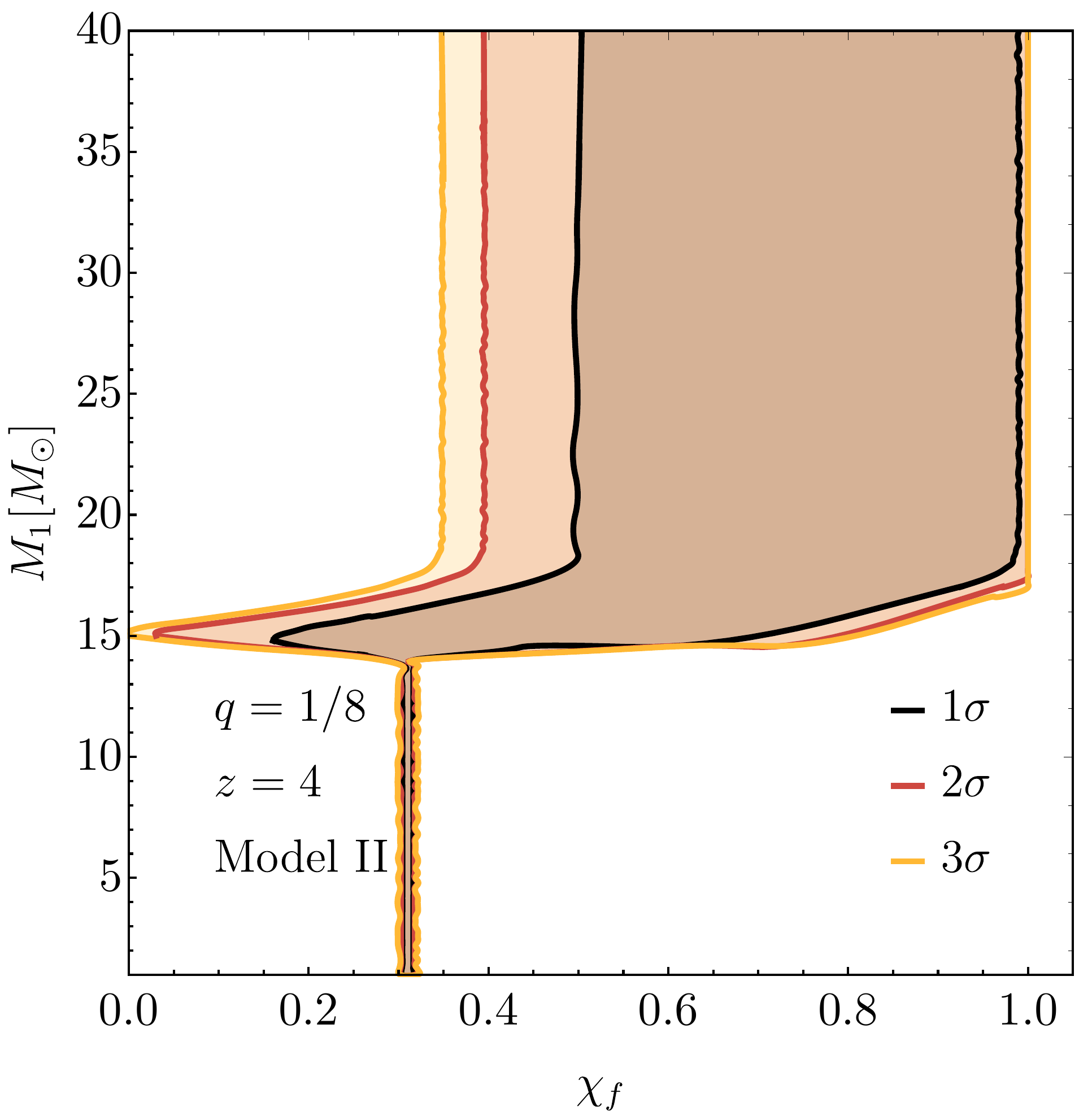}

	\vspace{.2cm}
	
	\includegraphics[width=0.39 \linewidth]{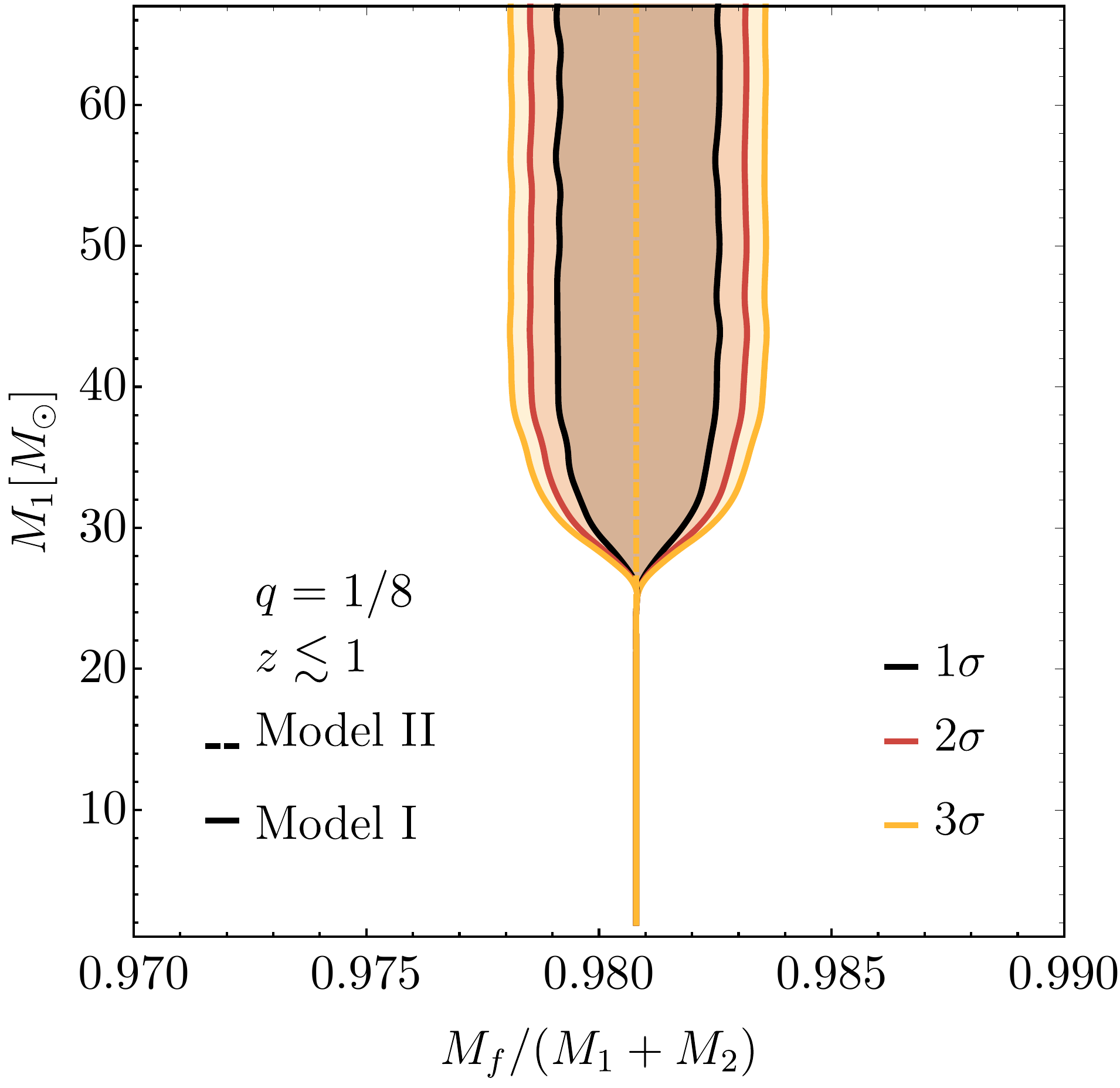}
    	\hspace{.25cm}
	\includegraphics[width=0.39 \linewidth]{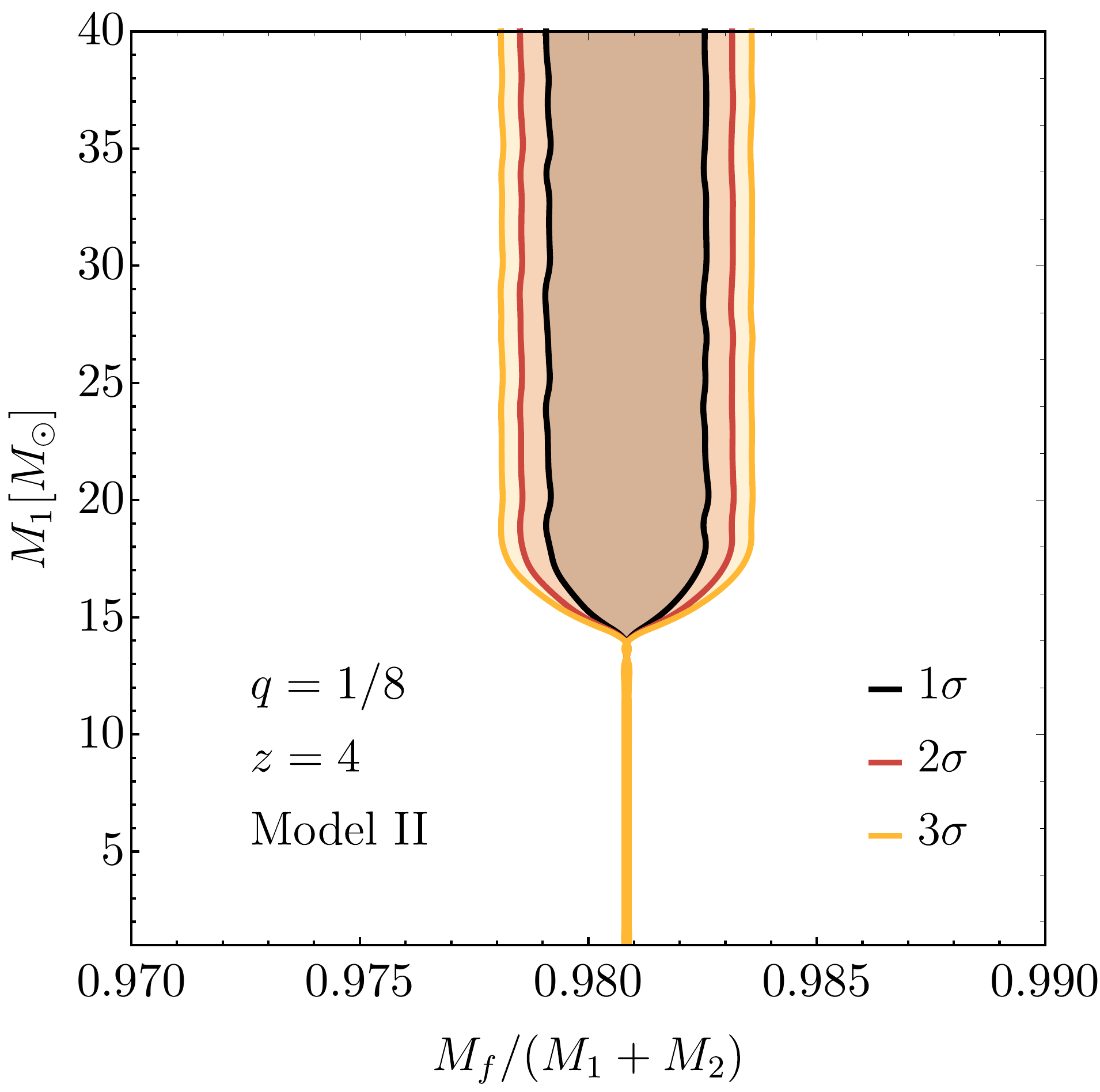}
    \caption{\it
    Same as in Fig.~\ref{pdf-Mchi} but for $M_1= 8 M_2$ (i.e., mass ratio $q=1/8$). 
    We have not plotted the observed data for $z \lsim 1$ because current events have a larger mass ratios.
	    }
    \label{pdf-Mchiq8}
   \end{figure}

In Figs.~\ref{pdf-Mchi}, \ref{pdf-Mchiq2}, and \ref{pdf-Mchiq8} we  provide the confidence intervals for given mass 
$M_1$ at 68\% ($1 \sigma$), 95\% ($2 \sigma$) and 99\% ($3 \sigma$) CL for the final mass, spin, and for the effective 
spin parameter of the merging binary.
We construct these distributions by fixing the mass ratio\footnote{The distribution of the mass ratio for different 
mass functions is discussed in Appendix~\ref{appmf}, see Fig.~\ref{pdfq}.} ($q=1$ 
in Fig.~\ref{pdf-Mchi}, $q = 1/2$ in Fig.~\ref{pdf-Mchiq2}, and $q = 1/8$ in Fig.~\ref{pdf-Mchiq8}, respectively) and, 
for each value of redshift $z$ and of 
the mass $M_1$ of the primary component of the binary, we draw the individual spin directions from an 
isotropic distribution whereas the spin magnitudes are obtained as $\chi_{i}=\chi_{i}(M_i,z)$ from the right panels 
of Figs.~\ref{Spin-I} and \ref{Spin-II} for Model~I and Model~II, respectively. 
The probabilities are evaluated at a fixed redshift $z \lsim 1$ for both Model~I and Model~II for the left panels and 
at $z = 4$ for Model II for the right panels.
We have also reported the data (with error bars) corresponding to the GW events detected so far in the first two 
observation runs of LIGO and Virgo (only for the plots with $q = 1$ and $q = 1/2$ which are consistent with the observed mass ratios). In particular, the blue data points refer to events with high 
statistical significance~\cite{LIGOScientific:2018mvr}, whereas green and red data points refer to the events 
discovered in Refs.~\cite{Zackay:2019tzo,Venumadhav:2019lyq}, which have a lower statistical significance. The red data 
points refer to GW151216 and GW170403, for which the measured value of $\chi_\text{\tiny eff}$ is 
significantly affected by the priors, in particular whether one assumes a uniform 
prior on $\chi_\text{\tiny eff}$ or an isotropic distribution for the directions of the 
individual spins with uniform magnitude of the latter~\cite{Huang:2020ysn}. The red and green data points in 
Fig.~\ref{pdf-Mchi} refer to the former assumption.

For Model I the results in Fig.~\ref{pdf-Mchi} show the general tendency of having peaked distributions for masses smaller than $\mathcal{O}(30)M_\odot$ and broader ones for bigger masses once
the transition from low to high spins has been reached, see Figs.~\ref{Spin-I} and \ref{Spin-II} for details on the transition. Even though we have chosen $z \lsim 1$ (left panels), such a tendency is maintained for all redshifts smaller than $z \sim 10$.
Interestingly, especially for $q\approx1$ (for $q \approx 1/2$ only low mass data and the corresponding values of the final spin are excluded as one can see from Fig.~\ref{pdf-Mchiq2}), the predicted distributions in case of accreting PBHs are consistent 
with the observed distribution of the GW events detected so far \cite{byrnes}, especially when including those obtained in 
Refs.~\cite{Zackay:2019tzo,Venumadhav:2019lyq}, which tend to have large effective spin for some high-mass binaries.
It is harder to predict such distribution in the case of binaries of astrophysical origin, although binaries formed 
in stellar clusters by dynamical capture might have larger $\chi_\text{\tiny eff}$ for larger total 
masses~\cite{Safarzadeh:2020mlb,safa}. For example, the predicted distribution in this case is in tension with a 
measurement $M_1+M_2\approx 60 M_\odot$ and $\chi_\text{\tiny eff}\approx 0.8$ such as that reported in 
Ref.~\cite{Zackay:2019tzo}.

For Model II one observes a similar behaviour for all the probabilities of the three parameters, but at 
higher redshift (we stress that the left and right panels of Figs.~(\ref{pdf-Mchi}-\ref{pdf-Mchiq8}) refer to $z \lsim 1$ 
and $z=4$, respectively). As previously discussed, for Model~II the effect of accretion on the spin in the present epoch for the range of masses of interest is small, 
which explains why in the left panels of Figs.~(\ref{pdf-Mchi}-\ref{pdf-Mchiq8}) ($z \lsim 1$) the distributions for 
Model~II appear as thin curves on the scale of the plots.
In the right panels we have instead chosen  the representative 
redshift $z = 4$, which corresponds to a transition mass $\mathcal{O}(15)M_\odot$, in order to highlight the tendency 
in the probability.
In Fig.~\ref{pdf-Mchiq2} the  same tendency shows up with an additional transition point at higher values of $M_1$ due 
to the crossing point from low to high spins of the second mass $M_2 = M_1/2$.

For low mass ratios, $q \approx 1/8$ in Fig.~\ref{pdf-Mchiq8}, there is a general tendency for which the mass and the 
spin of the lighter component of the binary
play a minor role in the determination of the spin parameters. In particular, $\chi_{\text{\tiny eff}}$ is mainly affected by $\chi_1$ and attains values close to zero and $\pm 1$ for the lightest and largest mass, respectively (the sign depending upon the spin orientation). This tendency manifests itself by having a much broader confidence interval in the high mass portion of Fig.~\ref{pdf-Mchiq8}. The final spin of the remnant BH is mainly inherited by the primary constituent of mass $M_1$ of the binary, and for small initial spins we find $\chi_f \sim 3.4 q -10 q^2 + \mathcal{O}(q^3)$ in the limit of small $q$ from Eq.~\eqref{eq:general}. This explains the shift of $\chi_f$ towards zero for smaller mass ratios. 
On the other hand, the distribution shifts towards higher values of $\chi_f$ in the case of a highly spinning primary, 
$\chi_f \in [1-6.2 q + \mathcal{O}(q^2), 1]$ for small $q$.
Finally, the rescaled final mass tends towards unity as $M_f/(M_1 + M_2) \sim 1-0.2 q + 
\mathcal{O}(q^2)$.

Overall, the effect of having higher binary-component spins is to make the distributions of 
$\chi_\text{\tiny eff}$, $M_f$, and $\chi_f$ broader (see Fig.~\ref{pdf}). Thus, our main results are robust 
against the value of the maximum spin reached through accretion, in particular they would be qualitatively the same 
also if the maximum value is $\chi \sim 0.9$, as suggested by magnetohydrodynamic simulations of relatively thick
disks~\cite{Gammie:2003qi}. In this case we expect the distributions to be slightly less broad.
   
\section{Conclusions}\label{secconcl}

In this paper we have discussed the cosmological evolution of the mass and spin of PBHs. Our results can be relevant in 
two contexts:
\begin{itemize}
 \item For the merger events detected so far by LIGO-Virgo, the effective spin parameter of the binary is compatible to 
zero, except possibly for few high-mass 
events~\cite{LIGOScientific:2018jsj,LIGOScientific:2018mvr,Zackay:2019tzo,Venumadhav:2019lyq,Huang:2020ysn}. We have 
shown that a primordial origin of these BH binaries could naturally explain this distribution, especially in the 
likely scenario in which accretion is quenched at $z\lesssim10$ due to structure formation~\cite{Ricotti:2007au}. 
Indeed, due to the redshift dependence of the accretion rate, PBHs with masses below ${\cal O}(30)M_\odot$ are likely 
non-spinning at any redshift, whereas heavier BHs can be nearly extremal up to redshift $z\sim10$, resulting in a 
broader distribution of the effective spin parameter, which is compatible with the observed distribution of the GW 
events detected so far. 
On the contrary, it is more challenging to explain such distribution and 
$(M$-$\chi_{\text{\tiny eff}})$ correlation in the case of binaries of astrophysical origin~\cite{Safarzadeh:2020mlb}.
\item Current bounds on PBH abundance assume that the mass distribution at the present epoch is the same as that at 
formation in the early universe. We have shown that accretion might significantly modify the mass and spin 
distributions in a redshift-dependent fashion. The implications of this effect for current constraints on PBHs will be 
discussed in a forthcoming work~\cite{followup}.
\end{itemize}
Upcoming results from LIGO-Virgo third observation run might reinforce 
or weaken these predictions, in particular whether light binaries
(mass of the binary components $M\lesssim 30M_\odot$) can have large
effective spin parameter or not.

Future detections will provide measurements not only of the effective spin of the binary, but also of the individual 
spins, with $30\%$ accuracy~\cite{TheLIGOScientific:2016pea}. This will allow to constrain the primordial formation 
scenario more accurately and possibly distinguish between different formation mechanisms and different accretion models. %
Indeed, the physics of accretion in the early universe is very rich~\cite{Ricotti:2007jk,Ricotti:2007au}. A natural 
extension of our work would be to refine the accretion model, for example considering also (relatively) thick accretion 
disks and other models of the accretion flow.
Overall, our results suggest that it would be crucial to correlate the mass and spin distributions with the 
redshift of the source, since the transition between non-spinning and highly-spinning BHs occurs at redshift $z\sim 10$. 
This will be possible with future GW detectors, such as ET, that will detect binary BHs up $z\lesssim 
100$~\cite{Hild:2010id}.


\acknowledgments
\noindent
We thank E. Barausse, V. Desjacques, F. K$\ddot {\rm u}$hnel, M. Maggiore, M. Ricotti, H. Veerm$\ddot{\rm a}$e and A. 
Zimmerman for interesting discussions, and E. Berti for 
useful comments on the draft.  
V.DL., G.F. and A.R. are supported by the Swiss National Science Foundation 
(SNSF), project {\sl The Non-Gaussian Universe and Cosmological Symmetries}, project number: 200020-178787.
P.P. acknowledges financial support provided under the European Union's H2020 ERC, Starting 
Grant agreement no.~DarkGRA--757480, under the MIUR PRIN and FARE programmes (GW-NEXT, CUP:~B84I20000100001), and 
support from the Amaldi Research Center funded by the MIUR program `Dipartimento di 
Eccellenza" (CUP:~B81I18001170001).


\appendix

\renewcommand\theequation{\Alph{section}.\arabic{equation}}

\section{Binary merger rates}
\label{app:formation}

In this Appendix we review the formation of PBH binaries and their related merging rate, see Ref.~\cite{sasaki} for a 
review.
There are two main formation mechanisms for PBH binaries, one taking place in the early universe, especially before 
matter-radiation equality, and the other taking place in the late time universe in the present-day halos.
For simplicity, we provide the estimates for equal masses binaries.

\subsection{Early-time PBH binaries}
A pair of neighboring PBHs of masses $M$ separated by a physical distance $x$ can decouple from the Hubble flow provided that their gravitational interaction is strong enough, i.e. $M x^{-3}  (z)> \rho (z)$, where $\rho$ represents the background cosmic energy density.
Expressing the quantities with respect to the ones at the  matter-radiation equality $z_{\rm eq}$, one finds that the decoupling occurs at $z_\text{\tiny dec}$ if
\be\label{zdec}
\frac{1+z_\text{\tiny dec}}{1+z_\text{\tiny eq}}=f_\PBH {\left( \frac{\bar x}{x} \right)}^3 -1 > 0,
\ee
where we denoted with $f_\PBH$ the fraction of PBHs in DM at that time, and the PBH physical mean separation ${\bar x}$ 
reads
\be
\label{meandis}
{\bar x} (z_\text{\tiny e})={\left( \frac{M}{\rho_\text{\tiny PBH}(z_\text{\tiny eq})} \right)}^{1/3}
=\frac{1}{(1+z_\text{\tiny eq}) f_\PBH^{1/3}} 
{\left( \frac{8\pi G}{3H_0^2} \frac{M}{\Omega_\text{\tiny DM}} \right)}^{1/3}.
\ee
Eq.~\eqref{zdec} shows that the characteristic formation redshift is of the order $z_\text{\tiny dec} > 10^4$ for the masses and $f_\PBH$ considered \cite{Ali-Haimoud:2017rtz}.

The initial infall motion of the PBHs can be affected by the surrounding, and especially the closest, PBHs, which can 
exert tidal forces and give angular momentum to the system, forming therefore a binary.
The semi-major axis $a$ and the eccentricity $e$ of the binary are given by 
\be
a=\frac{\alpha}{f_\PBH} \frac{x^4}{{\bar x}^3},~~~~~e=\sqrt{1-\beta^2{\left( \frac{x}{y} \right)}^6}, 
\label{ellipse}
\ee
where $y$ is the physical distance to the third PBH at $z_{\rm eq}$. In the following we will assume $\alpha = \beta =1$ \cite{ Nakamura:1997sm,Sasaki:2016jop} (for a more precise quantitative estimate of these coefficients see Ref.~\cite{Ioka:1998nz}).
Imposing the geometrical condition $x< y<{\bar x}$ one gets an upper bound on the eccentricity as
\be
e_{\rm max}=\sqrt{1-f_\PBH^{3/2} {\left( \frac{a}{\bar x} \right)}^{3/2}}. \label{emax}
\ee
Assuming a uniform probability distribution for both 
$x$ and $y$ in three dimensional space and converting it in terms of $a$ and $e$, one obtains
\be
\d P=\frac{3}{4} f_\PBH^{3/2} {\bar x}^{-3/2} a^{1/2} e {(1-e^2)}^{-3/2} \d a \d e. \label{pdf-ae}
\ee
Once the PBHs form a binary, their distance gradually shrinks due to the energy loss through GWs radiation
and eventually merge with a coalescence time given by \cite{Peters:1963ux, Peters:1964zz}
\be
t=Q a^4 {(1-e^2)}^{7/2},~~~~~Q=\frac{3}{170} {(GM)}^{-3}.  
\label{coalescing-t}
\ee
We can convert the probability distribution above in terms of the coalescence time and eccentricity, and then
integrate over $e$ for fixed time $t$. The probability that the coalescence occurs in the time interval $(t,t+\d t)$ can 
be then connected to the merger rate $R(t)$ through the PBH number density $n_\text{\tiny PBH}$ as (see also 
\cite{Raidal:2017mfl, Raidal:2018bbj, Wang:2019kaf})
\be
R (t)=\begin{cases}
	\frac{3}{58} n_\text{\tiny PBH} \bigg[ -{\left( \frac{t}{T} \right)}^{3/8}
	+{\left( \frac{t}{T} \right)}^{3/37} \bigg] \frac{1}{t}~~~~~~~~~~~~{\rm for}~t<t_c\\
	\frac{3}{58} n_\text{\tiny PBH} {\left( \frac{t}{T} \right)}^{\frac{3}{8}} \bigg[
	-1+{\left( \frac{t}{t_c} \right)}^{-\frac{29}{56}} f_\PBH^{-\frac{29}{8}} \bigg] 
	\frac{1}{t}~~~~~{\rm for}~t\ge t_c,
\end{cases}\label{dpt}
\ee
where $t_c=Q {\bar x}^4 f_\PBH^{\frac{25}{3}}$ 
and 
$T \equiv {\bar x}^4 Q/f^4_\PBH$. The corresponding merging rate is shown in Fig.~\ref{mrates}.

\subsection{Late-time PBH binaries}
\begin{figure}[h!]
	\centering
	\includegraphics[width=0.6 \linewidth]{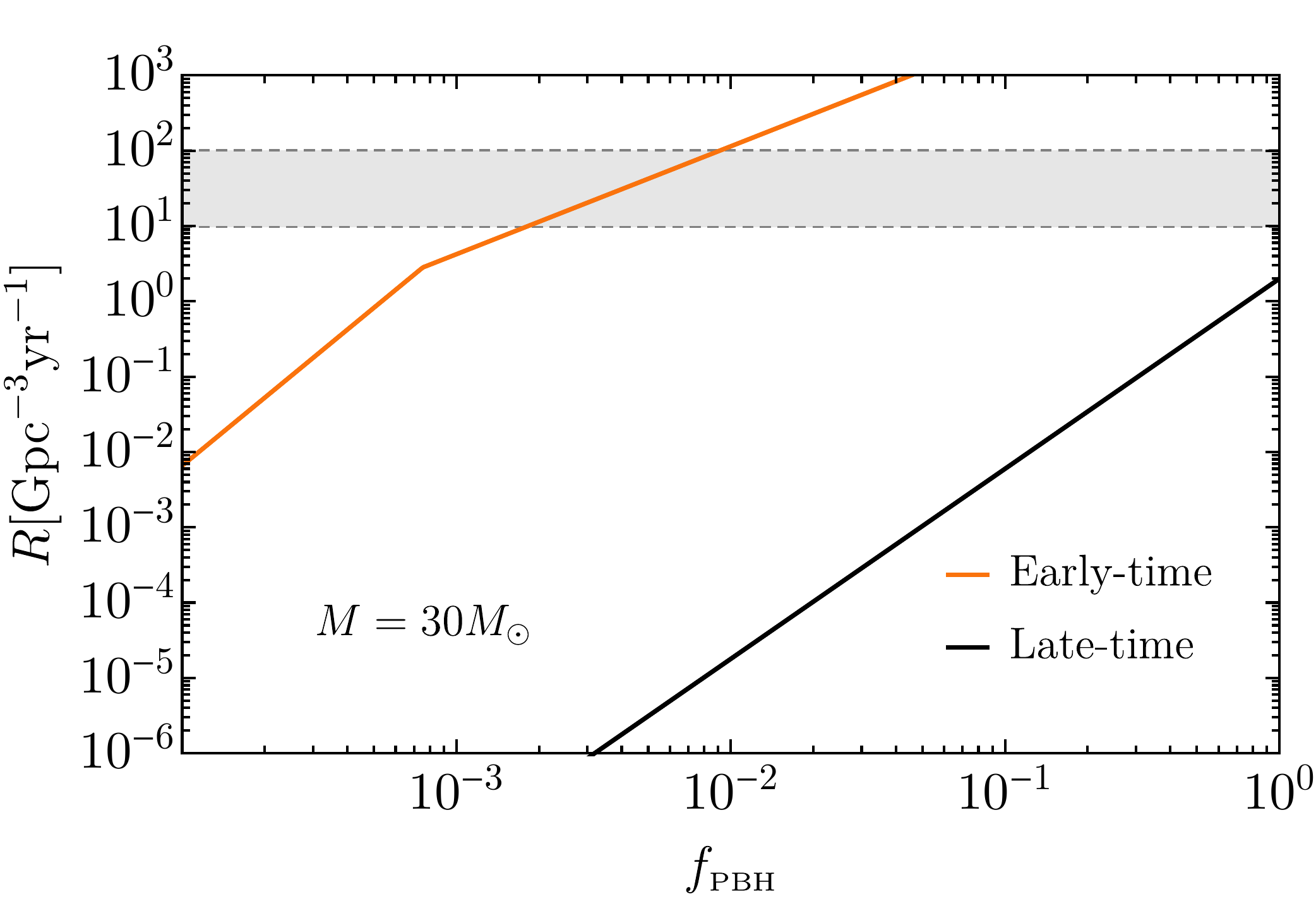}
	\caption{\it Estimate for the early- and late-time universe merger rate as a function of $f_\PBH$ for an equal mass merger with individual masses $M=30 M_\odot$. The grey band indicates the merger rate at $90\%$ CL observed by LIGO-Virgo collaboration \cite{LIGOScientific:2018mvr}.}
	\label{mrates}
\end{figure}

A second mechanism of formation of PBH binaries can take place in the present-day halos \cite{shapi,Bird:2016dcv}. If a 
PBH moving at a given velocity $v$ passes close to another PBH, the energy loss due to the sudden GW emission can make 
the former loose its kinetic energy becoming bound to the latter. The energy loss during the encounter can be estimated 
to be
\begin{equation}
\Delta E = \frac{85 \pi \sqrt{ G M} G^3 M^4}{12 r_\text{\tiny p}^{7/2}}\,,
\end{equation}
where $r_\text{\tiny p}$ is the periastron. Using the Newtonian approximation, for which the impact parameter is 
$b(r_\text{\tiny p}) = \sqrt{r_\text{\tiny p}^2 + 2 G M r_\text{\tiny p}/v^2}$, one finds the cross section for a 
binary 
formation 
\begin{equation}
\sigma_\text{\tiny bin} \simeq \lp\frac{85\pi }{3} \rp ^{2/7} \frac{\pi \lp2 G M \rp^2}{v^{18/7}}.
\end{equation}
Once formed, such a binary can merge in less than the age of the universe. The merger rate for a halo of mass $M_h$ can be computed as
\begin{equation}
R_h(M_h) = \int _0 ^{R_\text{\tiny vir}} \d r  2 \pi r^2 \lp \frac{\rho_\PBH(r)}{M}\rp ^2 \langle \sigma_\text{\tiny bin} v\rangle
\end{equation}
where $R_\text{\tiny vir}$ identifies the virial radius, $\rho_\PBH(r)$ is the PBH local density profile (typically taken to be the Navarro-Frenk-White profile) and the brackets stand for the mean value of the combination $\sigma_\text{\tiny bin} v$ computed using the Maxwell-Boltzmann velocity distribution. Finally, the total merger rate is found to be
\begin{equation}
R (t) = \int _{M_\text{\tiny min }} \d M_h\frac{\d  n }{\d M_h}R_h(M_h) 
\end{equation}
where $M_\text{\tiny min }$ is the minimum halo  mass \cite{sasaki}. The result is sensitive to the mass function $\d n 
/ \d M_h$ which can be estimated using Press-Schechter formalism \cite{Press:1973iz} or based on numerical simulations 
as in Refs.~\cite{sim1,sim2}. As shown in Fig.~\ref{mrates}, the final merger rate is orders of magnitude lower than 
its early universe counterpart, but one should take into account various uncertainties. Furthermore, for future 
experiments like ET which will have a much larger statistics, late-time universe binary mergers will be relevant.

\section{Mass accretion rate}
\label{accretionapp}
In the case of a dark halo clothing, we need to take into account the dark halo mass given in Eq.~\eqref{halo mass} 
and, if the typical size of the halo is smaller than the Bondi radius, 
then the accretion rate is the same as the one for a PBH of point mass $M_h$.
We can define the parameter
\be
\kappa \equiv \frac{r_\text{\tiny B}}{r_h} = 0.22\lp \frac{1+z}{1000}\rp \lp \frac{M_h}{M_\odot}\rp^{2/3} \lp 
\frac{v_\text{\tiny eff}}{{\rm km \, s^{-1}}} \rp^{-2}\,.
\ee
Different behaviors occurs when $\kappa\geq2$ or $\kappa<2$. In the former case the dark halo behaves the same as a 
point mass $M_h$ in terms of accretion rate, sonic 
radius and viscosity, with accretion rate given by~\cite{Ricotti:2007au}
\be
\dot{m} \equiv \frac{\dot{M}_\text{\tiny B}}{\dot{M}_\text{\tiny Edd}} = 0.023 \lambda \lp \frac{1+z}{1000}\rp \lp 
\frac{M}{M_\odot}\rp \lp \frac{v_\text{\tiny eff}}{5.74 \, {\rm km \, s^{-1}}} \rp^{-3},
\ee
where
\be
\lambda = {\rm exp} \lp \frac{9/2}{3 + \hat{\beta}^{0.75}} \rp x_{\rm cr}^2,
\ee
in terms of the sonic radius 
\be
x_{\rm cr} \equiv \frac{r_{\rm cr}}{r_\text{\tiny B}}= \frac{-1 + (1+ \hat{\beta})^{1/2}}{ \hat{\beta}}
\ee
and the gas viscosity parameter $\hat{\beta}$ given by
\be
 \hat{\beta} = \lp \frac{M}{10^4 M_\odot} \rp \lp \frac{1+z}{1000}\rp^{3/2} \lp \frac{v_\text{\tiny eff}}{5.74 \, {\rm 
km \, s^{-1}}} \rp^{-3} \llp 0.257 + 1.45 \lp \frac{x_e}{0.01}\rp \lp \frac{1+z}{1000}\rp^{5/2} \rrp,
\ee
as a function of the redshift, the PBH mass, effective velocity, and ionization fraction of the cosmic gas $x_e$.

If $\kappa <2$ one has instead to correct the quantities with respect to the naked case as
\be
\hat{\beta}^{h} \equiv \kappa^{\frac{p}{1-p}} \hat{\beta}, \quad \lambda^{h} \equiv \bar\Upsilon^{\frac{p}{1-p}} 
\lambda 
(\hat{\beta}^{h}), \quad r_{\rm cr}^h \equiv \lp \frac{\kappa}{2} \rp^{\frac{p}{1-p}} r_{\rm cr},
\ee
where $p = 2- \alpha$ and
\be
\bar\Upsilon = \lp 1 + 10 \hat{\beta}^h \rp^{\frac{1}{10}} {\rm exp} (2 - \kappa) \lp \frac{\kappa}{2} \rp^2.
\ee
The proper motion of PBHs strongly affects the dynamics of the accretion process and depends on the amplitude of the 
inhomogeneities of the DM and baryon fluids. Following Ref.~\cite{Ricotti:2007au}, one can estimate the 
relative velocity of PBHs with respect to the accreting baryonic matter assuming PBHs to behave like DM particles 
and identifying two main regimes of interest, the linear and non-linear regime.

In the linear regime before the decoupling redshift $z_{\rm dec}$, the Silk damping acts suppressing the growth of 
inhomogeneities on small scales, such that the PBH peculiar velocity is of order of the gas sound speed; for $z<z_{\rm 
dec}$, the gas flow lags behind the DM with a relative velocity $v_{\rm rel} = v_\text{\tiny DM}-v_{\text{\tiny b}}$, such 
that the PBH peculiar velocity follows a Maxwellian distribution with variance $\sigma = \langle v_{\rm rel} \rangle$ 
and has expectation value given by~\cite{Ricotti:2007au}
\begin{align}
	\langle v_\text{\tiny eff} \rangle_\text{\tiny A} &\sim c_s \lp \frac{16}{\sqrt{2\pi}} {\cal M}^3 \rp^{\frac{1}{6}} \theta 
({\cal M}-1) + c_s \lp 1 + {\cal M}^2 \rp^{\frac{1}{2}} \theta (1- {\cal M}),\nonumber \\
	\langle v_\text{\tiny eff} \rangle_\text{\tiny B} &\sim c_s {\cal M} \llp \sqrt\frac{2}{\pi} {\rm ln}\lp \frac{2}{e}{\cal M} 
\rp \rrp^{-\frac{1}{3}} \theta ({\cal M}-1) + c_s \lp 1 + {\cal M}^2 \rp^{\frac{1}{2}} \theta (1- {\cal M}),
	\end{align}
in terms of the Mach number ${\cal M} = \langle v_{\rm rel} \rangle/c_s$. Here the scenario A refers to a 
low efficient accretion rate, $\dot m < 1$, characterised by a spherical geometry, while scenario B refers 
to an efficient accretion rate, $\dot m > 1$, which supports the presence of an accretion disk, see 
Sec.~\ref{secacc}.

In the non-linear regime, the non-linear perturbations (halos) can prevent PBHs to accrete gas from the intergalactic 
medium, since they make PBHs fall in their potential wells with an enhanced velocity, which leads to a huge suppression 
of the gas accretion rate for an increasingly large fraction of the PBH population. To account for this possibility, in 
the main text we assumed two opposite and extreme models, dubbed as Model~I (in which accretion at $z<10$ is suppressed) 
and 
Model~II (in which accretion is sustained also when $z<10$).


\section{Effects of second-generation mergers and PBH mass functions}
\label{appmer}
The analysis presented in the main text ignores the possibility of second-generation mergers. Namely, a PBH binary 
might merge into a new BH which, at a later time, might undergo a further coalescence with another PBH, forming a 
new PBH binary which is eventually detected. We report here the formalism to compute the secondary mergers rates 
following the approach in Ref.~\cite{Liu:2019rnx,Wu:2020drm}\footnote{The analysis we adopt neglects the suppression factor on the merging rates due to the disruption of the binaries from the surrounding PBHs, which is found to be effective  only for $f_\PBH\gsim 0.01$ \cite{Raidal:2018bbj,Vaskonen:2019jpv}. Taking into account this effect would make the conclusion of this Appendix slightly stronger.}.

In the following we will assume that the fraction of PBHs as DM, $f_\PBH$, is bigger than a critical value $f_c$, above 
which the effects of the linear density perturbations are negligible on the merger rate of PBH binaries. The critical 
value reads
\be
f_c = 1.63 \times 10^{-4} \lp \frac{M_c}{M_\odot} \rp^{\frac{5}{21}} \lp \frac{t}{t_0} \rp^{\frac{1}{7}},
\ee
and is at most of order $f_c \sim 10^{-3}$ for the relevant range of masses at the present time $t = t_0$. Here $M_c$ 
is the reference mass associated to the scale re-entering the horizon.

\begin{figure}[t!]
    \centering
    \includegraphics[width=0.48 \linewidth]{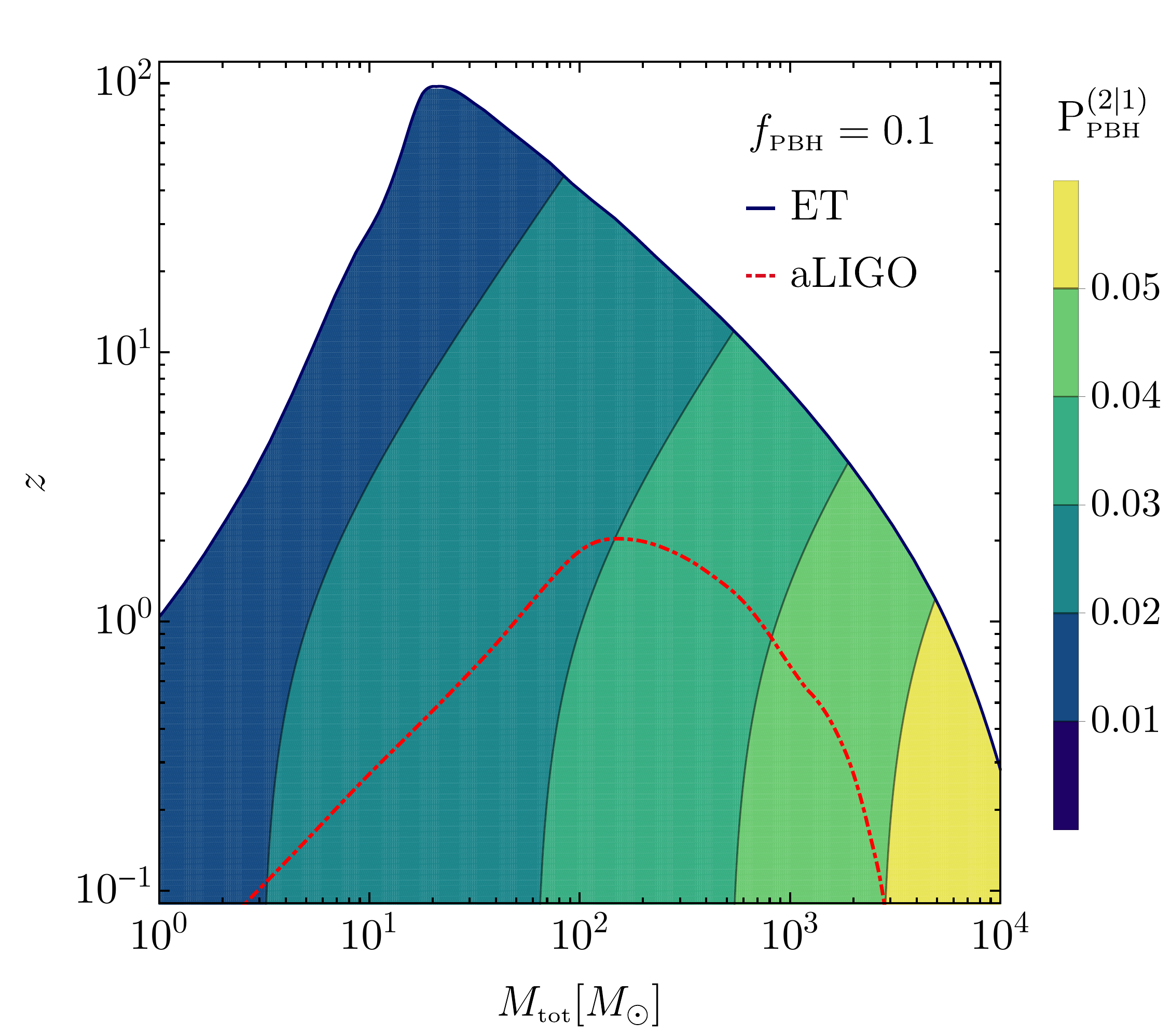}
        \hspace{.2 cm}
        \includegraphics[width=0.492 \linewidth]{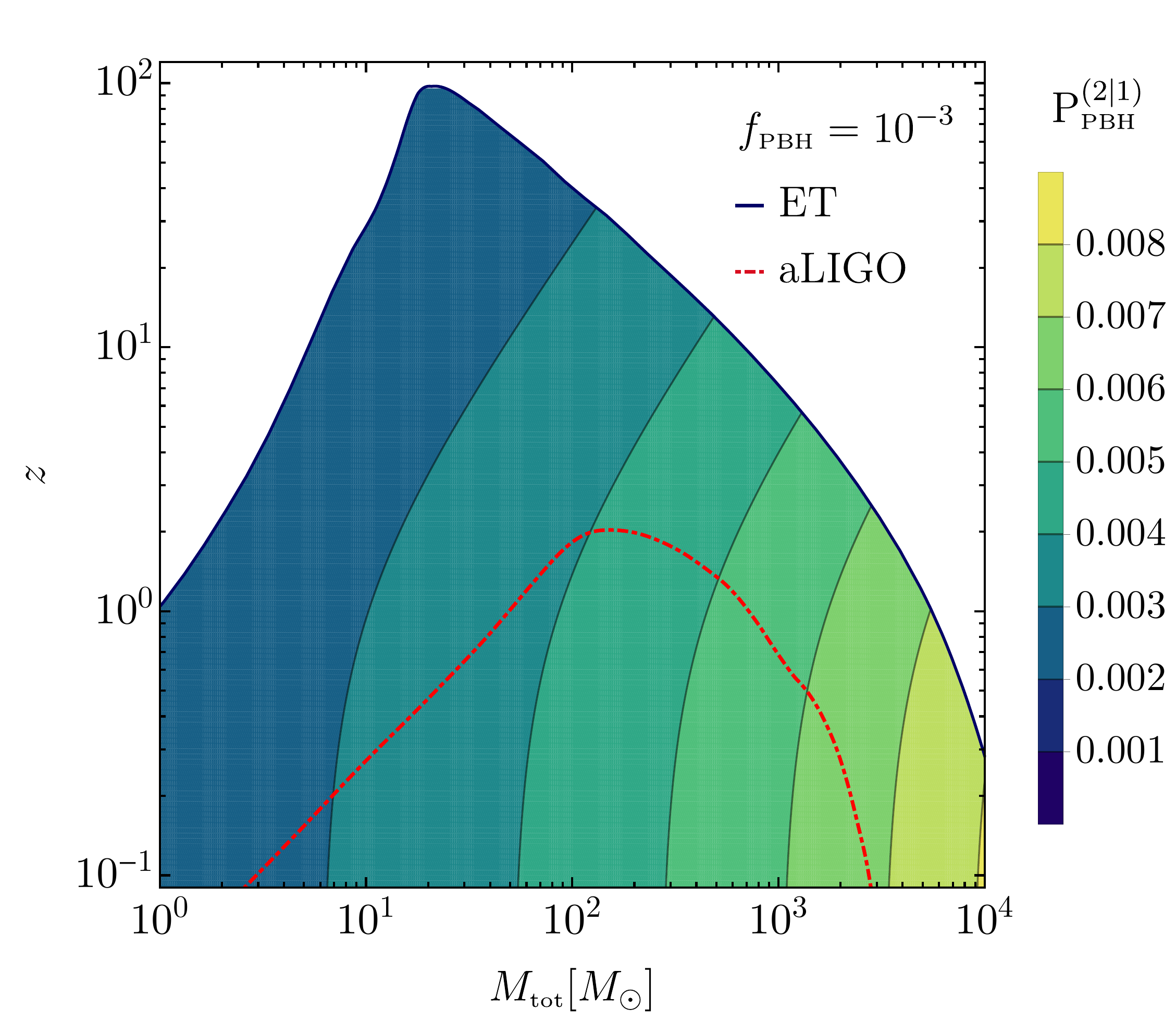}
    \caption{\it Total fraction of PBH binaries formed out of PBHs that have already merged before time 
$t(z)$ in terms of the total mass $M_\text{\tiny tot}$ of the binary and the redshift $z$, with fixed $f_\text{\tiny \rm PBH}$. We 
superimpose the estimated curves identifying the horizon within which aLIGO (red line) and ET (blue line) will be 
able to detect a merger.}
    \label{history}
   \end{figure}

We define the mass function identifying the fraction of PBHs with mass in the range $(M,M +\d M)$ as
\begin{equation}
\psi (M) = \frac{1}{\rho_\PBH} \frac{\d \rho_\PBH(M)}{M},
\end{equation}
normalised such that 
\begin{equation}
	\int \d M \,\psi (M) = 1.
\end{equation}
The fraction of the present average number density of PBHs with mass $M$ with respect to the total average is given by 
the expression
\be
F(M) = \frac{\psi (M)}{M} \llp \int \d {\rm ln} M' \, \psi (M') \rrp^{-1},
\ee
and the fraction of PBHs that have undergone a merging event before the time $t$ is given by~\cite{Liu:2019rnx}
\begin{align}
P_{\PBH}^{(1)} (t) &= 1.34 \times 10^{-2} \lp \frac{M_c}{M_\odot} \rp^{\frac{5}{37}} \lp \frac{t}{t_0} 
\rp^{\frac{3}{37}} f_\PBH^{\frac{16}{37}} \Upsilon_1,
\end{align}
where all the dependence on the shape of the mass function is given by the adimensional  factor
\begin{equation}
\Upsilon_1 =\lp \int \d {\rm ln} x\, \tilde \psi (x) \rp^{\frac{16}{37}}
 \int \d x_i \d x_j  \d x_l
\tilde F (x_i) \tilde F (x_j) \tilde F (x_l)
(x_i + x_j)^{\frac{36}{37}} x_i^{\frac{3}{37}} x_j^{\frac{3}{37}} x_l^{-\frac{21}{37}},
\end{equation}
where $\tilde F (x=M/M_c)= M_c F (M,M_c)$ and $\tilde \psi (x=M/M_c) = M_c \psi(M,M_c) $.
The fraction of PBHs that have merged in a  second-merger process at time $t$ is instead given by \cite{Liu:2019rnx}
\begin{align}
P_{\PBH}^{(2)} (t) &= 1.21 \times 10^{-4} \lp \frac{M_c}{M_\odot} \rp^{\frac{10}{37}} \lp \frac{t}{t_0} 
\rp^{\frac{6}{37}} f_\PBH^{\frac{32}{37}} \Upsilon_2,
\end{align}
in terms of the factor 
\begin{equation}
\Upsilon_2 =\lp \int \d {\rm ln} x\, \tilde \psi (x) \rp^{\frac{32}{37}}
 \int \d x_i \d x_j \d x_k \d x_l
\tilde F (x_i) \tilde F (x_j) \tilde F (x_k) \tilde F (x_l)
 (x_i + x_j)^{\frac{6}{37}} x_k^{\frac{6}{37}} x_l^{-\frac{42}{37}} (x_i + x_j + x_k)^{\frac{72}{37}}.
\end{equation}
The conditional probability that PBHs which have merged  are the results of a second-merger process at time $t$ is given by 
\begin{align}
P_{\PBH}^{(2|1)} (t) &= 9 \times 10^{-3} \lp \frac{M_c}{M_\odot} \rp^{\frac{5}{37}} \lp \frac{t}{t_0} 
\rp^{\frac{3}{37}} f_\PBH^{\frac{16}{37}} \Upsilon 
\qquad 
\text{with}
\qquad
\Upsilon = \frac{\Upsilon_2}{\Upsilon_1}.
\end{align}
Typically, various shapes of the mass fraction are considered in literature, corresponding to the most common formation 
scenarios of PBHs. In the following we will consider a critical, spiky, lognormal, and power-law mass functions. 
\begin{itemize}
\item {\bf Critical scaling mass function:} the mass function given by the one for the critical collapse as 
\cite{Niemeyer:1997mt, Yokoyama:1998xd} 
\be
\psi (M) = \frac{3.2}{M} \lp \frac{M}{M_c} \rp^{3.85} e^{- \lp \frac{M}{M_c} \rp^{2.85}},
\ee
and therefore 
\be
F(M) = \frac{2.85}{M} \lp \frac{M}{M_c} \rp^{2.85} e^{- \lp \frac{M}{M_c} \rp^{2.85}}.
\ee
The resulting value for the rescaled merger fraction is $\Upsilon = 6$.
\item{\bf Monochromatic mass function:} PBHs which have a monochromatic mass function are distributed as
\be
\psi (M) = \delta \lp M-M_c \rp,
\ee
i.e. they have all the same mass $M_c$.
The resulting value for the rescaled merger fraction is $\Upsilon = 4.8$. This case is the simplest 
configuration but it is unrealistic since even a monochromatic power spectrum of curvature perturbations
gives rise to a larger mass function (i.e. the ``critical scaling mass function" presented above) due to the dynamics 
of the critical collapse.

\item {\bf Lognormal mass function:} this represents a frequent parametrisation which describes the case of a PBHs population arising from a symmetric peak in the primordial power spectrum, see for example Ref.~\cite{Carr:2017jsz} and references  therein, being
\be
\psi (M) = \frac{1}{\sqrt{2\pi}\sigma M} {\rm exp} \lp - \frac{{\rm log}^2 (M/M_c)}{2 \sigma^2} \rp.
\ee
The resulting value for the rescaled merger fraction is $\Upsilon\sim (5 \div 15)$, depending on the width of the 
distribution $\sigma$.

 \item {\bf Power-law mass function:}  which is typically obtained when considering the time evolution for models with a broad power spectrum of the curvature perturbations, see Ref.~\cite{DeLuca:2020ioi},
 \be
\psi (M) =\frac{1}{2} \frac{M_c^{1/2}}{M^{3/2}} \Theta(M-M_c),
\ee
The resulting value for the rescaled merger fraction is $\Upsilon=4.75$.
\end{itemize} 
In Fig.~\ref{history} we plotted  $P_\text{\tiny PBH}^{(2|1)}(z)$ in terms of the total mass $M_\text{\tiny tot}$ of the 
observed binary and the redshift assuming a certain value for the fraction of PBHs as DM $f_\PBH$. We superimpose the 
curves representing the horizon at which aLIGO and ET experiments will be able to detect a merger at that total mass 
and redshift, assuming equal-mass and non-spinning binaries, see  Refs.~\cite{Sathyaprakash:2019nnu,Maggiore:2019uih}.  
The plotted results are found assuming a critical scaling mass function. Other mass functions 
typically used in literature would give rise to ${\cal O} (1)$ corrections to the result.

One can conclude that only a tiny fraction of BHs  detectable through their merger at aLIGO or ET have suffered a 
previous merger. 
This implies that the spin of each binary component is the one at primordial formation, plus possibly its accretion 
contribution, as discussed in the main text.


\section{The evolution of the mass function}
\label{appmf}
\begin{figure}[t!]
 \centering
 \includegraphics[width=0.45 \linewidth]{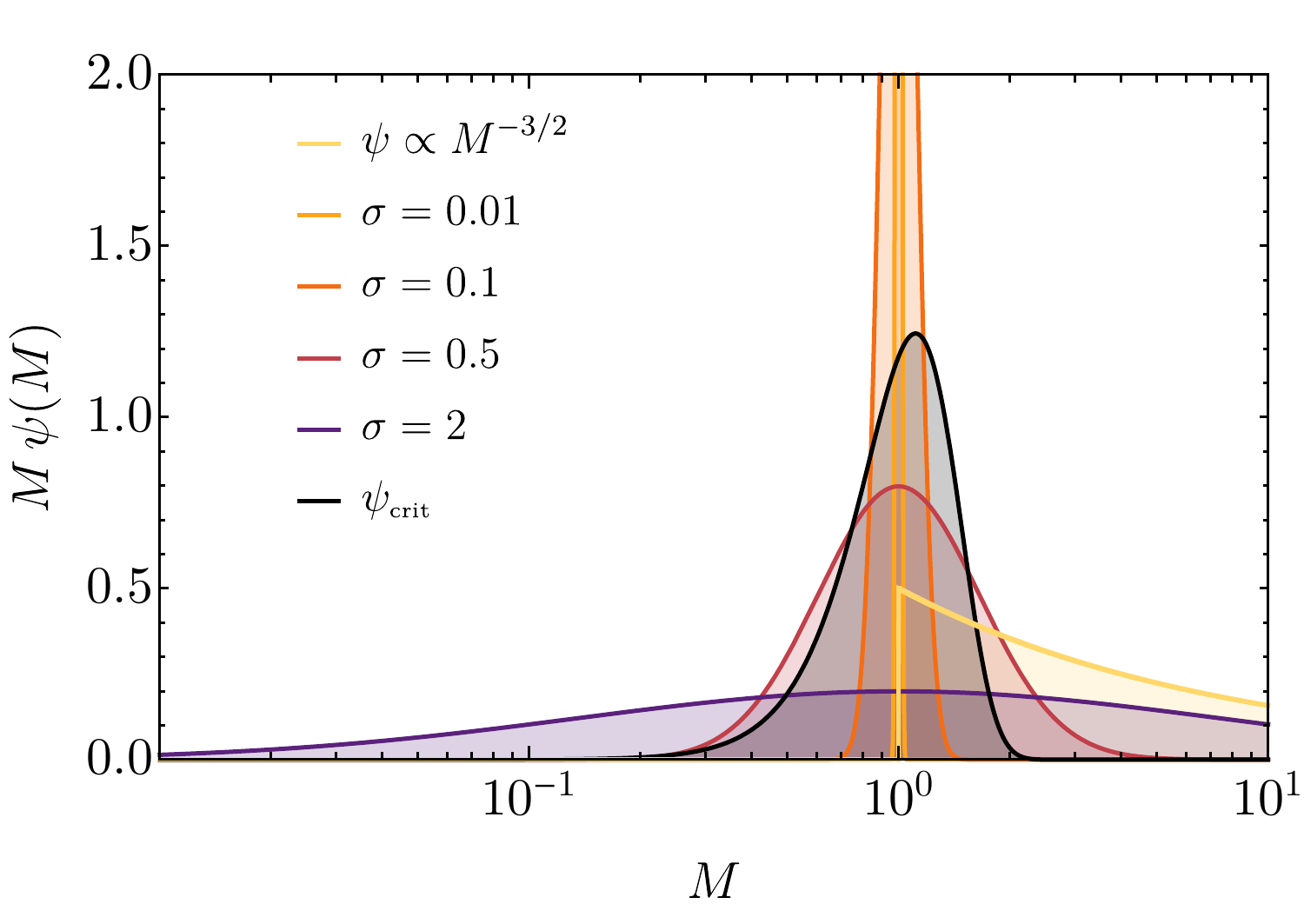}
     \hspace{.2 cm}
    \includegraphics[width=0.44 \linewidth]{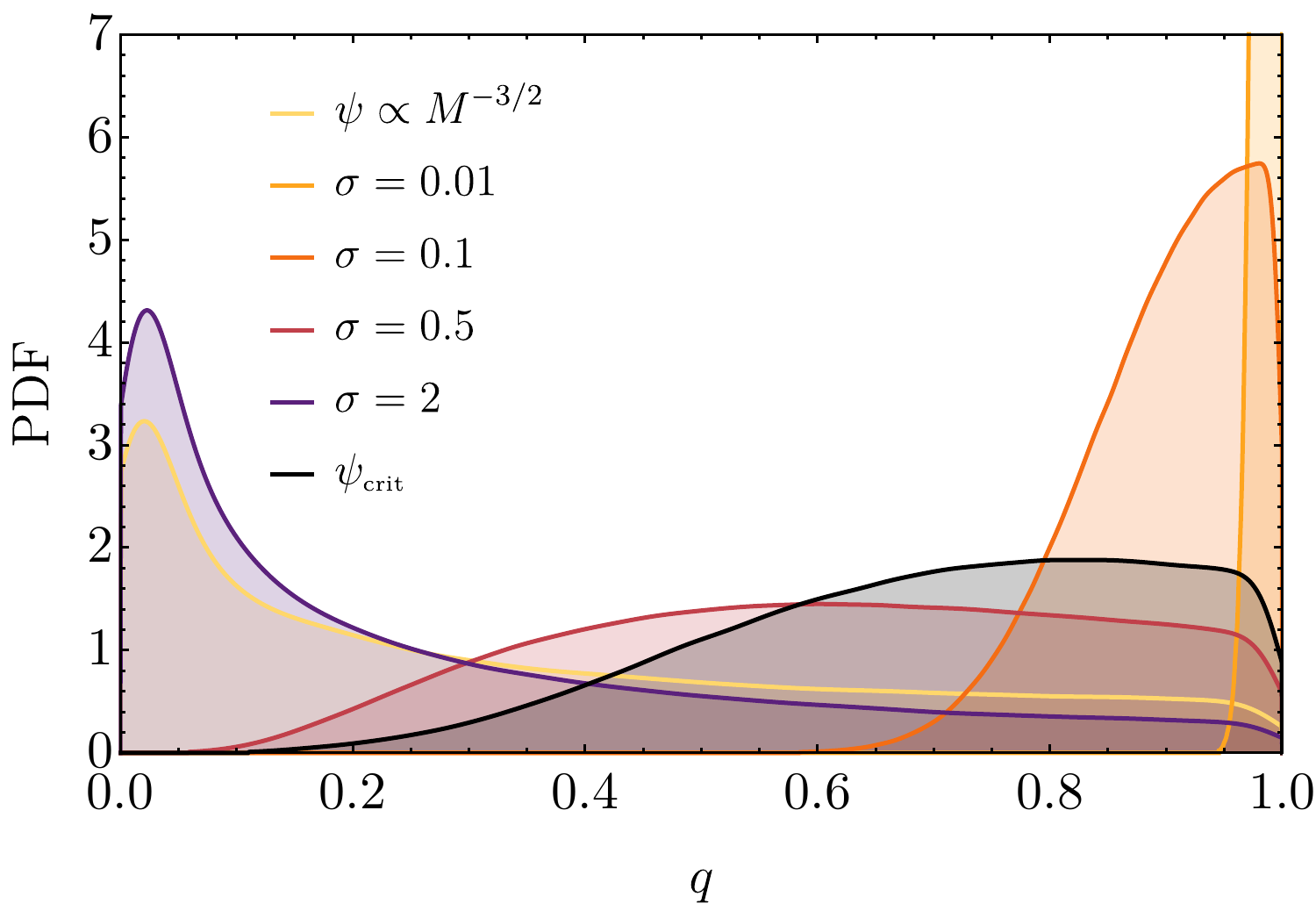}
    \caption{\it Left: Various mass functions $\psi(M)$ considered in this work. We set $M_c=1$.
    Right: Corresponding distributions of the mass ratio $q = M_2/M_1$ of a PBH binary.
    }
    \label{pdfq}
   \end{figure}

In this appendix we discuss how the mass function changes when the mass evolution is taken into account. To give a 
quantitative estimate of this effect we will consider some of the mass functions analysed in the previous appendix, 
which are summarised in the left panel of Fig.~\ref{pdfq}.
The resulting distribution of the mass ratio $q=M_2/M_1\leq1$ for a random pair of masses ($M_1$, $M_2$) in each case is 
plotted in the right panel of Fig.~\ref{pdfq}. One can notice how peaked mass functions give a distribution for the 
mass ratio peaked close to unity, while broader mass functions favour mass ratios closer to zero.

The evolution of the mass function can be computed analytically once the mass evolution is taken into 
account. 
Following the evolution of the mass accretion rate (see Fig.~\ref{Spin-I} for Model I and 
Fig.~\ref{Spin-II} for 
Model II) one can start with a certain mass at high redshift and evolve it into a different mass as
\begin{equation}
M \rightarrow f(M,z).
\end{equation} 
Correspondingly, defining $\psi(M)$ as the mass function at the formation time, the resulting mass function at redshift 
$z$ will be
\begin{equation}
\psi(M,z) = \psi(M') \lp \frac{\d f(M',z)}{\d M'}\rp^{-1} \bigg | _{M' \to f^{-1} (M,z)}.
\end{equation}
One can check explicitly that the unitary normalisation is maintained. 

\begin{figure}[t!]
    \centering
    \includegraphics[width=0.35 \linewidth]{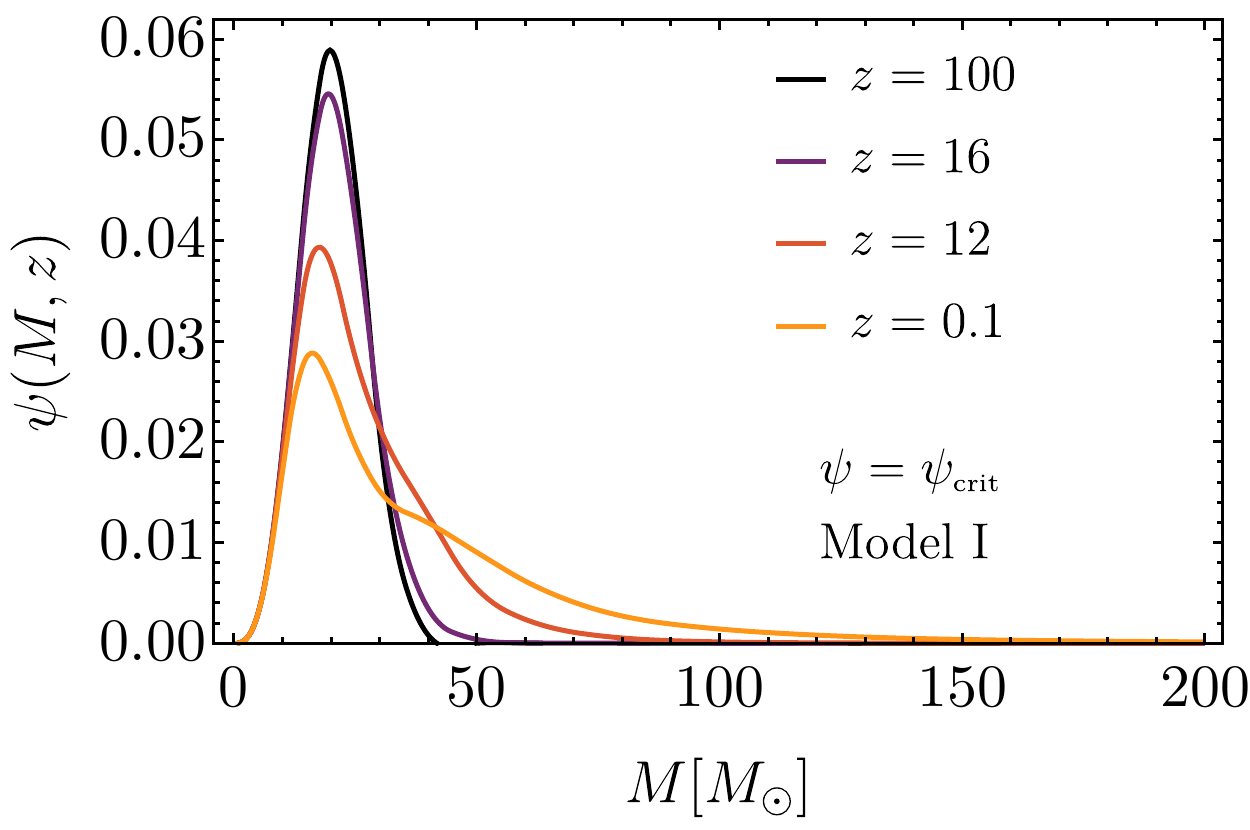}
    \hspace{.2 cm}
    \includegraphics[width=0.35 \linewidth]{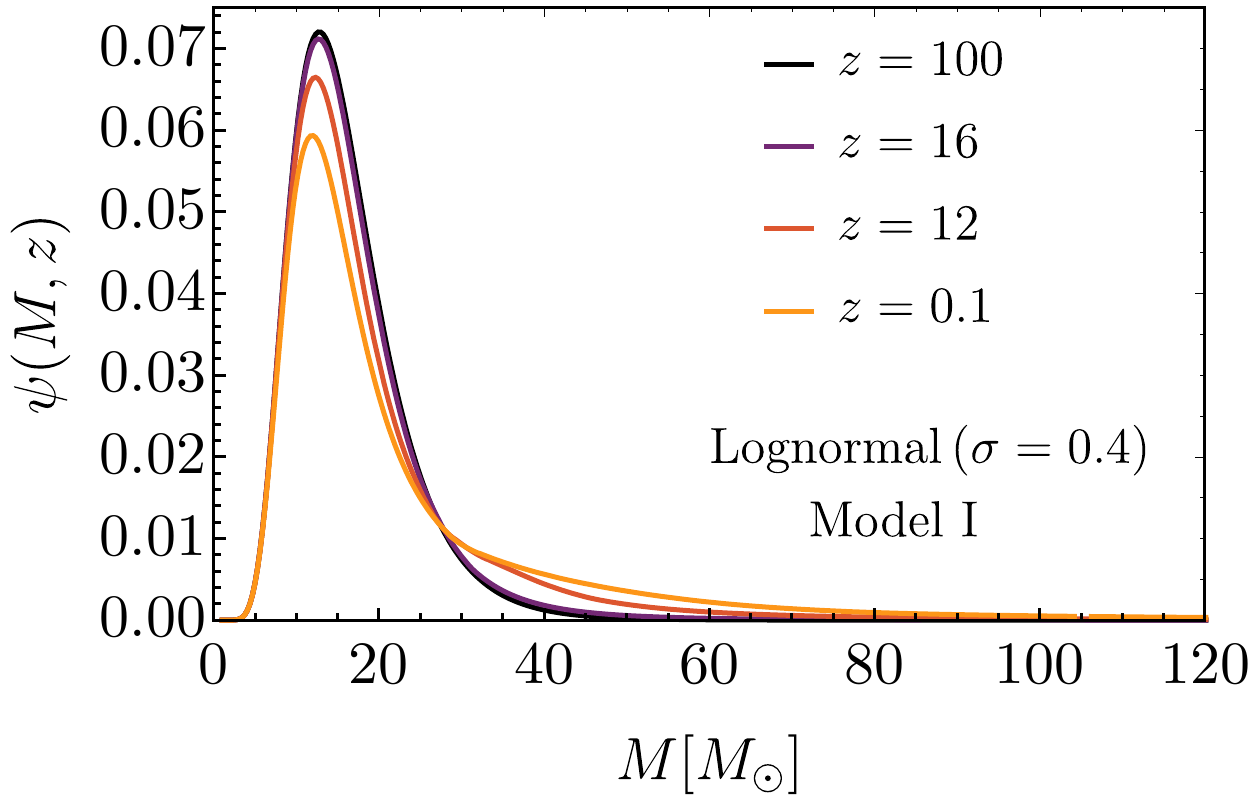}
    
    \vspace{.2cm}
    
    \includegraphics[width=0.35 \linewidth]{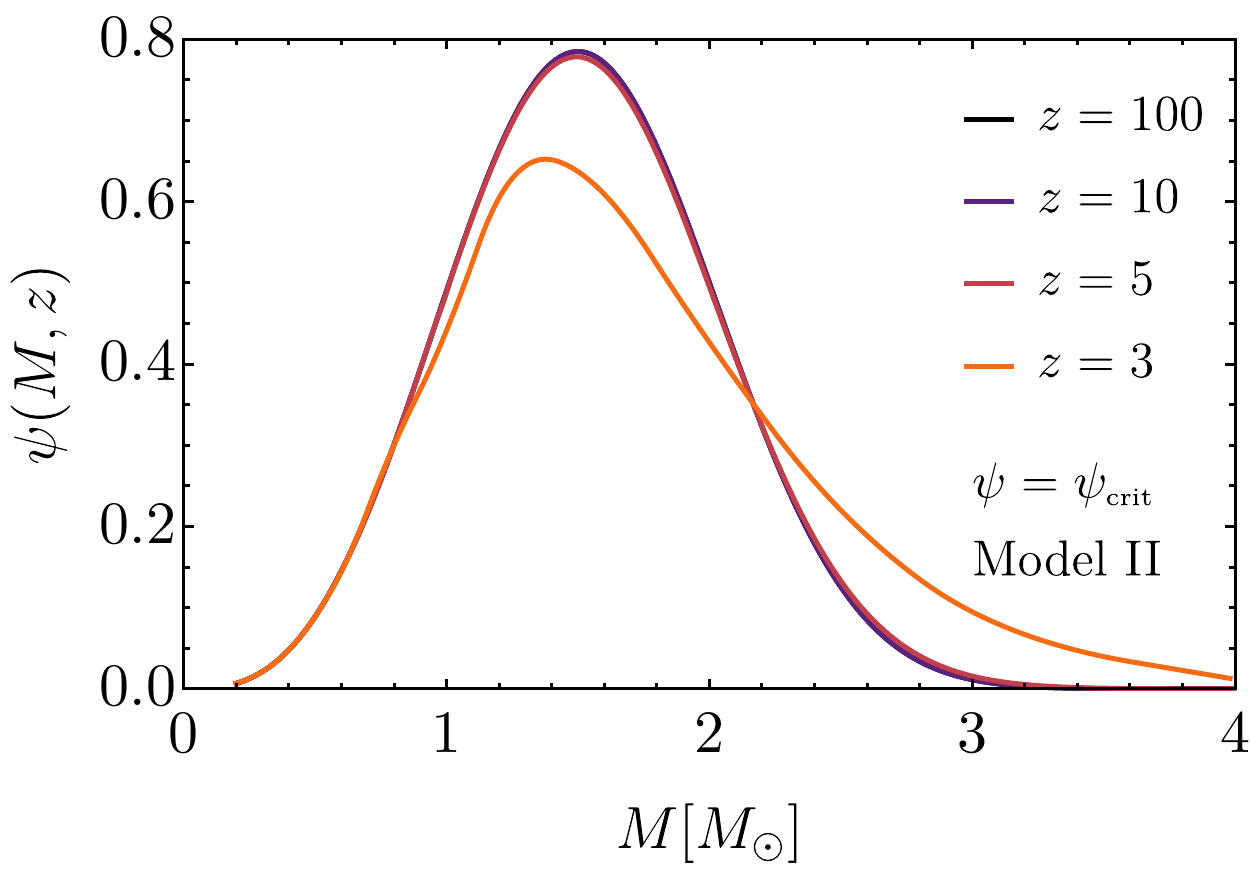}
        \hspace{.2 cm}
    \includegraphics[width=0.35 \linewidth]{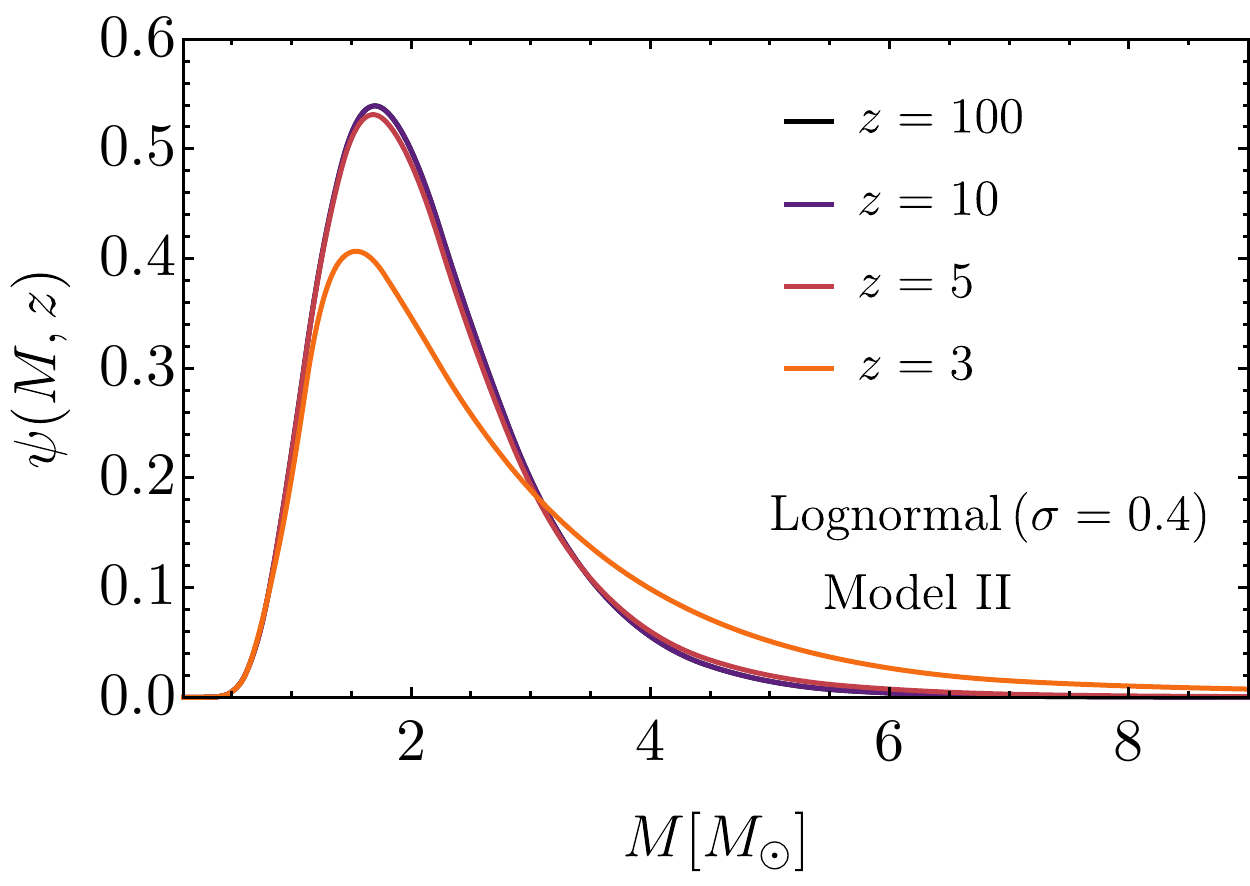}
    \caption{\it Top: Examples of evolution of the mass function in Model I.
    Bottom: Examples of evolution of the mass function in Model II.
    }
    \label{psiev-IeII}
   \end{figure}

Fig.~\ref{psiev-IeII} shows the evolution of the mass function for Model I (top panels) and Model II (bottom panels), 
for the choices of a critical and lognormal shape at the formation time. As the evolution proceeds, the height of the 
mass function decreases with a consequent increase of the tail of the distribution due to the effect of the accretion 
processes, even though the peak is maintained at almost the same position in the mass range.


\end{document}